\PassOptionsToPackage{table}{xcolor}
\documentclass[
twocolumn,
 amsmath,amssymb,
 aps,
]{revtex4-2}

\usepackage{caption}
\usepackage{enumitem}
\usepackage{tabularx}
\usepackage{placeins}
\usepackage{booktabs}
\usepackage{graphicx}
\usepackage{capt-of} % for \captionof
\usepackage{dcolumn}% Align table columns on decimal point
\usepackage{bm}% bold math
\usepackage{xcolor}

\usepackage{tcolorbox}
\tcbuselibrary{skins,breakable}

\usepackage{tikz}
\usepackage{ragged2e}
\newcolumntype{M}[1]{>{\centering\arraybackslash}m{#1}}
\newcolumntype{C}[1]{>{\centering\arraybackslash}p{#1}}
\definecolor{strong}{HTML}{C6EFCE} % Light green
\definecolor{moderate}{HTML}{FFEB9C} % Light yellow
\definecolor{weak}{HTML}{FFC7CE} % Light red
\newcounter{infobox}
\renewcommand{\theinfobox}{\Roman{infobox}}

\newcommand{\lgel}[1]{\textcolor{black}{#1}}

\begin{document}
   
\preprint{APS/123-QED}

\title{Data-driven discovery of dynamical models in biology}% Force line breaks with \\

\author{Bartosz Prokop}
\author{Lendert Gelens}%
 \email{lendert.gelens@kuleuven.be}
\affiliation{Laboratory of Dynamics in Biological Systems, Department of Cellular and Molecular Medicine, KU Leuven, Herestraat 49, 3000 Leuven, Belgium
}%

\date{\today}% It is always \today, today,

\begin{abstract}
Dynamical systems theory provides a mathematical framework for describing how interacting biological components evolve over time and space, from molecular oscillators to large-scale biological patterns. Such systems often involve nonlinear feedbacks, delays, and multiscale interactions, making mechanistic model construction increasingly challenging as experimental measurements become richer and higher-dimensional. This has motivated the development of data-driven approaches that infer model structure directly from data, offering alternative routes to constructing dynamical models.
In this review, we discuss and compare data-driven approaches for model discovery in biological dynamical systems, focusing on three major methodological families: regression-based methods, network-based architectures, and decomposition techniques. We compare how these approaches address three core objectives: forecasting future behavior, identifying interactions between system components, and characterizing qualitative dynamical solutions such as steady states, oscillations, and transitions between them. To enable a direct comparison, representative methods are applied to a common benchmark—the Oregonator model—a minimal nonlinear oscillator that captures shared design principles of chemical and biological systems. By highlighting practical strengths, limitations, and degrees of interpretability, this review aims to guide researchers in selecting appropriate tools for analyzing complex, nonlinear, and high-dimensional biological dynamics.
\end{abstract}

\maketitle

\section{Introduction}
\label{sec:general_introduction}
\begin{figure*}
    \centering
    \includegraphics[width=\linewidth]{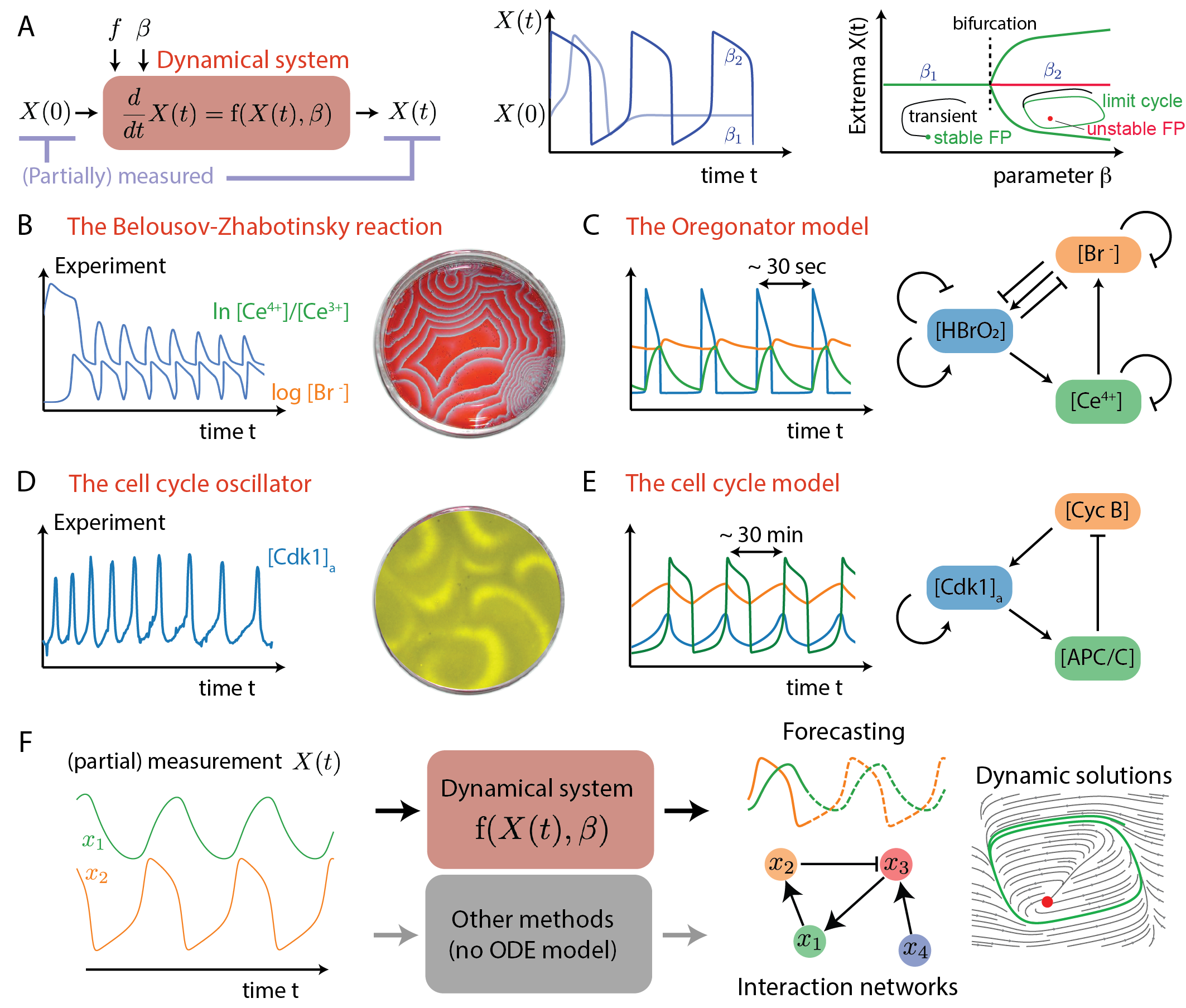}
    \caption[From data to dynamical systems in chemistry and biology]{\textbf{From data to dynamical systems in chemistry and biology.}
    \textbf{A.} General representation of a dynamical system. The state vector $X(t)$ evolves according to governing function $f$ and parameters $\beta$. Small parameter changes can lead to qualitatively different time series, summarized in a bifurcation diagram where a stable fixed point loses stability and gives rise to a limit cycle.  
    \textbf{B–C.} The Belousov–Zhabotinsky (BZ) reaction: experimental recordings (Adapted with permission from \cite{Field1972}) and spatial patterns (Reproduced with permission from \cite{morris2009belousov-zhabotinsky-reaction}) (B), alongside the Oregonator model (C), which reduces the chemistry to three interacting intermediates ($[\mathrm{Br}^-]$, $[\mathrm{HBrO}_2]$, $[\mathrm{Ce}^{4+}]$).  
    \textbf{D–E.} The embryonic cell cycle oscillator: experimental observations of Cdk1 dynamics (Adapted with permission from \cite{rombouts2025mechanistic}) and mitotic (spiral) waves (Adapted with permission from \cite{Cebrian-Lacasa2024}) (D), together with a reduced three-variable model \cite{Parra-Rivas2023} (E) capturing the interactions among cyclin B, active cyclin B–Cdk1 (Cdk1$_a$), and APC/C.  
    \textbf{F.} From data to understanding. Ideally, one recovers the full governing equations $f(X(t),\beta)$, enabling \textbf{forecasting}, reconstruction of \textbf{interaction networks}, and identification of \textbf{dynamic solutions}. Alternatively, data-driven approaches can aim directly at these three complementary outcomes.}
    \label{fig:understandingdynamicalsystems}
\end{figure*}

Dynamic behavior, expressed as changes over time and space, is fundamental to natural systems.
From chemical reactions to biological regulation, feedback and nonlinearity give rise to a rich repertoire of dynamics, including switching, adaptation, oscillations, and spatial waves.
Mathematical models of dynamical systems provide a unifying language to describe and analyze such behaviors across disciplines.
A dynamical system specifies how a state vector $X(t)$, representing the variables that define the system’s current state, evolves according to governing rules $\mathrm{f}(X(t), \beta)$ and parameters $\beta$ (Fig.~\ref{fig:understandingdynamicalsystems}A).
Small parameter changes can qualitatively alter system trajectories, for example by changing a stable steady state into sustained oscillations through a bifurcation~\cite{Strogatz2000}.

Biological systems exhibit diverse dynamical behaviors across scales~\cite{tyson2003sniffers,goldbeter2012systems,gelens2014spatial,beta2017intracellular,deneke2018chemical,volpert2009reaction,goldbeter2018dissipative}.
Purely temporal examples include the cell cycle oscillator that governs cell division~\cite{ferrell2011modeling,Novak1993}, circadian clocks that synchronize physiology with the day--night cycle~\cite{dunlap1999molecular,bell2005circadian}, intracellular calcium oscillations that regulate fertilization and neuronal signaling~\cite{berridge1988cytosolic,borghans1997complex}, glycolytic oscillations in yeast metabolism~\cite{ghosh1964oscillations,chandra2011glycolytic}, p53 oscillations in the DNA damage response~\cite{geva2006oscillations,batchelor2008recurrent,purvis2012p53}, and neuronal excitability, where action potential pulses arise from threshold-triggered ionic currents~\cite{hodgkin1939action,HodgkinHuxley,izhikevich2007dynamical}.
Spatiotemporal examples include mitotic waves in early embryonic development~\cite{chang2013mitotic,Nolet2020,afanzar2020nucleus,Puls2024,Huang2024TriggerWaves}, cyclic AMP waves in social amoebae during aggregation~\cite{devreotes1983quantitative,tyson1989cyclic}, and traveling waves of Erk activity in regenerating tissues~\cite{de2021control}.
Across these systems, nonlinear feedbacks, time delays, and spatial coupling jointly generate complex patterns in both time and space. Spatial coupling can transform local oscillations into traveling wave patterns, providing additional constraints and structure that are visible in spatiotemporal measurements (Fig.~\ref{fig:understandingdynamicalsystems}B,D).

To study these phenomena in tractable settings, researchers often turn to simplified systems that capture the same underlying dynamical principles.
Oscillations and waves arise in both chemical and cellular contexts, providing canonical examples in which temporal rhythms and spatial patterning are tightly linked.
Two well-known oscillatory systems illustrate this connection (Fig.~\ref{fig:understandingdynamicalsystems}B–E).
In both cases, oscillatory dynamics observed in time are accompanied by spatial wave patterns that reveal how the same underlying feedback architecture coordinates behavior across extended domains.
In the Belousov–Zhabotinsky (BZ) reaction, periodic color changes and traveling waves emerge from nonlinear chemical feedbacks \cite{Winfree1984,Field1972,morris2009belousov-zhabotinsky-reaction}.
The minimal Oregonator model developed by Field and Noyes \cite{field1974oscillations} and normalized by Tyson \cite{tyson1985quantitative} reduces this chemistry to three interacting intermediates, whose mutual activation and inhibition generate rhythmic oscillations.
A closely analogous dynamical architecture governs the eukaryotic cell cycle oscillator: in early \textit{Xenopus} embryos, interlinked positive and negative feedbacks involving cyclin~B, Cdk1, and APC/C drive mitotic oscillations and propagating mitotic waves.
Detailed model formulations are provided in the \ref{eqn:BZ}.
Despite their distinct molecular components, both systems share a common dynamical blueprint, generating limit cycle oscillations through a Hopf bifurcation \cite{Strogatz2000,novak2008design,tyson2020dynamical,de2021modular,Parra-Rivas2023}.
When such oscillatory systems are spatially extended, the same underlying feedback interactions can give rise to spatial patterns and waves through diffusion-driven instabilities or wave-propagation mechanisms. 
These examples highlight that understanding a dynamical system involves several complementary objectives (Fig.~\ref{fig:understandingdynamicalsystems}F):
\begin{description}
\item[\textbf{Forecasting}] Predicting future system states from current measurements.
\item[\textbf{Interactions}] Identifying the (directed) network of variables and feedbacks responsible for observed behavior.
\item[\textbf{Dynamic solutions}] Characterizing the system’s repertoire of solutions, including steady states, limit cycles, and transitions between them.
\end{description}
Inferring the full governing equations $\mathrm{f}(X(t),\beta)$ would, in principle, address all three objectives.
In practice, however, this becomes infeasible for high-dimensional biological systems with incomplete observability and uncertain interactions.
Modern data-driven approaches therefore aim to reconstruct one or more aspects of system dynamics directly from time-resolved data, providing a bridge between mechanistic modeling and empirical inference.
Historically, the study of dynamical systems relied primarily on equation-based modeling, in which governing laws are derived from first principles or mechanistic intuition, as in Newton’s laws or kinetic models of chemical oscillations.
This approach remains powerful for low-dimensional, well-characterized systems, but becomes increasingly intractable for large biological networks \cite{Arnold1978,Strogatz2000}.
Advances in experimental measurement and computation have enabled a complementary paradigm: inferring dynamical structure directly from data \cite{Brunton2022book}.
Machine-learning and statistical methods, ranging from nonlinear time-series reconstruction \cite{Takens1981,Abarbanel1990,Casdagli1989,Mindlin1992,GRASSBERGER2012,Bucci2023} to modern neural-network–based discovery \cite{Alber2019,Krizhevsky2012}, can now extract predictive structure without fully specified mechanistic models.
In biology, these techniques are increasingly applied to live-cell imaging, single-cell dynamics, and omics time courses \cite{Hey2009,Alber2019}.

\lgel{
In this review, we focus on methods that learn directly from time-resolved biological data to uncover the governing dynamics of complex systems, with an emphasis on oscillatory processes as canonical examples of nonlinear biological behavior.
Although many
biological systems are governed by spatially extended
(e.g. reaction-diffusion) models, in this review we focus on approaches for discovering effective low-dimensional dynamical models (ODEs and latent-state
representations) from time-resolved data, rather than on
the identification of partial differential equations (PDEs).}
We compare representative regression-based, network-based, and decomposition-based approaches, highlighting their respective strengths and limitations using canonical biological models.
Most comparisons are illustrated using the Oregonator model (Supplementary Note~2), which captures essential nonlinear oscillatory behavior while remaining analytically and computationally tractable.
Supplementary Notes~3–4 provide additional details, and all code used in this review is available on GitLab~\cite{gitlab_review2025}. 
Supplementary Note~3 examines the robustness and identifiability of regression- and network-based methods under increasingly realistic data conditions, while Supplementary Note~4 focuses on explicit Koopman operator inference and decomposition applied to the same benchmark system.

Our aim is to provide researchers in biology, chemistry, and related fields with an accessible overview of available tools, guidance on their applicability, and perspectives on future directions for data-driven model discovery in the natural sciences.
We focus on deterministic methods and only briefly mention probabilistic frameworks \cite{Kwiatkowska2008,JianyongSun2012}, such as Bayesian inference, when relevant.
In particular, we highlight complementary approaches that emphasize dynamical structure, interpretability, and connections to Koopman theory—methods that remain less familiar to many life science researchers despite their growing potential.

\begin{figure*}[t!]
	\centering
	\includegraphics[width=1\linewidth]{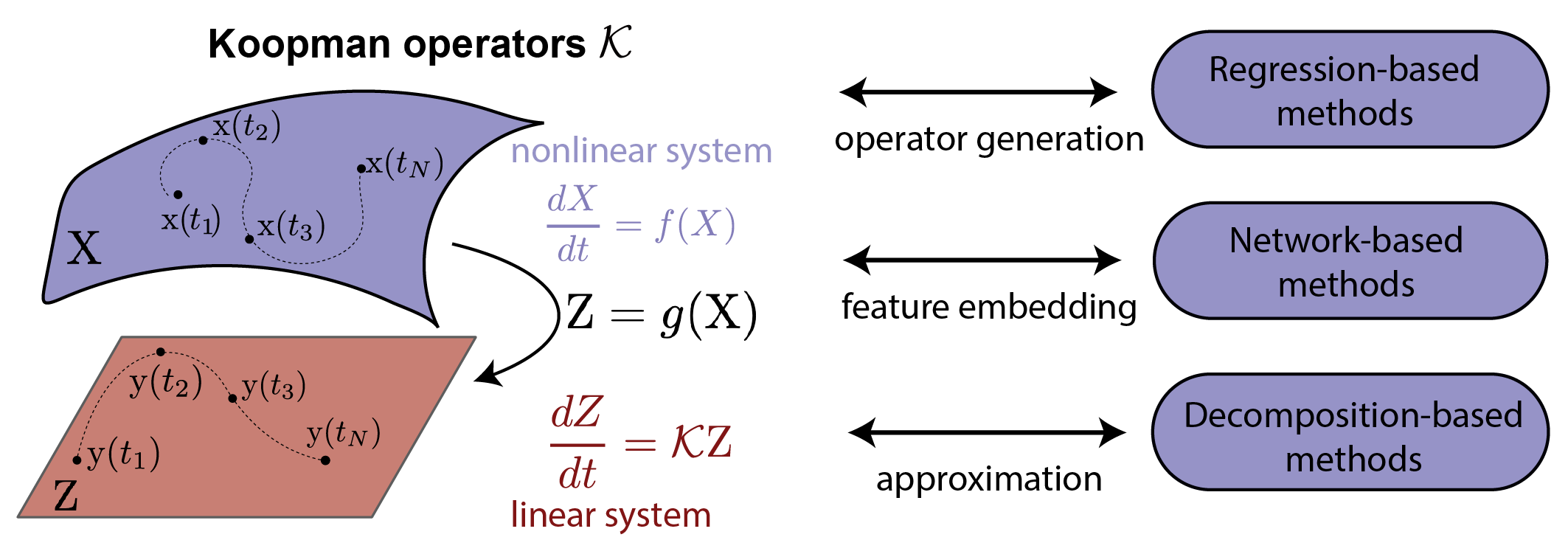}
	\caption[Koopman operator theory and its connection to data-driven methods for dynamical systems]{\textbf{Koopman operator theory and its connection to data-driven methods for dynamical systems.} By finding lifting functions \(g(X)\) corresponding to Koopman eigenfunctions \(\varphi\), a nonlinear system $X_t=\mathrm{f}(X(t),\beta)$ can be represented linearly. Many data-driven methods can be viewed as approximating the Koopman operator, using different underlying methodologies.}
	\label{fig:koopmantheory}
\end{figure*}

\section[A unifying framework for data-driven methods in biological dynamical systems]{A unifying framework for data-driven methods in biological dynamical systems}
\label{sec:unifying_framework}

\subsection{From linear systems to operator thinking}
When studying a dynamical system in the form $\frac{dX(t)}{dt}=\mathrm{f}(X(t),\beta)$, linear systems are the simplest to analyze because they can be explicitly solved and decomposed into independent modes.  
In vector form:
\begin{equation}
    \frac{dX(t)}{dt} = \mathrm{A} X(t),
\end{equation}
with solution
\begin{equation}
    X(t+t_0) = \mathrm{V} e^{\Lambda t} \mathrm{V}^{-1} X(t_0),
\end{equation}
where $\mathrm{A} = \mathrm{V} \Lambda \mathrm{V}^{-1}$, $\Lambda$ contains the eigenvalues $\lambda_i$, and $\mathrm{V}$ the eigenvectors $v_i$.  
Transforming to the eigenvector coordinates $\mathrm{z} = \mathrm{V}^{-1} X$ decouples the dynamics:
\begin{equation}
    z_{t,j}(t) = \lambda_j z_j(t),
\end{equation}
so each mode evolves independently. The real parts of $\lambda_i$ determine stability; imaginary parts set oscillation frequencies.

Most biological systems are nonlinear and cannot be diagonalized in this way.  
However, Koopman operator theory offers a unifying perspective: any nonlinear system can be recast as a linear system in a (possibly infinite-dimensional or Hilbert) space of observables.  
Originally developed by Koopman and von Neumann in the 1930s~\cite{Koopman1931,Koopman1932}, this approach defines functions $g(X)$ whose evolution is governed by a linear operator $\mathcal{K}$:
\begin{equation}
    \frac{dZ}{dt}  = \mathcal{K} \mathrm{Z}, \quad \mathrm{Z} = g(X).
\end{equation}
If a finite set of observables captures all relevant dynamics, we can apply the same eigen-analysis as for a purely linear system.  
The central challenge for data-driven approaches is to discover such observables automatically from data.  Figure~\ref{fig:koopmantheory} illustrates this transformation in schematic form, and highlights how many modern algorithms can be interpreted as finite-dimensional approximations to $\mathcal{K}$ (for examples see~\ref{box:koopman_example}).

\subsection{From operator thinking to method categories}
\newcommand{\capHi}{\textbf{\large$\bullet\bullet\bullet$}}
\newcommand{\capMed}{\textbf{\large$\bullet\bullet\circ$}}
\newcommand{\capLo}{\textbf{\large$\bullet\circ\circ$}}

\newcommand{\btri}{\tikz[baseline=0ex]\filldraw (0,0)--(1.5ex,0)--(0.75ex,1.5ex)--cycle;}
\newcommand{\wtri}{\tikz[baseline=0ex]\draw (0,0)--(1.5ex,0)--(0.75ex,1.5ex)--cycle;}
\newcommand{\reqHi}{\btri\;\btri\;\btri}
\newcommand{\reqMed}{\btri\;\btri\;\wtri}
\newcommand{\reqLo}{\btri\;\wtri\;\wtri}

\begin{table*}[t!]
\begin{ruledtabular}
\centering
\caption[Overview of data-driven methods for dynamical systems]{
\textbf{Overview of data-driven methods for the study of dynamical systems.} 
Methods are grouped by approach and rated for \textbf{Forecasting}, \textbf{Interactions}, \textbf{Dynamic solutions}, \textbf{Prior Knowledge}, and \textbf{Interpretability}.
\textbf{Notation:} Circles indicate capability of the method: \capHi = strong, \capMed = medium, \capLo = weak.  
Triangles indicate a requirement: \reqHi = high requirement (worse), \reqMed = medium, \reqLo = low requirement (better).  
Background shading (green/yellow/red) gives a quick visual cue from better to worse.
\textit{Undefined acronyms used in the table:} NARMAX – Nonlinear AutoRegressive Moving Average with eXogenous inputs; 
AI-Feynman – Physics-inspired symbolic regression method; 
GOBI – Gene network inference with ODE-based modeling; 
CLINE – Computational Learning and Identification of Nullclines.
\textit{Note:} The qualitative ratings shown here are informed by quantitative benchmarks for the different method classes reported in the literature (see Table~\ref{tab:overview_benchmarks}).
}

\begin{tabular}{l p{0.9cm} p{0.7cm} ccccc}
\label{tab:data_driven_methods_overview}
\textbf{Method} & \textbf{Year} & \textbf{Ref.} &
Forecasting & Interactions & Solutions & Prior Knowledge & Interpretability \\
\hline

\rowcolor{gray!15}\multicolumn{8}{l}{\textbf{Classical methods}} \\
Ordinary differential Eqs. (ODEs) & 18\textsuperscript{th}c. & \cite{Newton1729} &
\cellcolor{green!30}\capHi & \cellcolor{green!30}\capHi & \cellcolor{green!30}\capHi &
\cellcolor{red!30}\reqHi & \cellcolor{green!30}\capHi \\

\rowcolor{gray!15}\multicolumn{8}{l}{\textbf{Regression-based methods}} \\
Granger Causality & 1969 & \cite{Granger1969} &
\cellcolor{red!30}\capLo & \cellcolor{green!30}\capHi & \cellcolor{red!30}\capLo &
\cellcolor{green!30}\reqLo & \cellcolor{yellow!30}\capMed \\
NARMAX & 1986 & \cite{CHEN1989} &
\cellcolor{yellow!30}\capMed & \cellcolor{yellow!30}\capMed & \cellcolor{red!30}\capLo &
\cellcolor{yellow!30}\reqMed & \cellcolor{yellow!30}\capMed \\
Support Vector Machine (SVM and SVR) & 1997 & \cite{vapnik1996support} &
\cellcolor{green!30}\capHi & \cellcolor{yellow!30}\capMed & \cellcolor{red!30}\capLo &
\cellcolor{yellow!30}\reqMed & \cellcolor{yellow!30}\capMed \\
Gaussian Process Regression (GPR) & 2006 & \cite{williams2006gaussian} &
\cellcolor{green!30}\capHi & \cellcolor{yellow!30}\capMed & \cellcolor{red!30}\capLo &
\cellcolor{yellow!30}\reqMed & \cellcolor{yellow!30}\capMed \\
Symbolic Regression (SR) & 2009 & \cite{Schmidt2009} &
\cellcolor{yellow!30}\capMed & \cellcolor{yellow!30}\capMed & \cellcolor{red!30}\capLo &
\cellcolor{yellow!30}\reqMed & \cellcolor{yellow!30}\capMed \\
Sparse Ident. of Nonlinear Dyn. (SINDy) & 2016 & \cite{Brunton2016} &
\cellcolor{yellow!30}\capMed & \cellcolor{yellow!30}\capMed & \cellcolor{red!30}\capLo &
\cellcolor{yellow!30}\reqMed & \cellcolor{yellow!30}\capMed \\
AI-Feynman & 2019 & \cite{Udrescu2020} &
\cellcolor{yellow!30}\capMed & \cellcolor{yellow!30}\capMed & \cellcolor{red!30}\capLo &
\cellcolor{yellow!30}\reqMed & \cellcolor{yellow!30}\capMed \\
GOBI & 2023 & \cite{Park2023} &
\cellcolor{yellow!30}\capMed & \cellcolor{green!30}\capHi & \cellcolor{red!30}\capLo &
\cellcolor{yellow!30}\reqMed & \cellcolor{green!30}\capHi \\

\rowcolor{gray!15}\multicolumn{8}{l}{\textbf{Network-based methods}} \\
Feed-forward Neural Networks (FFNNs) & 1969 & \cite{Minsky1969} &
\cellcolor{yellow!30}\capMed & \cellcolor{red!30}\capLo & \cellcolor{yellow!30}\capMed &
\cellcolor{green!30}\reqLo & \cellcolor{red!30}\capLo \\
Recurrent Neural Networks (RNNs) & 1985 & \cite{Rumelhart1987} &
\cellcolor{green!30}\capHi & \cellcolor{red!30}\capLo & \cellcolor{yellow!30}\capMed &
\cellcolor{green!30}\reqLo & \cellcolor{red!30}\capLo \\
Autoencoders (AEs) & 1987 & \cite{Plaut1987} &
\cellcolor{red!30}\capLo & \cellcolor{red!30}\capLo & \cellcolor{yellow!30}\capMed &
\cellcolor{green!30}\reqLo & \cellcolor{yellow!30}\capMed \\
Reservoir Computing & 2001 & \cite{jaeger2001echo} &
\cellcolor{green!30}\capHi & \cellcolor{red!30}\capLo & \cellcolor{red!30}\capLo &
\cellcolor{green!30}\reqLo & \cellcolor{red!30}\capLo \\
Variational Autoencoders (VAEs) & 2013 & \cite{Kingma2013} &
\cellcolor{yellow!30}\capMed & \cellcolor{red!30}\capLo & \cellcolor{yellow!30}\capMed &
\cellcolor{green!30}\reqLo & \cellcolor{yellow!30}\capMed \\
Neural ODEs & 2018 & \cite{Chen2018} &
\cellcolor{green!30}\capHi & \cellcolor{yellow!30}\capMed & \cellcolor{red!30}\capLo &
\cellcolor{yellow!30}\reqMed & \cellcolor{yellow!30}\capMed \\
Physics-informed NNs (PINNs) & 2018 & \cite{Raissi2018} &
\cellcolor{green!30}\capHi & \cellcolor{yellow!30}\capMed & \cellcolor{red!30}\capLo &
\cellcolor{yellow!30}\reqMed & \cellcolor{yellow!30}\capMed \\
CLINE & 2025 & \cite{Prokop2025} &
\cellcolor{red!30}\capLo & \cellcolor{green!30}\capHi & \cellcolor{red!30}\capLo &
\cellcolor{green!30}\reqLo & \cellcolor{yellow!30}\capMed \\

\rowcolor{gray!15}\multicolumn{8}{l}{\textbf{Decomposition methods}} \\
Tucker Decomposition & 1966 & \cite{Tucker1966} &
\cellcolor{red!30}\capLo & \cellcolor{green!30}\capHi & \cellcolor{green!30}\capHi &
\cellcolor{yellow!30}\reqMed & \cellcolor{yellow!30}\capMed \\
Nonnegative Matrix Factorization  & 1999 & \cite{Lee1999} &
\cellcolor{red!30}\capLo & \cellcolor{yellow!30}\capMed & \cellcolor{green!30}\capHi &
\cellcolor{yellow!30}\reqMed & \cellcolor{green!30}\capHi \\
Dynamic Mode Decomp. (DMD) & 2009 & \cite{Schmid2010} &
\cellcolor{red!30}\capLo & \cellcolor{yellow!30}\capMed & \cellcolor{green!30}\capHi &
\cellcolor{yellow!30}\reqMed & \cellcolor{yellow!30}\capMed \\
extended DMD & 2015 & \cite{Williams2015} &
\cellcolor{red!30}\capLo & \cellcolor{yellow!30}\capMed & \cellcolor{green!30}\capHi &
\cellcolor{yellow!30}\reqMed & \cellcolor{yellow!30}\capMed \\
Hankel alt. View On Koopman (HAVOK) & 2017 & \cite{Brunton2017} &
\cellcolor{red!30}\capLo & \cellcolor{yellow!30}\capMed & \cellcolor{green!30}\capHi &
\cellcolor{yellow!30}\reqMed & \cellcolor{yellow!30}\capMed \\

\rowcolor{gray!15}\multicolumn{8}{l}{\textbf{Pseudotime-based methods}} \\
Monocle (pseudotime ordering) & 2014 & \cite{Trapnell2014} &
\cellcolor{red!30}\capLo & \cellcolor{yellow!30}\capMed & \cellcolor{green!30}\capHi &
\cellcolor{yellow!30}\reqMed & \cellcolor{yellow!30}\capMed \\
Wishbone (branching trajectories) & 2016 & \cite{Setty2016} &
\cellcolor{red!30}\capLo & \cellcolor{yellow!30}\capMed & \cellcolor{green!30}\capHi &
\cellcolor{yellow!30}\reqMed & \cellcolor{yellow!30}\capMed \\
\rowcolor{gray!15}\multicolumn{8}{l}{\textbf{Hybrid methods}} \\
Universal differential Eqs. (UDEs) & 2020 & \cite{Rackauckas2020} &
\cellcolor{green!30}\capHi & \cellcolor{yellow!30}\capMed & \cellcolor{yellow!30}\capMed &
\cellcolor{yellow!30}\reqMed & \cellcolor{yellow!30}\capMed \\
Symbolic Deep Learning & 2020 & \cite{Cranmer2019} &
\cellcolor{yellow!30}\capMed & \cellcolor{yellow!30}\capMed & \cellcolor{red!30}\capLo &
\cellcolor{yellow!30}\reqMed & \cellcolor{green!30}\capHi \\
SINDy + AE & 2023 & \cite{Bakarji2023} &
\cellcolor{green!30}\capHi & \cellcolor{green!30}\capHi & \cellcolor{yellow!30}\capMed &
\cellcolor{yellow!30}\reqMed & \cellcolor{yellow!30}\capMed \\
CLINE–SINDy/SR & 2025 & \cite{Prokop2025} &
\cellcolor{yellow!30}\capMed & \cellcolor{green!30}\capHi & \cellcolor{red!30}\capLo &
\cellcolor{green!30}\reqLo & \cellcolor{green!30}\capHi \\

\end{tabular}
\end{ruledtabular}
\end{table*}

Koopman theory provides a conceptual bridge between classical linear analysis and modern data-driven approaches.  
Most methods for studying nonlinear dynamical systems can be interpreted as different strategies for approximating the action of $\mathcal{K}$, whether explicitly, implicitly, or in restricted subspaces of observables.

In this review, we categorize data-driven methods primarily by their underlying methodology:  
\textbf{regression-based approaches}, \textbf{network-based architectures}, and \textbf{linearization or decomposition techniques}.  
This categorization is intentional. Other classification schemes exist, such as distinguishing between \textit{white-box} (mechanistic) and \textit{black-box} (phenomenological) models, or between supervised and unsupervised learning. However, these often miss the methodological constraints and opportunities that come with approximating $\mathcal{K}$. In white-box models, parameters and structure have a clear physical or biochemical meaning, allowing for mechanistic interpretation. In black-box models, parameters are tuned for predictive accuracy, often at the expense of interpretability. Supervised learning maps known inputs to outputs, while unsupervised learning instead searches for latent structure or dynamics in raw data.

Moreover, interpretability is not determined solely by whether a model is labeled as white-box or black-box.  
For example, a sparse regression model (often considered white-box) can yield uninterpretable results if applied to high-dimensional, noisy data with an ill-chosen basis; conversely, a neural-network-based approach (typically called black-box) can be made interpretable by constraining architectures or incorporating physics-informed features \cite{Shiguihara2021,Kitson2023}.  
Thus, by focusing on methodology rather than philosophical labels, we gain a clearer view of each approach's strengths and limitations (see Table~\ref{tab:data_driven_methods_overview}) in addressing the core aspects introduced before: forecasting, interaction inference, and 
the identification of dynamic solutions (Fig.~\ref{fig:understandingdynamicalsystems}F).

\begin{description}
    \item [Regression-based methods]  use regression to test whether symbolic model equations or inferred connections between variables can explain observed data. Unlike classical modeling, where equations are specified \textit{a priori}, these methods select or refine equations during the fitting process. Their symbolic form often aids interpretability, but performance can degrade in high-dimensional settings due to the curse of dimensionality.  

    \item [Network-based methods] employ architectures such as neural or Bayesian networks \cite{Shiguihara2021,Kitson2023} to capture complex nonlinear relationships. They often learn suitable observables to approximate $\mathcal{K}$ in high-dimensional latent spaces, producing accurate forecasts, but sometimes limiting direct interpretability.  

    \item [Decomposition methods] apply mathematical transformations to extract dominant spatio-temporal modes directly from data, without requiring explicit governing equations. Many of these modes correspond to Koopman modes, enabling low-dimensional or linear representations of otherwise complex, nonlinear systems.  
\end{description} 

\lgel{
While the landscape of data-driven modeling is vast, the evaluation of new methods typically centers on a small set of canonical dynamical systems.
These include low-dimensional oscillators (e.g., Duffing, FitzHugh-Nagumo), chaotic systems (e.g., Lorenz), and PDEs describing transport or turbulence (e.g., Burgers, Korteweg-de~Vries, and Kuramoto-Sivashinsky).
Our heuristic ratings in Table~\ref{tab:data_driven_methods_overview} are based on a structured, literature-informed comparison of forecasting accuracy, interpretability, and dynamical characterization across methods, drawing on benchmarks reported in prior studies (Table~\ref{tab:overview_benchmarks}).
A single unified numerical benchmark across all evaluation criteria is often precluded by an \emph{incommensurability of objectives}:
methods designed for symbolic discovery (such as SINDy or AI-Feynman) prioritize parsimony and structural fidelity, whereas approaches such as reservoir computing are optimized for high-fidelity forecasting of complex or chaotic dynamics.
As a result, a ``weak'' rating in forecasting for a decomposition-based method does not indicate algorithmic failure, but rather reflects a deliberate design trade-off in which latent-state identification or dynamical structure is favored over long-term temporal extrapolation.
To enable a meaningful comparison under these constraints, we examine representative methods with comparable objectives from each methodological category, applied to synthetic data generated by the Oregonator model introduced earlier.
The Oregonator provides a well-studied (bio)chemical oscillator whose dimensionality and nonlinear structure capture key challenges common to biological systems, including hidden variables, multiple coupled time scales, and nonlinear feedbacks (see Supplementary Notes~2--4 for details).}

\section{Regression-based methods}
\label{sec:regression_based_methods}

Regression-based approaches are among the most widely used data-driven tools for identifying dynamical systems.  
They aim to infer how changes in one measured variable can be explained by others through explicit functional relationships~\cite{Brunton2022book}.  
In general, the regression problem can be expressed as:
\begin{equation}
	Y = f(X, \beta),
\end{equation}
where \(Y\) are observations, \(X\) the independent variables (often past values of system states), and \(\beta\) model parameters.  
The task is to select a model structure \(f\) and parameters \(\beta\) such that \(f(X)\) best predicts the observed data (see ~\ref{box:regression_primer}).

\begin{figure*}
	\centering
\includegraphics[width=1.0\linewidth]{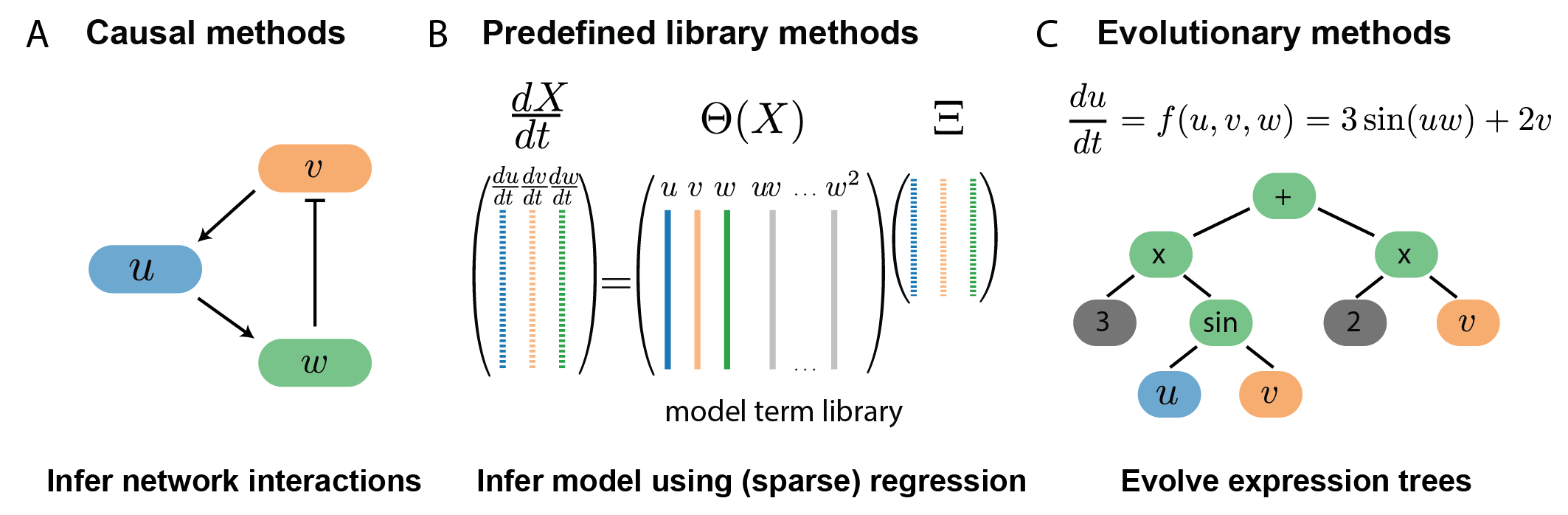}
	\caption{\textbf{Overview of regression-based methods.} Regression-based approaches can be grouped into three classes: causality methods (e.g., Granger causality, GOBI), library-based sparse regression methods (NARMAX, SINDy), and evolutionary approaches (Symbolic Regression, AI-Feynman).}
	\label{fig:regression_methods}
\end{figure*}

Next, we outline the main regression-based approaches used for data-driven model discovery in dynamical systems.
Such methods can be grouped into three broad classes:
those emphasizing \textbf{causal inference},
those based on \textbf{predefined candidate libraries with sparse regression},
and those employing \textbf{evolutionary algorithms}
(see Fig.~\ref{fig:regression_methods}).
Despite methodological differences, all regression-based approaches share two important characteristics. 
First, they typically produce \emph{symbolic and interpretable} models, either in the form of interaction networks or explicit differential equations. 
Second, they cannot on their own identify whether all relevant state variables are included in the data, i.e., whether $X$ truly spans the system’s underlying state space. 
These shared strengths and limitations shape how regression-based methods can be applied in biological contexts, as we illustrate below.

\subsection{Causal methods}

Causality approaches aim to infer whether one variable helps explain the future of another~\cite{WISDOM1960}.  
In biology, this is often used to reconstruct interaction networks from time series, such as gene expression or ecological dynamics~\cite{Saint-Antoine2020}.  

A classical method is \textit{Granger Causality} (GC)~\cite{Granger1969}, which compares prediction accuracy with and without including past values of another variable.  
Formally, a baseline model predicts future $Y$ from its own history:
\begin{equation}
	Y_t = \alpha_0 + \alpha_1 Y_{t-1} + \alpha_2 Y_{t-2}+ \dots + \eta_t,
\end{equation}
while an extended model adds past values of $X$:
\begin{equation}
	Y_t = \beta_0 + \beta_1 Y_{t-1} + \dots + \gamma_1 X_{t-1} + \gamma_2 X_{t-2} + \mu_t.
\end{equation}
If prediction improves, $X$ is said to “Granger cause” $Y$.  

Although simple and widely applied~\cite{Runge2019}, GC performs poorly in nonlinear, oscillatory, or synchronously coupled systems, where it can yield spurious full networks~\cite{Nawrath2010, Guo2008, Stokes2017}.

\lgel{Recent alternatives address these issues.  
\textit{General ODE-Based Inference} 
(GOBI)~\cite{Park2023}, for example, infers causal regulation by testing whether observed time series are consistent with \emph{sign-consistent regulatory effects} of the form
\begin{equation}
    \frac{dY}{dt} = f(X_1,\ldots,X_N),
\end{equation}
using regulation-detection scores based on time- and derivative-difference statistics to prune false positives and to identify higher-order regulatory motifs (e.g., $u \to w \leftarrow v$).
From a Koopman perspective, both regression-based and constraint-based causal inference methods can be viewed as comparing how well different sets of observables capture the evolution of the system. In Granger-type approaches, this comparison is explicit, as one evaluates predictors built from $X$ versus $X,Y$. In GOBI, it is implicit: adding or removing a candidate regulator $X$ changes the set of observables (here constructed from time and derivative differences) on which the Koopman generator is tested, and causality is inferred by whether the corresponding Koopman-invariant structure satisfies the sign-consistency constraints imposed by the assumed ODE form~\cite{Rupe2024, DiAntonio2024}.}

\subsection{Sparse regression with predefined libraries}
\lgel{
Sparse regression methods reconstruct governing equations by selecting a (small) number of active terms from a predefined library of candidate functions.
The defining feature of this class is the specification of a functional basis \emph{a priori}, rather than the restriction to any particular functional form.
Two representative approaches are \textit{Nonlinear Autoregressive Moving Average models with Exogenous Inputs} (NARMAX) and \textit{Sparse Identification of Nonlinear Dynamics} (SINDy).}

\lgel{
\textit{NARMAX.}  
Originally developed in control theory, NARMAX models describe discrete-time dynamics using predefined functional expansions~\cite{Billings1998}.
The general form is:
% \begin{equation}
% 	X_k = F\!\left( X_{k-1}, \dots, X_{k-n_x},\ Y_{k-1}, \dots, Y_{k-n_y},\ \eta_{k-1}, \dots, \eta_{k-n_\eta} \right) + \eta_k,
% \end{equation}
\begin{align}
X_k &= F\!\Big( X_{k-1}, \dots, X_{k-n_x}, \nonumber \\
    &\qquad\; Y_{k-1}, \dots, Y_{k-n_y}, \nonumber \\
    &\qquad\; \eta_{k-1}, \dots, \eta_{k-n_\eta} \Big) + \eta_k
\end{align}
where $F$ is expanded in polynomial, rational, or radial basis functions~\cite{Chen1990, Hong2008}.  
NARMAX prioritizes predictive accuracy and controllability over interpretability.  
It has been widely used in biomechanics, physiology, and biochemical signal prediction~\cite{Kukreja2003,Khodadadi2023,Mendes1998,Krishnanathan2012}.}

\lgel{
\textit{SINDy.}  
In contrast to fixed polynomial regression, SINDy emphasizes parsimony and interpretability by performing sparse regression over a user-defined candidate library~\cite{Brunton2016}.
It constructs a library $\Theta(X)$ of candidate functions and identifies a minimal set of active terms:
\begin{equation}
	\frac{dX}{dt}  = \Theta(X)\xi.
\end{equation}
The library $\Theta(X)$ is not restricted to polynomials and may include rational functions, trigonometric terms, exponentials, or other problem-specific basis functions.
The method assumes parsimony, meaning that most coefficients $\xi$ are zero, identifying only a minimal subset.  
This can be interpreted as approximating the Koopman generator, with $\Theta$ serving as lifting functions~\cite{Brunton2022book}.
SINDy has been applied in ecology~\cite{Hirsh2022,Fasel2022,Sandoz2023,Anjos2023,Gutierrez-Vilchis2023,Messenger2024}, cellular dynamics~\cite{Mangan2016,Messenger2022,Brummer2023,Naozuka2025}, and other domains, though mostly on synthetic rather than experimental data.  }

In summary, library-based sparse regression methods like SINDy provide interpretable models and are particularly powerful when domain knowledge informs the basis function selection. 
Yet, they still depend on predefined libraries, which can limit discovery in highly nonlinear biological systems. 
A further limitation is their sensitivity to data quality: because regression is typically performed on estimated derivatives, noise or insufficient temporal resolution can severely distort the results. 
Several studies have proposed strategies to mitigate these issues, such as improved differentiation schemes and noise-aware extensions of SINDy~\cite{Delahunt2022,Cortiella2023,Prokop2024,Messenger2021}. 
We illustrate these challenges in more detail in ~\ref{box:SINDy_oregonator}, where SINDy is tested on the Oregonator model under increasingly realistic experimental conditions.

\subsection{Evolutionary methods}

Library-based sparse regression methods rely on predefined functional libraries,
which can limit their ability to capture unknown dynamics.
Evolutionary algorithms, in particular \textit{Symbolic Regression} (SR), overcome this by simultaneously discovering both the structure and parameters of candidate models~\cite{eiben2015introduction,Schmidt2009}. 
In SR, models are encoded as hierarchical \textit{expression trees} (see Fig.~\ref{fig:regression_methods}C), where internal nodes represent operators (e.g., $+, \times, \sin$) and leaves correspond to variables or constants. 
These trees evolve over successive generations through three main operations: \textit{mutation} (replacing a subtree with a randomly generated one), \textit{recombination} (swapping subtrees between two parents), and \textit{selection} (retaining models with lower prediction error or higher parsimony). 
This evolutionary search enables the discovery of both novel structures and parameter values without relying on a fixed functional library.

The flexibility of SR makes it powerful, but also computationally expensive, prone to overfitting, and sensitive to noise due to the vast search space. 
Recent developments have sought to mitigate these challenges by incorporating domain-specific constraints. A prominent example is \textit{AI-Feynman}~\cite{Udrescu2020}, which uses physical priors such as dimensional consistency, parsimony, compositionality, symmetry, and separability to restrict the search. By narrowing the solution space in this way, AI-Feynman was able to rediscover dozens of canonical physics equations directly from data, demonstrating how carefully chosen constraints can both accelerate convergence and enhance interpretability.  

In biology, such universal rules are harder to define, but analogous constraints can be introduced by exploiting biochemical priors, conservation laws, or known regulatory motifs. Applications of SR in biology already span gene regulatory and biochemical networks~\cite{Schmidt2011,Beauregard2019,Zhang2019}, ecological dynamics~\cite{Chen2019,Haldar2024}, \textcolor{red}{cellular interaction networks \cite{Daniels2015, Daniels2015b}} and mechanistic tissue models~\cite{Hou2024}, though the success of these approaches often depends on constraining the search with system-specific knowledge. Conceptually, SR can be seen as approximating Koopman lifting functions in a highly flexible but relatively unstructured way, trading generality for computational efficiency. When guided by appropriate priors, however, evolutionary methods hold particular promise for interpretable model discovery in complex biological systems.

\subsection{Limitations in biological applications}

Across subclasses, regression-based methods face three recurring challenges:
\begin{enumerate}
    \item \textbf{Strict data requirements:} \lgel{methods that rely on explicit numerical differentiation amplify noise}; dense, high-resolution time series are often essential. Biological data often violate these assumptions.  
    \item \textbf{Dependence on prior knowledge:} success depends on access to all relevant state variables and informed choices of libraries or operator sets. Without such priors, spurious or uninterpretable models often result.  
    \item \textbf{Scaling issues:} high-dimensional biological networks exacerbate the curse of dimensionality and limit applicability.  
\end{enumerate}

\lgel{Recent weak-form and integral formulations of sparse regression (including WSINDy, WENDy, and related approaches) partially alleviate this limitation by avoiding explicit numerical differentiation and instead enforcing the governing equations in an integral or variational sense \cite{Schaeffer2017,Messenger2021,Messenger2024,LyonsDukicBortz2025,WangHuanGarikipati2019,Pantazis2019}. These methods substantially improve robustness to noise and sparse sampling, and have enabled successful equation discovery in several experimental and biological settings. However, while they mitigate noise amplification, they still share the dependence on library design and observability characteristic of regression-based methods, and they additionally introduce sensitivities to the choice of test functions and to the modeling of noise and correlations in the weak residuals, which can limit their reliability in complex biological data.}

These challenges are not just theoretical but are consistently observed in practice. For instance, applying SINDy to the Oregonator model shows that under ideal conditions—complete observability, sufficiently rich libraries, and low-noise data—the method can recover the governing equations exactly. However, as conditions become more realistic, with limited observability, noisy measurements, or incomplete prior knowledge, SINDy rapidly fails to reproduce the dynamics \cite{Prokop2024}. 

This underscores that successful regression-based discovery in biology hinges critically on data quality, careful feature selection, and substantial domain expertise;
%(for more details see Table~\ref{tab:rb_methods_adv_disadv}).
Causality methods are simple to apply and interpretable as they directly infer networks, yet struggle with nonlinearity and oscillatory dynamics resulting in limited forecasting abilities.
Library-based sparse regression methods provide explicit equations that are immediately interpretable especially for low-dimensional systems but are sensitive to noise, require full state observability and informed term library formulation.
Evolutionary methods are highly flexible and unbiased, and are able to rediscover known laws but result in large search space making them computationally heavy and prone to overfitting.

\lgel{Thus, while regression-based discovery methods that rely on explicit differentiation have excelled in proof-of-concept and low-dimensional systems, even modern weak-form variants remain challenged by limited observability, model misspecification, and data heterogeneity in realistic biological settings.} Hybrid approaches that combine regression with mechanistic priors or dimensionality reduction hold promise for overcoming these limitations.

\subsection{Other regression-based methods}  
Beyond polynomial expansions and sparse regression, several other machine learning techniques have been applied to dynamical systems. 
Kernel methods such as \textit{support vector machines} (SVMs) or kernel-based NARMAX extend linear regression into nonlinear feature spaces via kernel functions (e.g., Gaussian or polynomial kernels) \cite{suykens2001nonlinear,suthaharan2016support, Chen2004}.  
In this way, SVMs or other kernel methods can model nonlinear state-to-state mappings \cite{Gilpin2020} and have been used for forecasting and classification of dynamical regimes \cite{Ben-Hur2008}, including inference of specifically gene regulatory \cite{Gillani2014} or generally biological (metabolic, gene-gene interaction etc.) networks \cite{Vert2007}.  
Similarly, \textit{Gaussian process regression} (GPR) provides a flexible, nonparametric Bayesian framework for learning dynamical systems from data \cite{williams2006gaussian}.  
GPR models offer probabilistic predictions with uncertainty quantification and have been applied in systems biology to capture noisy dynamics of gene networks and signaling pathways \cite{gao2008gaussian,duvenaud2014automatic,frigola2014variational}.  
However, both SVMs and GPR typically act as black-box predictors rather than interpretable dynamical models, and they scale poorly to very high-dimensional or long time-series data.  
As such, they are often used as complementary tools for model identification tasks.

\section{Network-based methods}
\label{sec:network_based_methods}

Regression-based methods (Sec.~\ref{sec:regression_based_methods}) rely on explicit symbolic structures to represent dynamics. 
By contrast, network-based methods take a different approach: \textit{artificial neural networks} (ANNs) serve as flexible, high-dimensional function approximators capable of capturing nonlinear dynamics directly from data. 
Their strength lies in expressivity: they can approximate arbitrarily complex mappings without predefined equations, making them well-suited for systems where governing rules are unknown or partially observed. 
This flexibility comes at a cost, however, since neural networks typically act as ``black boxes,'' demanding large amounts of data and offering limited interpretability.  
Where regression methods produce explicit equations (e.g. sparse symbolic equations identified by SINDy), neural networks discover latent representations that may or may not align with interpretable observables.  
As in the regression case, this distinction can be understood through Koopman operator theory: neural networks can be viewed as learning finite-dimensional approximations of Koopman embeddings, but often without explicit interpretability.

As a primer, \ref{box:nn_basics} introduces the mathematical foundations of neural networks. 
In the main text, we focus on architectures most relevant to dynamical systems: {feed-forward neural networks} (FFNNs), {recurrent neural networks} (RNNs), and {autoencoders} (AEs). 
These form the basis for specialized variants such as {physics-informed neural networks} (PINNs) or {biologically informed neural networks} (BINNs), which aim to balance predictive power with interpretability (see Fig.~\ref{fig:network_methods}).  

\subsection{Feed-forward neural networks}
\begin{figure*}
	\centering
	\includegraphics[width=\linewidth]{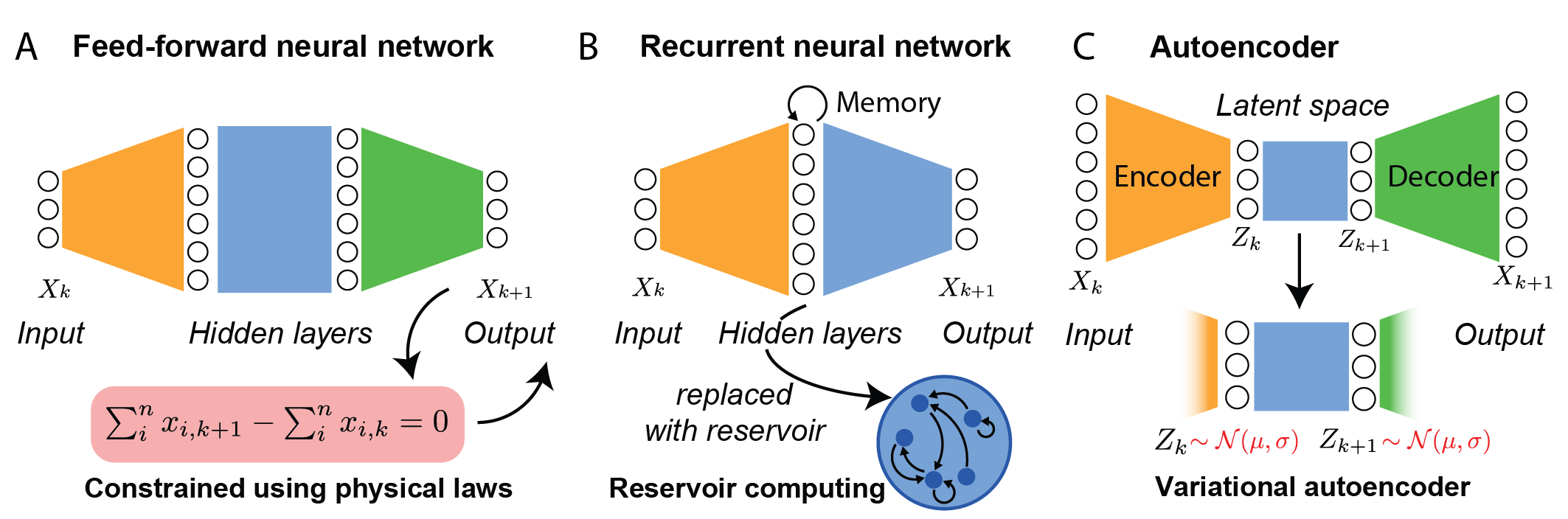}
	\caption[Overview of network-based methods for data-driven model identification/discovery]{\textbf{Neural network architectures for dynamical systems.} 
(A) Feed-forward neural networks (FFNNs) learn mappings $X_k \mapsto X_{k+1}$ and can be constrained by physical or biological laws (PINNs, BINNs).  
(B) Recurrent neural networks (RNNs) incorporate memory of past states; reservoir computing (RC) is a special case where recurrent weights are fixed.  
(C) \lgel{Autoencoders (AEs) compress dynamics into latent coordinates $Z_k$, providing reduced state-space representations for dynamical analysis. Variational autoencoders (VAEs) augment this framework with a probabilistic, generative latent space that is smoothly structured and can be coupled to Koopman or neural-ODE dynamics. }Together, these architectures illustrate how neural networks can be tailored or extended to balance predictive power with interpretability.
}
	\label{fig:network_methods}
\end{figure*}

Among neural architectures (Fig.~\ref{fig:network_methods}), the FFNN is the simplest, mapping a current state $X_k$ to the next state $X_{k+1}$ \cite{Narendra1992,Suykens1996}. 
Formally, the loss function is
\begin{equation}
\mathcal{L} = \left\| X_{k+1}-\mathit{NN}(X_k)\right\|_2^2,
\end{equation}
so that the network approximates the evolution operator $\mathrm{F}$.  

FFNNs have been applied to oscillatory and chaotic systems \cite{sandberg2001nonlinear,Chattopadhyay2020} and, in biology, to brain development timing \cite{Nagarajan2013}, gene regulatory networks \cite{Choo2020, Chen2022}, and intercellular signaling \cite{Nilsson2022}.  
From a Koopman perspective, FFNNs can be viewed as implicitly discovering nonlinear observables, akin to polynomial libraries in SINDy but in a black-box manner.  

A common limitation is error accumulation over long trajectories, sometimes leading to implausible predictions such as negative concentrations \cite{Wang2005, Pan2018, Zhang2021}.  
Extensions such as physics-informed (PINNs) \cite{Raissi2018,Raissi2019} and biologically informed NNs (BINNs) \cite{Yazdani2020,Lagergren2020BINN} mitigate this by embedding mechanistic constraints into the loss function (e.g. conservation of mass, implementation of Michaelis-Menten kinetics, etc.).  

Another direction uses FFNNs not for forecasting but for uncovering static geometric features of phase space.  
The CLINE method (Computational Learning and Identification of Nullclines) \cite{Prokop2025} trains a FFNN to approximate inverse relations of the form
\begin{align}
	\begin{split}
		\frac{du}{dt}  &= f(u,v) \rightarrow u = f_{u}^{-1}(v,\frac{du}{dt}),\\
		\frac{dv}{dt} &= g(u,v) \rightarrow u = g_{u}^{-1}(v,\frac{dv}{dt}),
	\end{split}
	\label{eq:null_inverse}
\end{align}
using measured trajectories and their derivatives.  
By evaluating conditions such as $\frac{du}{dt}= 0$ or $\frac{dv}{dt} = 0$, CLINE recovers nullcline structures directly from data, without requiring a mechanistic model.  
Because nullclines define the intersections and stability structure of a system’s phase portrait, this approach provides mechanistic insights into dynamics that go beyond time-series prediction.

\subsection{Recurrent neural networks}

RNNs extend FFNNs by introducing loops that provide memory, making them natural tools for temporal and sequential data \cite{Rumelhart1987}.  
Each hidden state $h_k$ depends on both the current input $X_k$ and the previous hidden state $h_{k-1}$:
\begin{subequations}
\begin{align}
  h_k &= f_{a,1}(W_x X_k + W_h h_{k-1} + b_h), \\
  X_{k+1} &= f_{a,2}(W_{\text{out}} h_k + b_{\text{out}}),
\end{align}
\end{subequations}
where $W_x$, $W_h$, and $W_{\text{out}}$ are trainable weight matrices and $f_{a}$ are activation functions.  
This recursive structure allows RNNs to capture dependencies across time, unlike FFNNs which only map one state to the next.  

RNNs have been applied to chaotic attractors \cite{Pathak2017}, oscillatory systems \cite{Cestnik2019}, and system identification tasks \cite{Vlachas2018}.  
In biology, they have been used to model gene regulatory dynamics \cite{Jia2019}, cell cycle progression \cite{Samarasinghe2017}, microbiome dynamics \cite{Baranwal2022}, and microbial population behavior \cite{Vidal-Saez2024,Vidal-Saez2025}.  

A widely used special case is {reservoir computing} (RC), in which the recurrent reservoir is fixed and only the output weights are trained:
\begin{subequations}
\begin{align}
  h_k &= f_a(W_{\text{in}} X_k + W_{\text{res}} h_{k-1}), \\
  X_{k+1} &= W_{\text{out}} h_k.
\end{align}
\end{subequations}
Because only $W_{\text{out}}$ is optimized, RC reduces computational cost and avoids instability while retaining expressive power \cite{jaeger2001echo,Tanaka2019,Gauthier2021}.  
From a Koopman perspective, both RNNs and RCs can be interpreted as flexible ways of encoding time-delay coordinates, analogous to explicit lag terms in regression-based approaches.

\subsection{Autoencoders}

{Autoencoders} (AEs) learn compressed latent representations by mapping the input $X$ into a reduced latent variable $Z$ \textcolor{red}{($X \in \mathcal{R}^n, Z \in \mathcal{R}^m, m\ll n$)}%($X = x_i, Z = z_j, j\ll i$)
and reconstructing the original state \cite{Plaut1987}:
\begin{subequations}
\begin{align}
  Z &= f_{\text{enc}}(X), \\
  \hat{X} &= f_{\text{dec}}(Z),
\end{align}
\end{subequations}
trained to minimize the reconstruction loss
\begin{equation}
  \mathcal{L} = \|X - \hat{X}\|^2_2.
\end{equation}

With linear activations, this reduces to a {principal component analysis} (PCA) \cite{Bourlard1988}; with nonlinear activations, it can uncover richer feature spaces \cite{Hinton2006}.  

In dynamical systems, AEs provide a way to reduce high-dimensional dynamics into low-dimensional latent coordinates, or to approximate Koopman eigenfunctions \cite{Otto2019,Sondak2021}.  
Applications in biology include cell fate dynamics \cite{Maizels2024}, microbial growth \cite{Baig2023}, molecular simulations \cite{Wehmeyer2018}, and high-dimensional omics data \cite{Liu2021}.  

\lgel{In this context, standard autoencoders are primarily used as representation-learning tools: they define low-dimensional state spaces in which the system’s dynamics can be analyzed or modeled. A prominent extension is the {variational autoencoder} (VAE) \cite{Kingma2013}, which augments the latent space with a probabilistic structure and enables generative modeling of the data distribution:}

\begin{equation}
  \mathcal{L}_{\text{VAE}} = \mathbb{E}_{q(Z|X)}[\|X - \hat{X}\|^2] + \beta \, D_{\mathrm{KL}}(q(Z|X)\|p(Z)),
\end{equation}
where $D_{\mathrm{KL}}$ is the Kullback–Leibler (KL) divergence.  
\lgel{This KL regularization encourages the latent distribution to remain close to a simple prior, which suppresses fragmented or highly curved embeddings and promotes smooth, approximately linear latent dynamics. This bias toward low-complexity evolution facilitates the discovery of approximately Koopman-invariant coordinates, in which nonlinear dynamics become near-linear. In addition, the probabilistic structure of VAEs enables generative sampling and promotes generalizable latent representations that are well suited for Koopman-based modeling \cite{Asperti2021,Shrivastava2025}. As with regression-based polynomial expansions, the latent variables discovered by AEs (for representation learning) or VAEs (for structured and generative latent dynamics) can be viewed as approximate lifting functions.}
The main advantage is that these are learned directly from data rather than specified a priori.  
The drawback is sensitivity to latent dimensionality: too few dimensions miss essential modes, while too many cause overfitting \cite{Lopez2020,Zeng2023}.  

%\clearpage

\subsection{Limitations and outlook}

Three major challenges persist across NN-based methods:
\begin{enumerate}
    \item \textbf{Lack of interpretability:} the high-dimensional embeddings rarely align with mechanistic biological meaning, in contrast to regression-derived symbolic models.
    \item \textbf{Dependence on architecture:} performance depends strongly on architecture and hyperparameters, often requiring costly trial-and-error.  
    \item \textbf{Limited generalization:} generalization beyond the training domain is weak, particularly with noisy or undersampled biological data.  
\end{enumerate}
To highlight these challenges in more detail, we tested the ability of a FFNN to capture the oscillatory dynamics of the Oregonator model in~\ref{box:ffnn_example}.

Domain-informed variants such as PINNs, BINNs, and Koopman-inspired VAEs offer one path forward, balancing flexibility with interpretability, though at the cost of introducing strong priors.  
Taken together, regression and network-based methods represent complementary ends of a spectrum: the former emphasizes symbolic interpretability, the latter emphasizes flexible approximation.  
Both can be unified under the Koopman framework, but each has distinct strengths and weaknesses for biological applications.

\subsection{Other neural methods} 
Beyond the architectures discussed above, several additional neural approaches are increasingly relevant for modeling dynamical systems.  
{Neural ODEs} replace discrete mappings with continuous-time formulations, embedding differential equations directly into the network structure and enabling end-to-end training of time-continuous models ~\cite{Chen2018}.  
{Graph neural networks} (GNNs) extend standard architectures to relational data, making them suitable for systems with explicit interaction graphs such as gene regulatory or protein–protein interaction networks ~\cite{sanchez2018graph, bronstein2021geometric}.  
Transformer-based architectures are also being explored for dynamical prediction, where self-attention mechanisms capture long-range temporal dependencies more effectively than recurrent units ~\cite{vaswani2017attention, zhang2023applications}.  
While these methods are more established in the machine learning community, their application to biology is still emerging.  
They therefore complement FFNNs, RNNs, and AEs by providing powerful alternatives for modeling dynamical systems, particularly when complex interactions or long-range dependencies play a key role.  

\section{Decomposition methods}
\label{sec:decomposition_methods}

After showing that NNs can approximate the Koopman operator via high-dimensional embeddings (Sec.~\ref{sec:network_based_methods}), we now turn to decomposition-based methods that allow for a more direct approximation of the Koopman operator from measured data.  
The most widely used is {Dynamic Mode Decomposition} (DMD), introduced by Schmid \textit{et al.}~\cite{Schmid2010} to extract spatio-temporally coherent structures from high-dimensional data, originally in fluid dynamics~\cite{Brunton2022book}. 
Conceptually, DMD is related to {proper orthogonal decomposition} (POD)~\cite{Berkooz1993}, which identifies dominant spatial structures, but DMD additionally encodes their temporal evolution.  
Rowley and Mezić \textit{et al.}~\cite{Rowley2009} showed that DMD can be interpreted as a finite-dimensional approximation of the Koopman operator using only measured state variables.

\subsection{Dynamic Mode Decomposition}

The core idea in DMD is to analyze how successive snapshots of the system state are related.  
Given two sets of snapshots,
\begin{eqnarray}
	\mathcal{X} &=& \begin{bmatrix}X(t_1) & X(t_2) & \dots & X(t_{m-1})\end{bmatrix}, \quad \\ 
	\mathcal{X}' &=& \begin{bmatrix}X(t_2) & X(t_3) & \dots & X(t_m)\end{bmatrix},
\end{eqnarray}
DMD seeks a linear operator $A$ such that
\begin{equation}
	\mathcal{X}' \approx A \mathcal{X}, 
\end{equation}
or, with equidistant sampling,
\begin{equation}
	 X_{k+1} \approx A X_k.
\end{equation}
The operator $A$ thus provides a finite-dimensional approximation of the Koopman operator $\mathcal{K}$.  
Computationally, one solves
\begin{equation}
	A = \operatorname*{argmin}_A \left\| \mathcal{X}' - A \mathcal{X} \right\|_F = \mathcal{X}' \mathcal{X}^\dagger,
\end{equation}
where $\mathcal{X}^\dagger$ is the Moore–Penrose pseudo-inverse.

For high-dimensional systems with $n \gg m$, computing the full $A \in \mathbb{R}^{n \times n}$ is intractable.  
Instead, one projects onto a reduced system $\tilde{A} \in \mathbb{R}^{m \times m}$ obtained from the {singular value decomposition} (SVD) of $\mathcal{X}$.  
Tu \textit{et al.}~\cite{Tu2014} demonstrated that $\tilde{A}$ faithfully captures the essential eigenvalues of $A$, making the method computationally feasible.  
The number of retained modes $r$ must be chosen carefully: too few lose dynamical information, too many amplify noise.

A key limitation of (exact) DMD is that it relies on linear observables, which rarely suffice for nonlinear dynamics.  
{Extended DMD} (eDMD)~\cite{Williams2015} overcomes this by lifting the data into a higher-dimensional feature space via nonlinear functions $g$, and applying DMD there:
\begin{equation}
	g(X_{k+1}) \approx K g(X_k),
\end{equation}
where $K$ provides a finite-dimensional approximation of the Koopman operator in the lifted space.  
This approach parallels the role of basis libraries in regression-based methods such as SINDy (Sec.~\ref{sec:regression_based_methods}).

In biological contexts, both DMD and eDMD have been applied to neural dynamics in healthy~\cite{BBrunton2016} and epileptic~\cite{KarabiberCura2021} states, microbial ecosystems~\cite{Balakrishnan2020}, metabolic dynamics~\cite{Skantze2023}, blood flow~\cite{Habibi2020}, and tumor growth~\cite{Bourantas2014}.  
These studies highlight the potential of decomposition approaches to reveal coherent spatio-temporal patterns from complex biological data.

\subsection{Limitations of decomposition methods}

Despite their appeal, decomposition approaches face two main challenges in biological settings:

\begin{enumerate}
	\item \textbf{Requirement for sufficiently rich dynamics.}  
	As DMD and eDMD fit global linear operators, the data must capture diverse dynamical regimes to yield meaningful approximations.  
	When dynamics are dominated by switching, noise, or short transients — common in biological datasets — these methods fail to generalize.  
	While noise-robust variants exist~\cite{Chen2012,Hemati2017}, they still require high temporal resolution rarely available in biology.

	\item \textbf{Dependence on prior knowledge for interpretability.}  
	The success of eDMD depends on choosing appropriate nonlinear lifting functions $g$, a challenge shared with regression approaches like SINDy.  
	Without careful selection, the resulting modes may lack mechanistic meaning, undermining biological interpretability.  
\end{enumerate}

We evaluate these challenges in more detail by applying eDMD to the Oregonator model in~\ref{box:edmd_example}.
In summary, decomposition methods such as DMD and eDMD provide a principled way to approximate the Koopman operator directly from data and have demonstrated utility in biological systems.  
However, their reliance on rich data and appropriate lifting functions limits their interpretability and robustness, particularly in noisy, nonlinear biological contexts.

\subsection{Other decomposition methods}

While DMD and eDMD dominate the literature, other decomposition techniques are also relevant for dynamical systems.  
POD provides low-rank spatial modes but ignores temporal evolution, making it useful as a preprocessing tool for dimensionality reduction \cite{Berkooz1993}.  
{Nonnegative matrix factorization} (NMF)~\cite{Lee1999} has been applied to extract interpretable, additive components in gene expression dynamics and imaging data \cite{devarajan2008nonnegative,stein2018enter}.  
Tensor decomposition methods, such as {Tucker decomposition} or the related CANDECOMP/PARAFAC \cite{Tucker1966}, extend these ideas to multiway data (space–time–condition) and are increasingly used for multi-omics trajectories \cite{kolda2009tensor}.  
These approaches complement DMD by offering different trade-offs between interpretability, temporal structure, and computational cost.

Closely related to DMD are decomposition techniques that not only aim to reconstruct the system’s dynamical behavior but also provide meaningful interpretations of the recovered dimensions, following the Takens’ embedding theorem \cite{Takens1981}.

According to the theorem, all the information of a dynamical system is encoded in each measurement, meaning that even unobserved dimensions can be unfolded and a unique measurement-based coordinate system defined, for example, using the {Hankel Alternative View of Koopman} (HAVOK) method \cite{Brunton2017, Champion2019}. Such approaches are particularly useful when obtaining closed representations of chaotic systems via eDMD proves challenging. In principle, they allow reconstruction of missing state variables from partial observations, a scenario frequently encountered in biological systems. Identifying suitable embeddings from experimental data, however, is nontrivial \cite{Tan2023}, though it has already facilitated the application of SINDy in studies of tumor growth dynamics \cite{Brummer2023}.

\section[Future of data-driven methods in biology]{Future of data-driven methods for dynamical systems in biology}
\label{sec:future_of_dd_methods}

We have provided an overview of existing data-driven methods for studying dynamical systems. 
Yet all of these approaches exhibit certain limitations that can hinder their application to real-world experimental data, particularly in biology. 
This raises important questions: what is the future of data-driven science in the study of (biological) dynamical systems? 
And what are the most promising recent developments?  

\begin{figure*}[t!]
	\centering
	\includegraphics[width=\linewidth]{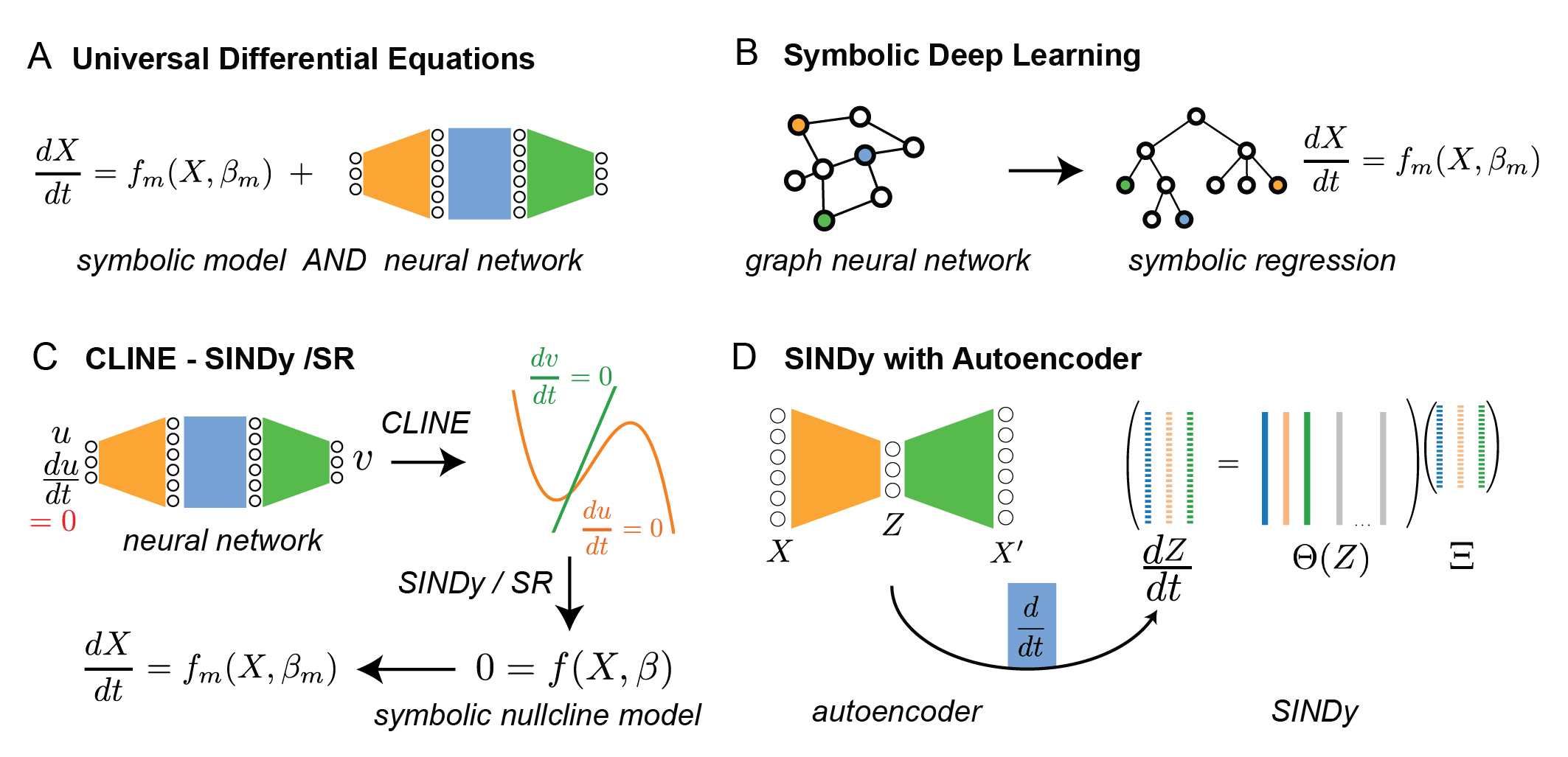}
	\caption[Selected hybrid methods combining regression- and network-based methodology]{Examples of hybrid methods integrating regression- and network-based approaches. 
	UDEs combine mechanistic equations with NNs; SDL translates NN structures into symbolic expressions; CLINE extracts nullclines which can be turned into equations via SINDy or SR; and SINDy with AEs balances interpretability with dimensionality reduction.}
	\label{fig:hybrid_methods}
\end{figure*}

\subsection{Experimental frontiers enabling data-driven modeling}
Some of the most pressing limitations arise not from methodology but from experimental constraints. 
Biological systems are often high-dimensional, noisy, and only partially observable. 
Advances in microscopy now offer ever higher spatio-temporal resolution, enabling detailed observations of cellular and subcellular processes \cite{Balasubramanian2023}. 
Novel sensors, including fluorescent reporters and FRET-based probes, further enhance the specificity and sensitivity of dynamic biochemical measurements \cite{Hickey2022,Specht2017}.  
Similarly, innovations in multi-omics promise richer molecular time series, although current techniques remain too slow and often destructive, complicating temporal reconstruction \cite{Bressan2023,Velten2023}. 
Recent pipelines such as TIMEOR (Trajectory Inference and Mechanism Exploration with Omics data in R) \cite{Conard2021}, MONFIT (multi-omics factorization-based integration of time-series data) \cite{Mihajlovic2024} or DANSE (Dynamic Analysis of multi-omics data) \cite{jansen2025danse} have begun to address this challenge by explicitly integrating time-resolved and multi-omics measurements into unified modeling frameworks (sometimes called omics-driven \cite{Espinel-Rios2025}), providing practical tools for high-dimensional dynamic inference \cite{Zhang2025}.
Another emerging direction is the use of synthetic surrogates: minimal artificial cells that reproduce aspects of biological dynamics while offering full experimental control \cite{Buddingh2017,Jiang2022, Wang2025}.  

These experimental advances will generate increasingly complex datasets. 
The challenge for data-driven methods is to use them effectively, motivating approaches that can both handle high-dimensional data and produce interpretable insight.

\subsection{Pseudotime-based analyses: strengths and caveats}

An additional class of approaches, widely used in single-cell biology, infers temporal orderings from static population-level data. 
{Pseudotime} methods assign cells a latent temporal coordinate based on transcriptomic similarity, thereby reconstructing trajectories of differentiation or oscillatory progression when true longitudinal measurements are unavailable \cite{Trapnell2014,Setty2016}. 
Extensions such as RNA velocity \cite{LaManno2018} and optimal-transport trajectory analysis \cite{Schiebinger2019} provide directionality and coarse rate information. 
These approaches have become indispensable for mapping cell states, branching topologies, and rare transitions, especially in developmental contexts.

For our purposes, pseudotime excels at identifying dynamical solutions in a phase space (Fig.~\ref{fig:understandingdynamicalsystems}F), but is much weaker for forecasting and characterizing interactions. 
Because pseudotime is not physical time, derivatives are ill-posed and regression-based or decomposition methods that rely on evenly spaced time series (e.g., SINDy, DMD) cannot be directly applied. \lgel{In practice, this limitation is compounded by the fact that pseudotime and trajectory inference are also highly sensitive to preprocessing choices, such as dimensionality reduction and clustering resolution—which define the effective state space. Recent bias-aware and stability-controlled methods such as scLENS \cite{Kim2024_scLENS} and scICE \cite{Kim2025_scICE} explicitly address this issue and can substantially improve the robustness of downstream dynamical analyses.}
Similarly, mechanistic inference is limited: pseudotime primarily recovers topology rather than explicit governing equations or reaction rates. 
Recently, it has been suggested that when facing oscillatory dynamics the predictive power of pseudotime methods can be severely impacted \cite{Vo2024}.
\lgel{More generally, purely geometric or optimal-transport–based trajectory reconstructions may reduce to connecting nearby cell states without capturing the underlying dynamical flow that governs biological progression. This has motivated a shift toward learning biologically meaningful vector fields on cell-state manifolds, enabling the reconstruction of disease progression, differentiation dynamics, and fate decisions from snapshot data \cite{Wijeratne2024,Hua2025,Sha2024}.}
Nevertheless, pseudotime can serve as a valuable complement by highlighting where dynamic changes occur, thereby guiding targeted acquisition of true time series data for downstream mechanistic modeling.

The field of pseudotime analysis is rapidly evolving, with close to a hundred tools developed to date just to mention the recent PseudoGA (cell pseudotime reconstruction based on genetic algorithm) \cite{Mondal2021}, CAPITAL (comparative analysis of pseudotime trajectory inference with tree alignment) \cite{Sugihara2022}, scTour \cite{Li2023} or CellRank \cite{Lange2022}. 
Saelens \textit{et al.}~\cite{Saelens2019} benchmarked 45 such methods (before 2019) across hundreds of real and synthetic datasets, concluding that no single approach is universally optimal. 
Instead, performance depends strongly on dataset size, dimensionality, and trajectory topology. 
Thus, while we do not evaluate pseudotime methods in the same way as regression-, network-, or decomposition-based approaches, we note that they provide a complementary entry point to dynamical analysis when destructive sampling prevents longitudinal measurements.

\subsection{Hybrid methods: combining strengths}

Recent work increasingly combines regression with neural architectures.
% As highlighted in Table~\ref{tab:dd_methods_advantages_disadvantages}, 
These two groups are complementary: regression emphasizes interpretability but struggles with high-dimensional data, while neural networks excel in flexibility but often act as black boxes. 
Hybrid methods aim to exploit the strengths of each.

Other combinations are possible, although explicit integrations with decomposition methods remain rare. 
Instead, decomposition is often embedded within network architectures such as autoencoders (Sec.~\ref{sec:decomposition_methods}), forming the basis of many modern hybrids \cite{Lusch2018,Iwata2020}. 
Initial applications to biological systems have shown promising results \cite{Yuan2021}.
Here we highlight several representative directions (Fig.~\ref{fig:hybrid_methods}).

\paragraph{Universal Differential Equations}
Universal differential equations (UDEs) \cite{Rackauckas2020} directly embed neural networks into mechanistic models. 
Known components of a system are described by classical differential equations $f_{\mathrm{m}}$, while unknown or poorly understood parts are represented by a NN:
\begin{equation}
	\frac{dX}{dt} = f_{\text{m}}(X,\beta_{\text{m}}) + \textit{NN}(X,\beta_{\text{NN}}).
\end{equation}
This allows training end-to-end while retaining temporal consistency. 
In biology, UDEs are attractive because many systems are partially characterized: known feedback loops can be modeled explicitly, while residual terms capture missing biology \cite{Philipps2024}. 
Applications range from gene regulatory networks \cite{Kana2021} to epidemic dynamics \cite{Rojas-Campos2023,Kuwahara2024} and ecological interactions \cite{Buckner2024}. 

\paragraph{Symbolic Deep Learning}
Symbolic deep learning (SDL) \cite{Cranmer2020} pursues a different strategy: rather than embedding NNs in equations, it translates trained networks into symbolic expressions. 
Typically, GNN first learns interaction rules; then symbolic regression (e.g., pySR) extracts interpretable equations. 
This approach has been applied to biochemical oscillations \cite{Boddupalli2023, AhmadiDaryakenari2024}, population dynamics \cite{Boddupalli2023}, and pharmacokinetics \cite{AhmadiDaryakenari2024}, combining the representational power of NNs with the interpretability of symbolic models. 

\paragraph{CLINE-SINDy/SR}
Recently, CLINE has been proposed as a model-free method to uncover geometric phase-space features such as nullclines \cite{Prokop2025}. 
Once identified, these structures can be converted into explicit equations via SINDy (with suitable libraries) or symbolic regression (requiring fewer priors). 
This reduces the task from full temporal fitting to geometric curve fitting, improving robustness. 
CLINE-SR has demonstrated the ability to identify strongly nonlinear functions using experimental biochemical data \cite{Prokop2025}, highlighting the potential of hybrid nullcline-based approaches. 

\paragraph{SINDy with Autoencoders}
SINDy struggles with high-dimensional data, while AEs excel at compressing dynamics into lower dimensions. 
Their combination \cite{Bakarji2022,Bakarji2023} leverages both: AEs discover reduced coordinates, and SINDy identifies interpretable governing equations within that latent space. 
This balance has been used, for example, to analyze \textit{C. elegans} locomotion \cite{Voina2024}.  
Additionally, this approach can be combined with methods aimed at uncovering missing state variables — for instance, delay embedding techniques suggested by Takens \cite{Takens1981} — which reconstruct the high-dimensional system from limited observations. 
Such integrations make SINDy--AE frameworks particularly attractive for biological and biochemical systems where measurements are sparse and incomplete.  

Beyond these frameworks, other emerging approaches combine neural architectures and dynamical landscapes. For example, neural flow maps extend neural ODEs toward improved time-series modeling in biological systems \cite{xu2025time}, while recent work on reconstructing Waddington-type landscapes directly from data \cite{cislo2025reconstructing,fontaine2025dynamic} highlights how high-dimensional learning can be linked to classical concepts in developmental biology.

\subsection{Future targets of data-driven methods in biology}
\label{sec:future_targets_dd_biology}
What, then, should be the goal of data-driven methods for dynamical systems in biology?
As discussed before, the primary objective is to enable a deeper understanding of dynamical systems in the context of underlying biological processes.
This understanding may relate to the ability to forecast system behavior, uncover hidden interactions, or determine the relevant state variables in a biological system.

Most of the methods presented throughout this review are capable of addressing one or more of these goals.
However, in biology, it is especially important to have tools that can holistically capture, describe, and integrate all the mechanisms that underlie the system of interest.

In our opinion, and as suggested by the comparative overview in Table~\ref{tab:data_driven_methods_overview}, this task can best be fulfilled by symbolic mathematical models in the form of differential equations. Such models explicitly describe interactions, mechanisms, and their influence on the system's state. When chosen appropriately, they are capable not only of forecasting system behavior but also of exploring \textit{what-if} scenarios. A ``good'' model is thus able to predict \textit{unknown} biology \textit{in silico} even before experiments are conducted. Of course, these predictions must ultimately be validated experimentally. This does not imply that such a model is ``correct'', as George Box noted in 1976: ``All models are wrong, but some are useful''. Indeed, there are numerous examples where mathematical modeling has been essential for uncovering biological mechanisms, and the importance of mathematics in biology has been widely recognized \cite{Gunawardena2014, Paluch2015, Goldstein2018, Phillips2015, Mogilner2006}.
\lgel{Differential equation models span a wide spectrum, from ODE-based descriptions that capture effective regulatory and network dynamics in state space to spatially extended (reaction-diffusion) PDEs that describe morphogenesis, tissue mechanics, and intracellular pattern formation. Data-driven discovery of such PDEs from spatio-temporal imaging has seen major advances, particularly in settings where dense, high-resolution fields are available \cite{Rudy_PDE_2017,Schaeffer_2017_RoySoc,LyonsDukicBortz2025, MessengerBortzIMA2024, MESSENGER2021110525,Messenger2021,Reinbold2020,Reinbold2021}. However, learning biological PDEs from experimental data remains a distinct and challenging inverse problem, as biological imaging often provides limited, noisy, and partially observed fields with complex and heterogeneous spatial coupling. 
For this reason, the present review has focused on inferring effective dynamical models from time-resolved biological measurements rather than on full-field PDE identification.}

Recent methodological advances illustrate how this vision is evolving.  
New approaches such as neural flow maps \cite{xu2025time} extend neural ODEs toward more expressive representations of time series, while pipelines like DANSE integrate multi-omics dynamics \cite{jansen2025danse}.  
At the same time, Waddington’s epigenetic landscape metaphor has led to machine learning methods that reconstruct developmental landscapes directly from data \cite{cislo2025reconstructing,fontaine2025dynamic}.  
Together, these efforts highlight how diverse research lines are converging on hybrid models that bridge mechanistic interpretability and data-driven flexibility.

Looking ahead, the future of data-driven biology will not be dominated by a single methodology but by hybrid approaches that combine the interpretability of symbolic equations, the flexibility of neural networks, and the robustness of probabilistic frameworks. 
Such methods may never yield a “correct” model in the absolute sense, but they can be profoundly {useful}.  
By distilling governing equations from data, uncovering hidden state variables, and integrating mechanistic priors with machine learning, the next generation of models will not only describe what is known but also predict the unknown, guiding experiments and shaping new biological theory.

\section*{Data availability statement}
The numerical codes to reproduce the figures in this study are openly available in \textsc{GitLab} \cite{gitlab_review2025}.

\section*{Acknowledgments}
L.G. acknowledges funding by the KU Leuven Research Fund (grant number C14/23/130 and IDN/25/007) and the Research Foundation - Flanders (FWO, grant number G074321N). 

\section*{Competing interests statement}
The authors declare no competing interests.

\bibliography{apssamp.bib}

@PREAMBLE{
 "\providecommand{\noopsort}[1]{}" 
 # "\providecommand{\singleletter}[1]{#1}%" 
}

@article{dunlap1999molecular,
  title={Molecular bases for circadian clocks},
  author={Dunlap, Jay C},
  journal={Cell},
  volume={96},
  number={2},
  pages={271--290},
  year={1999},
  publisher={Elsevier}
}

@article{hodgkin1939action,
author={A. L. {Hodgkin} and A. F. Huxley},
 year={1939},
title={Action Potentials Recorded from Inside a Nerve Fibre},
 journal={Nature},
 volume={144},
  pages={710--711},
xdoi={10.1038/144710a0},
 issn={0028-0836},
}

@article{geva2006oscillations,
  title={Oscillations and variability in the p53 system},
  author={Geva-Zatorsky, Naama and Rosenfeld, Nitzan and Itzkovitz, Shalev and Milo, Ron and Sigal, Alex and Dekel, Erez and Yarnitzky, Talia and Liron, Yuvalal and Polak, Paz and Lahav, Galit and others},
  journal={Molecular systems biology},
  volume={2},
  number={1},
  pages={2006--0033},
  year={2006},
  publisher={John Wiley \& Sons, Ltd Chichester, UK}
}

@article{ferrell2011modeling,
  title={Modeling the cell cycle: why do certain circuits oscillate?},
  author={Ferrell, James E and Tsai, Tony Yu-Chen and Yang, Qiong},
  journal={Cell},
  volume={144},
  number={6},
  pages={874--885},
  year={2011},
  publisher={Elsevier}
}

@article{yang2013cdk1,
  title={The Cdk1--APC/C cell cycle oscillator circuit functions as a time-delayed, ultrasensitive switch},
  author={Yang, Qiong and Ferrell Jr, James E},
  journal={Nature cell biology},
  volume={15},
  number={5},
  pages={519--525},
  year={2013},
  publisher={Nature Publishing Group UK London}
}

@article{newport1984regulation,
  title={Regulation of the cell cycle during early Xenopus development},
  author={Newport, John W and Kirschner, Marc W},
  journal={Cell},
  volume={37},
  number={3},
  pages={731--742},
  year={1984},
  publisher={Cell Press}
}

@article{king1996proteolysis,
  title={How proteolysis drives the cell cycle},
  author={King, Randall W and Deshaies, Raymond J and Peters, Jan-Michael and Kirschner, Marc W},
  journal={Science},
  volume={274},
  number={5293},
  pages={1652--1659},
  year={1996},
  publisher={American Association for the Advancement of Science}
}

@article{pomerening2005systems,
  title={Systems-level dissection of the cell-cycle oscillator: bypassing positive feedback produces damped oscillations},
  author={Pomerening, Joseph R and Kim, Sun Young and Ferrell, James E},
  journal={Cell},
  volume={122},
  number={4},
  pages={565--578},
  year={2005},
  publisher={Elsevier}
}

@article{de2021control,
  title={Control of osteoblast regeneration by a train of Erk activity waves},
  author={De Simone, Alessandro and Evanitsky, Maya N and Hayden, Luke and Cox, Ben D and Wang, Julia and Tornini, Valerie A and Ou, Jianhong and Chao, Anna and Poss, Kenneth D and Di Talia, Stefano},
  journal={Nature},
  volume={590},
  number={7844},
  pages={129--133},
  year={2021},
  publisher={Nature Publishing Group UK London}
}

@article{afanzar2020nucleus,
  title={The nucleus serves as the pacemaker for the cell cycle},
  author={Afanzar, Oshri and Buss, Garrison K and Stearns, Tim and Ferrell Jr, James E},
  journal={Elife},
  volume={9},
  pages={e59989},
  year={2020},
  publisher={eLife Sciences Publications, Ltd}
}

@article{deneke2016waves,
  title={Waves of Cdk1 activity in S phase synchronize the cell cycle in Drosophila embryos},
  author={Deneke, Victoria E and Melbinger, Anna and Vergassola, Massimo and Di Talia, Stefano},
  journal={Developmental cell},
  volume={38},
  number={4},
  pages={399--412},
  year={2016},
  publisher={Elsevier}
}

@article{chang2013mitotic,
  title={Mitotic trigger waves and the spatial coordination of the Xenopus cell cycle},
  author={Chang, Jeremy B and Ferrell Jr, James E},
  journal={Nature},
  volume={500},
  number={7464},
  pages={603--607},
  year={2013},
  publisher={Nature Publishing Group UK London}
}

@article{goldbeter2018dissipative,
  title={Dissipative structures in biological systems: bistability, oscillations, spatial patterns and waves},
  author={Goldbeter, Albert},
  journal={Philosophical Transactions of the Royal Society A: Mathematical, Physical and Engineering Sciences},
  volume={376},
  number={2124},
  pages={20170376},
  year={2018},
  publisher={The Royal Society Publishing}
}

@article{volpert2009reaction,
  title={Reaction--diffusion waves in biology},
  author={Volpert, Vitaly and Petrovskii, Sergei},
  journal={Physics of life reviews},
  volume={6},
  number={4},
  pages={267--310},
  year={2009},
  publisher={Elsevier}
}

@article{deneke2018chemical,
  title={Chemical waves in cell and developmental biology},
  author={Deneke, Victoria E and Di Talia, Stefano},
  journal={Journal of Cell Biology},
  volume={217},
  number={4},
  pages={1193--1204},
  year={2018},
  publisher={The Rockefeller University Press}
}

@article{beta2017intracellular,
  title={Intracellular oscillations and waves},
  author={Beta, Carsten and Kruse, Karsten},
  journal={Annual Review of Condensed Matter Physics},
  volume={8},
  number={1},
  pages={239--264},
  year={2017},
  publisher={Annual Reviews}
}

@article{gelens2014spatial,
  title={Spatial trigger waves: positive feedback gets you a long way},
  author={Gelens, Lendert and Anderson, Graham A and Ferrell Jr, James E},
  journal={Molecular biology of the cell},
  volume={25},
  number={22},
  pages={3486--3493},
  year={2014},
  publisher={The American Society for Cell Biology}
}

@article{novak2008design,
  title={Design principles of biochemical oscillators},
  author={Nov{\'a}k, B{\'e}la and Tyson, John J},
  journal={Nature reviews Molecular cell biology},
  volume={9},
  number={12},
  pages={981--991},
  year={2008},
  publisher={Nature Publishing Group UK London}
}

@article{tyson2003sniffers,
  title={Sniffers, buzzers, toggles and blinkers: dynamics of regulatory and signaling pathways in the cell},
  author={Tyson, John J and Chen, Katherine C and Novak, Bela},
  journal={Current opinion in cell biology},
  volume={15},
  number={2},
  pages={221--231},
  year={2003},
  publisher={Elsevier}
}

@article{goldbeter2012systems,
  title={Systems biology of cellular rhythms},
  author={Goldbeter, Albert and G{\'e}rard, Claude and Gonze, Didier and Leloup, J-C and Dupont, Genevieve},
  journal={FEBS letters},
  volume={586},
  number={18},
  pages={2955--2965},
  year={2012},
  publisher={Elsevier}
}

@article{chandra2011glycolytic,
  title={Glycolytic oscillations and limits on robust efficiency},
  author={Chandra, Fiona A and Buzi, Gentian and Doyle, John C},
  journal={science},
  volume={333},
  number={6039},
  pages={187--192},
  year={2011},
  publisher={American Association for the Advancement of Science}
}

@article{berridge1988cytosolic,
  title={Cytosolic calcium oscillators},
  author={Berridge, Michael J and Galione, Antony},
  journal={The FASEB Journal},
  volume={2},
  number={15},
  pages={3074--3082},
  year={1988},
  publisher={Wiley Online Library}
}

@book{izhikevich2007dynamical,
  title={Dynamical systems in neuroscience},
  author={Izhikevich, Eugene M},
  year={2007},
  publisher={MIT press}
}

@article{HodgkinHuxley,
author={Hodgkin, Alan L. and Huxley, Andrew F.},
 year={1952},
title={A quantitative description of membrane current and its application to
  conduction and excitation in nerve},
 journal={J. Physiol.},
 volume={117},
 number={4},
  pages={500},
 publisher={Wiley-Blackwell},
}

@article{purvis2012p53,
  title={p53 dynamics control cell fate},
  author={Purvis, Jeremy E and Karhohs, Kyle W and Mock, Caroline and Batchelor, Eric and Loewer, Alexander and Lahav, Galit},
  journal={Science},
  volume={336},
  number={6087},
  pages={1440--1444},
  year={2012},
  publisher={American Association for the Advancement of Science}
}

@article{batchelor2008recurrent,
  title={Recurrent initiation: a mechanism for triggering p53 pulses in response to DNA damage},
  author={Batchelor, Eric and Mock, Caroline S and Bhan, Irun and Loewer, Alexander and Lahav, Galit},
  journal={Molecular cell},
  volume={30},
  number={3},
  pages={277--289},
  year={2008},
  publisher={Elsevier}
}

@article{devreotes1983quantitative,
  title={Quantitative analysis of cyclic AMP waves mediating aggregation in Dictyostelium discoideum},
  author={Devreotes, Peter N and Potel, Michael J and MacKay, Stephen A},
  journal={Developmental biology},
  volume={96},
  number={2},
  pages={405--415},
  year={1983},
  publisher={Elsevier}
}

@article{tyson1989cyclic,
  title={Cyclic AMP waves during aggregation of Dictyostelium amoebae},
  author={Tyson, John J and Murray, JD},
  journal={Development},
  volume={106},
  number={3},
  pages={421--426},
  year={1989},
  publisher={The Company of Biologists Ltd}
}

@article{ghosh1964oscillations,
  title={Oscillations of glycolytic intermediates in yeast cells},
  author={Ghosh, A and Chance, Britton},
  journal={Biochemical and biophysical research communications},
  volume={16},
  number={2},
  pages={174--181},
  year={1964},
  publisher={Elsevier}
}

@article{bell2005circadian,
  title={Circadian rhythms from multiple oscillators: lessons from diverse organisms},
  author={Bell-Pedersen, Deborah and Cassone, Vincent M and Earnest, David J and Golden, Susan S and Hardin, Paul E and Thomas, Terry L and Zoran, Mark J},
  journal={Nature Reviews Genetics},
  volume={6},
  number={7},
  pages={544--556},
  year={2005},
  publisher={Nature Publishing Group UK London}
}

@article{borghans1997complex,
  title={Complex intracellular calcium oscillations A theoretical exploration of possible mechanisms},
  author={Borghans, Jos{\'e}A M and Dupont, Genevi{\`e}ve and Goldbeter, Albert},
  journal={Biophysical chemistry},
  volume={66},
  number={1},
  pages={25--41},
  year={1997},
  publisher={Elsevier}
}

@book{Strogatz2000,
	author = {Strogatz, Steven H.},
	booktitle = {Nonlinear Dynamics and Chaos: With Applications to Physics, Biology, Chemistry, and Engineering},
	doi = {10.1201/9780429492563},
	isbn = {9780429961113},
	month = {may},
	pages = {1--513},
	publisher = {CRC Press},
	title = {{Nonlinear Dynamics and Chaos}},
	url = {https://www.taylorfrancis.com/books/mono/10.1201/9780429492563/nonlinear-dynamics-chaos-steven-strogatz https://www.taylorfrancis.com/books/9780429961113},
	year = {2000}
}

@article{Krizhevsky2012,
	author = {Krizhevsky, Alex and Sutskever, Ilya and Hinton, Geoffrey E.},
	doi = {10.1145/3065386},
	issn = {0001-0782},
	journal = {Communications of the ACM},
	month = {may},
	number = {6},
	pages = {84--90},
	title = {{ImageNet classification with deep convolutional neural networks}},
	url = {http://code.google.com/p/cuda-convnet/ https://dl.acm.org/doi/10.1145/3065386},
	volume = {60},
	year = {2017}
}

@article{Puls2024,
	author = {Puls, Owen and Ruiz-Reyn{\'{e}}s, Daniel and Tavella, Franco and Jin, Minjun and Kim, Yeonghoon and Gelens, Lendert and Yang, Qiong},
	doi = {10.1038/s41467-024-54752-7},
	issn = {2041-1723},
	journal = {Nature Communications 2024 15:1},
	month = {dec},
	number = {1},
	pages = {1--15},
	pmid = {39622792},
	publisher = {Nature Publishing Group},
	title = {{Spatial heterogeneity accelerates phase-to-trigger wave transitions in frog egg extracts}},
	url = {https://www.nature.com/articles/s41467-024-54752-7},
	volume = {15},
	year = {2024}
}

@book{Arnold1978,
	address = {New York (N.Y.)},
	author = {Arnold, Vladimir Igorevich and Vogtmann, K},
	isbn = {0-387-90314-3},
	publisher = {Springer},
	series = {Graduate texts in mathematics 60},
	title = {{Mathematical methods of classical mechanics }},
	year = {1978}
}

@article{Alber2019,
	archivePrefix = {arXiv},
	arxivId = {1910.01258},
	author = {Alber, Mark and {Buganza Tepole}, Adrian and Cannon, William R. and De, Suvranu and Dura-Bernal, Salvador and Garikipati, Krishna and Karniadakis, George and Lytton, William W. and Perdikaris, Paris and Petzold, Linda and Kuhl, Ellen},
	doi = {10.1038/s41746-019-0193-y},
	eprint = {1910.01258},
	issn = {2398-6352},
	journal = {npj Digital Medicine},
	month = {nov},
	number = {1},
	pages = {115},
	publisher = {Nature Publishing Group},
	title = {{Integrating machine learning and multiscale modeling—perspectives, challenges, and opportunities in the biological, biomedical, and behavioral sciences}},
	url = {https://www.nature.com/articles/s41746-019-0193-y},
	volume = {2},
	year = {2019}
}

@article{Casdagli1989,
	author = {Casdagli, Martin},
	doi = {10.1016/0167-2789(89)90074-2},
	issn = {0167-2789},
	journal = {Physica D: Nonlinear Phenomena},
	month = {may},
	number = {3},
	pages = {335--356},
	publisher = {North-Holland},
	title = {{Nonlinear prediction of chaotic time series}},
	volume = {35},
	year = {1989}
}

@article{Abarbanel1990,
	author = {Abarbanel, Henry D.I. and Brown, Reggie and Kadtke, James B.},
	doi = {10.1103/PhysRevA.41.1782},
	issn = {10502947},
	journal = {Physical Review A},
	month = {feb},
	number = {4},
	pages = {1782},
	publisher = {American Physical Society},
	title = {{Prediction in chaotic nonlinear systems: Methods for time series with broadband Fourier spectra}},
	url = {https://journals.aps.org/pra/abstract/10.1103/PhysRevA.41.1782},
	volume = {41},
	year = {1990}
}

@article{Mindlin1992,
	author = {Mindlin, Gabriel M. and Gilmore, R.},
	doi = {10.1016/0167-2789(92)90111-Y},
	issn = {0167-2789},
	journal = {Physica D: Nonlinear Phenomena},
	month = {sep},
	number = {1-4},
	pages = {229--242},
	publisher = {North-Holland},
	title = {{Topological analysis and synthesis of chaotic time series}},
	volume = {58},
	year = {1992}
}

@article{GRASSBERGER2012,
	author = {Grassberger, Peter and Schreiber, Thomas and Schaffrath, Carsten},
	doi = {10.1142/S0218127491000403},
	issn = {0218-1274},
	journal = {International Journal of Bifurcation and Chaos},
	month = {jan},
	number = {03},
	pages = {521--547},
	publisher = { World Scientific Publishing Company },
	title = {{Nonlinear time sequence analysis}},
	url = {https://www.worldscientific.com/worldscinet/ijbc},
	volume = {01},
	year = {2012}
}

@article{Bucci2023,
	author = {Bucci, Michele Alessandro and Semeraro, Onofrio and Allauzen, Alexandre and Chibbaro, Sergio and Mathelin, Lionel},
	doi = {10.1140/EPJE/S10189-023-00269-8},
	issn = {1292-895X},
	journal = {The European Physical Journal E 2023 46:3},
	month = {mar},
	number = {3},
	pages = {1--14},
	pmid = {36884147},
	publisher = {Springer},
	title = {{Curriculum learning for data-driven modeling of dynamical systems}},
	url = {https://link.springer.com/article/10.1140/epje/s10189-023-00269-8},
	volume = {46},
	year = {2023}
}

@book{Hey2009,
	author = {Hey, Tony and Tansley, Stewart and Tolle, Kristin and Gray, Jim},
	isbn = {978-0-9825442-0-4},
	month = {oct},
	publisher = {Microsoft Research},
	title = {{The Fourth Paradigm: Data-Intensive Scientific Discovery}},
	url = {https://www.microsoft.com/en-us/research/publication/fourth-paradigm-data-intensive-scientific-discovery/},
	year = {2009}
}

@article{Brunton2022,
	archivePrefix = {arXiv},
	arxivId = {2102.12086},
	author = {Brunton, Steven L. and Budi{\v{s}}i{\'{c}}, Marko and Kaiser, Eurika and Kutz, J. Nathan},
	doi = {10.1137/21M1401243},
	eprint = {2102.12086},
	issn = {0036-1445},
	journal = {SIAM Review},
	month = {may},
	number = {2},
	pages = {229--340},
	publisher = {Society for Industrial and Applied Mathematics},
	title = {{Modern Koopman Theory for Dynamical Systems}},
	url = {https://doi.org/10.1137/21M1401243 https://epubs.siam.org/doi/10.1137/21M1401243},
	volume = {64},
	year = {2022}
}

@article{Koopman1931,
	author = {Koopman, B. O.},
	doi = {10.1073/PNAS.17.5.315},
	issn = {0027-8424},
	journal = {Proceedings of the National Academy of Sciences},
	month = {may},
	number = {5},
	pages = {315--318},
	pmid = {16577368},
	publisher = {National Academy of Sciences},
	title = {{Hamiltonian Systems and Transformation in Hilbert Space}},
	url = {https://www-pnas-org.kuleuven.e-bronnen.be/content/17/5/315 https://www-pnas-org.kuleuven.e-bronnen.be/content/17/5/315.abstract},
	volume = {17},
	year = {1931}
}

@article{Koopman1932,
	author = {Koopman, B. O. and Neumann, J. v.},
	doi = {10.1073/pnas.18.3.255},
	issn = {0027-8424},
	journal = {Proceedings of the National Academy of Sciences},
	month = {mar},
	number = {3},
	pages = {255--263},
	title = {{Dynamical Systems of Continuous Spectra}},
	url = {https://pnas.org/doi/full/10.1073/pnas.18.3.255},
	volume = {18},
	year = {1932}
}

@article{Schmid2010,
	author = {Schmid, Peter J.},
	doi = {10.1017/S0022112010001217},
	issn = {0022-1120},
	journal = {Journal of Fluid Mechanics},
	month = {aug},
	pages = {5--28},
	publisher = {Cambridge University Press},
	title = {{Dynamic mode decomposition of numerical and experimental data}},
	url = {https://www.cambridge.org/core/journals/journal-of-fluid-mechanics/article/dynamic-mode-decomposition-of-numerical-and-experimental-data/AA4C763B525515AD4521A6CC5E10DBD4 https://www.cambridge.org/core/product/identifier/S0022112010001217/type/journal_article},
	volume = {656},
	year = {2010}
}

@article{Udrescu2020,
	archivePrefix = {arXiv},
	arxivId = {1905.11481},
	author = {Udrescu, Silviu Marian and Tegmark, Max},
	doi = {10.1126/SCIADV.AAY2631},
	eprint = {1905.11481},
	issn = {23752548},
	journal = {Science Advances},
	month = {apr},
	number = {16},
	pmid = {32426452},
	publisher = {American Association for the Advancement of Science},
	title = {{AI Feynman: A physics-inspired method for symbolic regression}},
	url = {/doi/pdf/10.1126/sciadv.aay2631?download=true},
	volume = {6},
	year = {2020}
}

@article{CHEN1989,
	author = {Chen, S and Billings, S. A. and Luo, W.},
	doi = {10.1080/00207178908953472},
	issn = {0020-7179},
	journal = {International Journal of Control},
	month = {nov},
	number = {5},
	pages = {1873--1896},
	publisher = {Taylor \& Francis},
	title = {{Orthogonal least squares methods and their application to non-linear system identification}},
	url = {https://doi.org/10.1080/00207178908953472 http://www.tandfonline.com/doi/abs/10.1080/00207178908953472},
	volume = {50},
	year = {1989}
}

@article{jaeger2001echo,
	title={The “echo state” approach to analysing and training recurrent neural networks-with an erratum note},
	author={Jaeger, Herbert},
	journal={Bonn, Germany: German national research center for information technology gmd technical report},
	volume={148},
	number={34},
	pages={13},
	year={2001},
	publisher={Bonn}
}

@incollection{Rumelhart1987,
	author = {Rumelhart, D. E and MacClelland, J. L and Hinton, Geoffrey E.},
	booktitle = {Parallel Distributed Processing: Explorations in the Microstructure of Cognition: Foundations},
    publisher = {MIT Press},
	isbn = {9780262291408},
	pages = {45--76},
	title = {{General Framework for Parallel Distribution Processing}},
	url = {https://ieeexplore.ieee.org/servlet/opac?bknumber=6276825},
	year = {1987}
}

@article{Plaut1987,
	author = {Plaut, David C. and Hinton, Geoffrey E.},
	doi = {10.1016/0885-2308(87)90026-X},
	issn = {0885-2308},
	journal = {Computer Speech \& Language},
	month = {mar},
	number = {1},
	pages = {35--61},
	publisher = {Academic Press},
	title = {{Learning sets of filters using back-propagation}},
	url = {https://www.sciencedirect.com/science/article/pii/088523088790026X?via%3Dihub},
	volume = {2},
	year = {1987}
}

@article{Prokop2025,
	archivePrefix = {arXiv},
    journal = {arXiv},
	arxivId = {2503.16240},
	author = {Prokop, Bartosz and Billen, Jimmy and Frolov, Nikita and Gelens, Lendert},
	eprint = {2503.16240},
	isbn = {2503.16240v1},
	month = {mar},
	title = {{Machine learning identifies nullclines in oscillatory dynamical systems}},
	url = {https://arxiv.org/abs/2503.16240v1 http://arxiv.org/abs/2503.16240},
	year = {2025}
}

@inproceedings{Chen2018,
	author = {Chen, Ricky T Q and Rubanova, Yulia and Bettencourt, Jesse and Duvenaud, David K},
	booktitle = {Advances in Neural Information Processing Systems},
	editor = {Bengio, S and Wallach, H and Larochelle, H and Grauman, K and Cesa-Bianchi, N and Garnett, R},
	publisher = {Curran Associates, Inc.},
	title = {{Neural Ordinary Differential Equations}},
	url = {https://proceedings.neurips.cc/paper_files/paper/2018/file/69386f6bb1dfed68692a24c8686939b9-Paper.pdf},
	volume = {31},
	year = {2018}
}

@article{WISDOM1960,
	author = {Wisdom, J O},
	doi = {10.1038/187092a0},
	issn = {0028-0836},
	journal = {Nature},
	month = {jul},
	number = {4732},
	pages = {92--92},
	title = {{Causation and Modern Science}},
	url = {https://doi.org/10.1038/187092a0 https://www.nature.com/articles/187092a0},
	volume = {187},
	year = {1960}
}

@article{Guo2008,
	author = {Guo, Shuixia and Seth, Anil K. and Kendrick, Keith M. and Zhou, Cong and Feng, Jianfeng},
	doi = {10.1016/j.jneumeth.2008.04.011},
	issn = {01650270},
	journal = {Journal of Neuroscience Methods},
	month = {jul},
	number = {1},
	pages = {79--93},
	pmid = {18508128},
	publisher = {J Neurosci Methods},
	title = {{Partial Granger causality-Eliminating exogenous inputs and latent variables}},
	url = {https://pubmed.ncbi.nlm.nih.gov/18508128/},
	volume = {172},
	year = {2008}
}

@article{Nawrath2010,
	author = {Nawrath, Jakob and Romano, M. Carmen and Thiel, Marco and Kiss, Istv{\'{a}}n Z. and Wickramasinghe, Mahesh and Timmer, Jens and Kurths, J{\"{u}}rgen and Schelter, Bj{\"{o}}rn},
	doi = {10.1103/PHYSREVLETT.104.038701},
	issn = {00319007},
	journal = {Physical Review Letters},
	month = {jan},
	number = {3},
	pages = {038701},
	publisher = {American Physical Society},
	title = {{Distinguishing direct from indirect interactions in oscillatory networks with multiple time scales}},
	url = {https://journals.aps.org/prl/abstract/10.1103/PhysRevLett.104.038701},
	volume = {104},
	year = {2010}
}

@article{Stokes2017,
	author = {Stokes, Patrick A. and Purdon, Patrick L.},
	doi = {10.1073/PNAS.1704663114},
	issn = {10916490},
	journal = {Proceedings of the National Academy of Sciences of the United States of America},
	month = {aug},
	number = {34},
	pages = {E7063--E7072},
	pmid = {28778996},
	publisher = {National Academy of Sciences},
	title = {{A study of problems encountered in Granger causality analysis from a neuroscience perspective}},
	url = {/doi/pdf/10.1073/pnas.1704663114?download=true},
	volume = {114},
	year = {2017}
}

@article{Runge2019,
	author = {Runge, Jakob and Bathiany, Sebastian and Bollt, Erik and Camps-Valls, Gustau and Coumou, Dim and Deyle, Ethan and Glymour, Clark and Kretschmer, Marlene and Mahecha, Miguel D. and Mu{\~{n}}oz-Mar{\'{i}}, Jordi and van Nes, Egbert H. and Peters, Jonas and Quax, Rick and Reichstein, Markus and Scheffer, Marten and Sch{\"{o}}lkopf, Bernhard and Spirtes, Peter and Sugihara, George and Sun, Jie and Zhang, Kun and Zscheischler, Jakob},
	doi = {10.1038/s41467-019-10105-3},
	issn = {2041-1723},
	journal = {Nature Communications 2019 10:1},
	month = {jun},
	number = {1},
	pages = {1--13},
	pmid = {31201306},
	publisher = {Nature Publishing Group},
	title = {{Inferring causation from time series in Earth system sciences}},
	url = {https://www.nature.com/articles/s41467-019-10105-3},
	volume = {10},
	year = {2019}
}

@article{Chen1990,
	author = {Chen, S. and Billings, S. A. and Cowan, C. F. and Grant, P. M.},
	doi = {10.1080/00207179008953599},
	issn = {13665820},
	journal = {International Journal of Control},
	number = {5},
	pages = {1327--1350},
	publisher = {Taylor & Francis Group},
	title = {{Practical identification of narmax models using radial basis functions}},
	url = {https://www.tandfonline.com/doi/abs/10.1080/00207179008953599},
	volume = {52},
	year = {1990}
}

@article{Hong2008,
	author = {Hong, X. and Mitchell, R. J. and Chen, S. and Harris, C. J. and Li, K. and Irwin, G. W.},
	doi = {10.1080/00207720802083018},
	issn = {00207721},
	journal = {International Journal of Systems Science},
	month = {oct},
	number = {10},
	pages = {925--946},
	publisher = {Taylor & Francis Group},
	title = {{Model selection approaches for non-linear system identification: A review}},
	url = {https://www.tandfonline.com/doi/pdf/10.1080/00207720802083018},
	volume = {39},
	year = {2008}
}

@article{Billings1998,
	author = {Billings, S. A. and Mao, K. Z.},
	doi = {10.1080/00207729808929516},
	issn = {14645319},
	journal = {International Journal of Systems Science},
	number = {3},
	pages = {223--231},
	publisher = {Taylor & Francis Group},
	title = {{Structure detection for nonlinear rational models using genetic algorithms}},
	url = {https://www.tandfonline.com/doi/pdf/10.1080/00207729808929516},
	volume = {29},
	year = {1998}
}

@article{Khodadadi2023,
	author = {Khodadadi, Vahid and Rahatabad, Fereidoun Nowshiravan and Sheikhani, Ali and Dabanloo, Nader Jafarnia},
	doi = {10.4103/JMSS.JMSS_3_22},
	issn = {22287477},
	journal = {Journal of Medical Signals and Sensors},
	month = {jan},
	number = {1},
	pages = {29--39},
	publisher = {Isfahan University of Medical Sciences(IUMS)},
	title = {{Prediction of Biceps Muscle Electromyogram Signal Using a NARX Neural Network}},
	url = {https://journals.lww.com/jmss/fulltext/2023/13010/prediction_of_biceps_muscle_electromyogram_signal.4.aspx},
	volume = {13},
	year = {2023}
}

@article{Kukreja2003,
	author = {Kukreja, Sunil L. and Galiana, Henrietta L. and Kearney, Robert E.},
	doi = {10.1109/TBME.2002.803507},
	issn = {00189294},
	journal = {IEEE Transactions on Biomedical Engineering},
	month = {jan},
	number = {1},
	pages = {70--81},
	pmid = {12617526},
	title = {{NARMAX representation and identification of ankle dynamics}},
	volume = {50},
	year = {2003}
}

@article{Krishnanathan2012,
	author = {Krishnanathan, Kirubhakaran and Anderson, Sean R. and Billings, Stephen A. and Kadirkamanathan, Visakan},
	doi = {10.1021/SB300009T},
	issn = {21615063},
	journal = {ACS Synthetic Biology},
	month = {aug},
	number = {8},
	pages = {375--384},
	pmid = {23651291},
	publisher = {American Chemical Society},
	title = {{A data-Driven framework for identifying nonlinear dynamic models of genetic parts}},
	url = {/doi/pdf/10.1021/sb300009t},
	volume = {1},
	year = {2012}
}

@article{Mendes1998,
	author = {Mendes, Pedro and Kell, Douglas B.},
	doi = {10.1093/BIOINFORMATICS/14.10.869,},
	issn = {13674811},
	journal = {Bioinformatics},
	number = {10},
	pages = {869--883},
	pmid = {9927716},
	publisher = {Oxford University Press},
	title = {{Non-linear optimization of biochemical pathways: Applications to metabolic engineering and parameter estimation}},
	url = {https://pubmed.ncbi.nlm.nih.gov/9927716/},
	volume = {14},
	year = {1998}
}

@article{MESSENGER2021110525,
title = {Weak SINDy for partial differential equations},
journal = {Journal of Computational Physics},
volume = {443},
pages = {110525},
year = {2021},
issn = {0021-9991},
doi = {https://doi.org/10.1016/j.jcp.2021.110525},
url = {https://www.sciencedirect.com/science/article/pii/S0021999121004204},
author = {Daniel A. Messenger and David M. Bortz}
}

@article{Messenger2022,
	archivePrefix = {arXiv},
	arxivId = {2204.14141},
	author = {Messenger, Daniel A. and Wheeler, Graycen E. and Liu, Xuedong and Bortz, David M.},
	doi = {10.1098/RSIF.2022.0412},
	eprint = {2204.14141},
	issn = {17425662},
	journal = {Journal of the Royal Society Interface},
	month = {oct},
	number = {195},
	publisher = {The Royal Society},
	title = {{Learning anisotropic interaction rules from individual trajectories in a heterogeneous cellular population}},
	url = {/doi/pdf/10.1098/rsif.2022.0412},
	volume = {19},
	year = {2022}
}

@article{Naozuka2025,
	author = {Naozuka, Gustavo T. and Rocha, Heber L. and Pereira, Thiago J. and Libotte, Gustavo B. and Almeida, Regina C.},
	doi = {10.1109/TCBBIO.2025.3565543},
	issn = {2998-4165},
	journal = {IEEE Transactions on Computational Biology and Bioinformatics},
	pages = {1--13},
	title = {{Connecting different approaches for cell cycle modeling: learning ordinary differential equations from individual-based models}},
	url = {https://ieeexplore.ieee.org/document/10979884/},
	year = {2025}
}

@book{eiben2015introduction,
	title={Introduction to evolutionary computing},
	author={Eiben, Agoston E and Smith, James E},
	year={2015},
	publisher={Springer}
}

@article{Hou2024,
	archivePrefix = {arXiv},
	arxivId = {2402.05238v1},
	author = {Hou, Jixin and Chen, Xianyan and Wu, Taotao and Kuhl, Ellen and Wang, Xianqiao},
	doi = {10.1016/j.actbio.2024.09.005},
	eprint = {2402.05238v1},
	journal = {Acta Biomaterialia},
	month = {feb},
	pages = {276--296},
	title = {{Automated Data-Driven Discovery of Material Models Based on Symbolic Regression: A Case Study on Human Brain Cortex}},
	url = {http://arxiv.org/abs/2402.05238 http://dx.doi.org/10.1016/j.actbio.2024.09.005},
	volume = {188},
	year = {2024}
}

@article{Haldar2024,
	archivePrefix = {arXiv},
	arxivId = {2410.16109},
	author = {Haldar, Swagatam and Stein-Thoeringer, Christoph and Borisov, Vadim},
	eprint = {2410.16109},
	month = {oct},
    journal={Arxiv},
	title = {{Interpreting Microbiome Relative Abundance Data Using Symbolic Regression}},
	url = {https://arxiv.org/pdf/2410.16109},
	year = {2024}
}

@article{Chen2019,
	author = {Chen, Yize and Angulo, Marco Tulio and Liu, Yang Yu},
	doi = {10.1002/BIES.201900069},
	issn = {15211878},
	journal = {BioEssays : news and reviews in molecular, cellular and developmental biology},
	month = {dec},
	number = {12},
	pages = {e1900069},
	pmid = {31617228},
	publisher = {John Wiley and Sons Inc.},
	title = {{Revealing Complex Ecological Dynamics via Symbolic Regression}},
	url = {https://pmc.ncbi.nlm.nih.gov/articles/PMC7339472/},
	volume = {41},
	year = {2019}
}

@article{Beauregard2019,
	author = {Beauregard, Nicole and Marten, Mark and Harris, Steven and Srivastava, Ranjan},
	journal = {Biochemical and Molecular Engineering XXI},
	month = {jul},
	title = {{Enhanced symbolic regression to infer biochemical network models}},
	url = {https://dc.engconfintl.org/biochem_xxi/26},
	year = {2019}
}

@article{Rupe2024,
    journal={Arxiv},
	archivePrefix = {arXiv},
	arxivId = {2410.10103},
	author = {Rupe, Adam and DeSantis, Derek and Bakker, Craig and Kooloth, Parvathi and Lu, Jian},
	eprint = {2410.10103},
	month = {oct},
	title = {{Causal Discovery in Nonlinear Dynamical Systems using Koopman Operators}},
	url = {https://arxiv.org/pdf/2410.10103 http://arxiv.org/abs/2410.10103},
	year = {2024}
}

@article{DiAntonio2024,
    journal={Arxiv},
	archivePrefix = {arXiv},
	arxivId = {2410.08708v2},
	author = {{Di Antonio}, Gabriele and Vinci, Gianni Valerio},
	eprint = {2410.08708v2},
	month = {oct},
	title = {{Non-linear correlations underlie linear response and causality}},
	url = {https://arxiv.org/pdf/2410.08708v2},
	year = {2024}
}

@article{Prokop2024,
	author = {Prokop, Bartosz and Gelens, Lendert},
	doi = {10.1016/J.ISCI.2024.109316},
	issn = {2589-0042},
	journal = {iScience},
	month = {apr},
	number = {4},
	pages = {109316},
	publisher = {Elsevier},
	title = {{From biological data to oscillator models using SINDy}},
	volume = {27},
	year = {2024}
}

@article{
Rudy_PDE_2017,
author = {Samuel H. Rudy  and Steven L. Brunton  and Joshua L. Proctor  and J. Nathan Kutz },
title = {Data-driven discovery of partial differential equations},
journal = {Science Advances},
volume = {3},
number = {4},
pages = {e1602614},
year = {2017},
doi = {10.1126/sciadv.1602614},
URL = {https://www.science.org/doi/abs/10.1126/sciadv.1602614},
eprint = {https://www.science.org/doi/pdf/10.1126/sciadv.1602614}}

@book{Minsky1969,
	address = {Oxford, England},
	author = {Minsky, Marvin and Papert, Seymour},
	booktitle = {Perceptrons.},
	publisher = {M.I.T. Press},
	title = {{Perceptrons.}},
	year = {1969}
}

@article{Hornik1989,
	author = {Hornik, Kurt and Stinchcombe, Maxwell and White, Halbert},
	doi = {10.1016/0893-6080(89)90020-8},
	issn = {08936080},
	journal = {Neural Networks},
	month = {jan},
	number = {5},
	pages = {359--366},
	publisher = {Pergamon},
	title = {{Multilayer feedforward networks are universal approximators}},
	url = {https://www.sciencedirect.com/science/article/pii/0893608089900208 https://linkinghub.elsevier.com/retrieve/pii/0893608089900208},
	volume = {2},
	year = {1989}
}

@article{Kingma2013,
	archivePrefix = {arXiv},
	arxivId = {1312.6114},
	author = {Kingma, Diederik P. and Welling, Max},
	doi = {10.61603/ceas.v2i1.33},
	eprint = {1312.6114},
	journal = {2nd International Conference on Learning Representations, ICLR 2014 - Conference Track Proceedings},
	month = {dec},
	publisher = {International Conference on Learning Representations, ICLR},
	title = {{Auto-Encoding Variational Bayes}},
	url = {https://arxiv.org/pdf/1312.6114 http://arxiv.org/abs/1312.6114},
	year = {2013}
}

@article{Yazdani2020,
	author = {Yazdani, Alireza and Lu, Lu and Raissi, Maziar and Karniadakis, George Em},
	doi = {10.1371/journal.pcbi.1007575},
	editor = {Hatzimanikatis, Vassily},
	isbn = {1111111111},
	issn = {1553-7358},
	journal = {PLOS Computational Biology},
	month = {nov},
	number = {11},
	pages = {e1007575},
	pmid = {33206658},
	publisher = {Public Library of Science},
	title = {{Systems biology informed deep learning for inferring parameters and hidden dynamics}},
	url = {https://journals.plos.org/ploscompbiol/article?id=10.1371/journal.pcbi.1007575 https://dx.plos.org/10.1371/journal.pcbi.1007575},
	volume = {16},
	year = {2020}
}

@article{Raissi2018,
	archivePrefix = {arXiv},
	arxivId = {1708.00588},
	author = {Raissi, Maziar and Karniadakis, George Em},
	doi = {10.1016/j.jcp.2017.11.039},
	eprint = {1708.00588},
	issn = {10902716},
	journal = {Journal of Computational Physics},
	month = {mar},
	pages = {125--141},
	publisher = {Academic Press Inc.},
	title = {{Hidden physics models: Machine learning of nonlinear partial differential equations}},
	volume = {357},
	year = {2018}
}

@article{Suykens1996,
	author = {Suykens, Johan A. K. and Vandewalle, Joos P. L. and {De Moor}, Bart L. R.},
	doi = {10.1007/978-1-4757-2493-6},
	journal = {Artificial Neural Networks for Modelling and Control of Non-Linear Systems},
	publisher = {Springer US},
	title = {{Artificial Neural Networks for Modelling and Control of Non-Linear Systems}},
	year = {1996}
}

@article{Narendra1992,
	author = {Narendra, Kumpati S. and Parthasarathy, Kannan},
	doi = {10.1016/0888-613X(92)90014-Q},
	issn = {0888-613X},
	journal = {International Journal of Approximate Reasoning},
	month = {feb},
	number = {2},
	pages = {109--131},
	publisher = {Elsevier},
	title = {{Neural networks and dynamical systems}},
	url = {https://www.sciencedirect.com/science/article/pii/0888613X9290014Q?utm_source=chatgpt.com},
	volume = {6},
	year = {1992}
}

@book{sandberg2001nonlinear,
	title={Nonlinear dynamical systems: feedforward neural network perspectives},
	author={Sandberg, Irwin W and Lo, James T and Fancourt, Craig L and Principe, Jose C and Katagiri, Shigeru and Haykin, Simon},
	year={2001},
	publisher={John Wiley \& Sons}
}

@article{Chattopadhyay2020,
	archivePrefix = {arXiv},
	arxivId = {2002.11167},
	author = {Chattopadhyay, Ashesh and Subel, Adam and Hassanzadeh, Pedram},
	doi = {10.1029/2020MS002084},
	eprint = {2002.11167},
	issn = {19422466},
	journal = {Journal of Advances in Modeling Earth Systems},
	month = {nov},
	number = {11},
	publisher = {Blackwell Publishing Ltd},
	title = {{Data-Driven Super-Parameterization Using Deep Learning: Experimentation With Multiscale Lorenz 96 Systems and Transfer Learning}},
	volume = {12},
	year = {2020}
}

@article{Nilsson2022,
	author = {Nilsson, Avlant and Peters, Joshua M. and Meimetis, Nikolaos and Bryson, Bryan and Lauffenburger, Douglas A.},
	doi = {10.1038/S41467-022-30684-Y},
	issn = {20411723},
	journal = {Nature Communications},
	month = {dec},
	number = {1},
	pages = {3069},
	pmid = {35654811},
	publisher = {Nature Research},
	title = {{Artificial neural networks enable genome-scale simulations of intracellular signaling}},
	url = {https://pmc.ncbi.nlm.nih.gov/articles/PMC9163072/},
	volume = {13},
	year = {2022}
}

@article{Chen2022,
	author = {Chen, Feng and Li, Chunhe},
	doi = {10.1093/NARGAB/LQAC068},
	issn = {26319268},
	journal = {NAR Genomics and Bioinformatics},
	month = {sep},
	number = {3},
	pages = {lqac068},
	pmid = {36110897},
	publisher = {Oxford University Press},
	title = {{Inferring structural and dynamical properties of gene networks from data with deep learning}},
	url = {https://pmc.ncbi.nlm.nih.gov/articles/PMC9469930/},
	volume = {4},
	year = {2022}
}

@article{Choo2020,
	author = {Choo, Sang Mok and Almomani, Laith M. and Cho, Kwang Hyun},
	doi = {10.3389/FPHYS.2020.594151/BIBTEX},
	issn = {1664042X},
	journal = {Frontiers in Physiology},
	month = {dec},
	pages = {594151},
	publisher = {Frontiers Media S.A.},
	title = {{Boolean Feedforward Neural Network Modeling of Molecular Regulatory Networks for Cellular State Conversion}},
	url = {www.frontiersin.org},
	volume = {11},
	year = {2020}
}

@article{Nagarajan2013,
	author = {Nagarajan, Radhakrishnan and Jonkman, Jeffrey N.},
	doi = {10.1371/JOURNAL.PONE.0053225,},
	issn = {19326203},
	journal = {PLoS ONE},
	month = {jan},
	number = {1},
	pmid = {23308165},
	publisher = {PLoS One},
	title = {{A Neural Network Model to Translate Brain Developmental Events across Mammalian Species}},
	url = {https://pubmed.ncbi.nlm.nih.gov/23308165/ https://pubmed.ncbi.nlm.nih.gov/23308165/?utm_source=chatgpt.com},
	volume = {8},
	year = {2013}
}

@article{Zhang2021,
	archivePrefix = {arXiv},
	arxivId = {1611.03530},
	author = {Zhang, Chiyuan and Bengio, Samy and Hardt, Moritz and Recht, Benjamin and Vinyals, Oriol},
	doi = {10.1145/3446776},
	eprint = {1611.03530},
	issn = {0001-0782},
	journal = {Communications of the ACM},
	month = {mar},
	number = {3},
	pages = {107--115},
	publisher = {Association for Computing Machinery},
	title = {{Understanding deep learning (still) requires rethinking generalization}},
	url = {https://arxiv.org/pdf/1611.03530 https://dl.acm.org/doi/10.1145/3446776},
	volume = {64},
	year = {2021}
}

@article{Pan2018,
	archivePrefix = {arXiv},
	arxivId = {1805.12547},
	author = {Pan, Shaowu and Duraisamy, Karthik},
	doi = {10.1155/2018/4801012},
	editor = {Chinesta, Francisco},
	eprint = {1805.12547},
	issn = {1076-2787},
	journal = {Complexity},
	month = {jan},
	number = {1},
	pages = {4801012},
	publisher = {John Wiley & Sons, Ltd},
	title = {{Long‐Time Predictive Modeling of Nonlinear Dynamical Systems Using Neural Networks}},
	url = {/doi/pdf/10.1155/2018/4801012 https://onlinelibrary.wiley.com/doi/abs/10.1155/2018/4801012 https://onlinelibrary.wiley.com/doi/10.1155/2018/4801012},
	volume = {2018},
	year = {2018}
}

@incollection{Wang2005,
	author = {Wang, Wen and {Van Gelder}, Pieter H. A. J. M. and Vrijling, J. K.},
	booktitle = {Lecture Notes in Computer Science (including subseries Lecture Notes in Artificial Intelligence and Lecture Notes in Bioinformatics)},
	doi = {10.1007/11550907_88},
	isbn = {978-3-540-28756-8},
	issn = {1611-3349},
	pages = {559--564},
	publisher = {Springer, Berlin, Heidelberg},
	title = {{Some Issues About the Generalization of Neural Networks for Time Series Prediction}},
	url = {https://link.springer.com/chapter/10.1007/11550907_88 http://link.springer.com/10.1007/11550907_88},
	volume = {3697 LNCS},
	year = {2005}
}

@article{Vlachas2018,
	archivePrefix = {arXiv},
	arxivId = {1802.07486},
	author = {Vlachas, Pantelis R. and Byeon, Wonmin and Wan, Zhong Y. and Sapsis, Themistoklis P. and Koumoutsakos, Petros},
	doi = {10.1098/RSPA.2017.0844},
	eprint = {1802.07486},
	issn = {14712946},
	journal = {Proceedings of the Royal Society A: Mathematical, Physical and Engineering Sciences},
	month = {may},
	number = {2213},
	publisher = {Royal Society Publishing},
	title = {{Data-driven forecasting of high-dimensional chaotic systems with long short-Term memory networks}},
	url = {/doi/pdf/10.1098/rspa.2017.0844},
	volume = {474},
	year = {2018}
}

@article{Lusch2018,
	archivePrefix = {arXiv},
	arxivId = {1712.09707},
	author = {Lusch, Bethany and Kutz, J. Nathan and Brunton, Steven L.},
	doi = {10.1038/S41467-018-07210-0},
	eprint = {1712.09707},
	issn = {20411723},
	journal = {Nature Communications},
	month = {dec},
	number = {1},
	pages = {1--10},
	pmid = {30470743},
	publisher = {Nature Publishing Group},
	title = {{Deep learning for universal linear embeddings of nonlinear dynamics}},
	url = {https://www.nature.com/articles/s41467-018-07210-0},
	volume = {9},
	year = {2018}
}

@article{Cestnik2019,
	archivePrefix = {arXiv},
	arxivId = {1904.03026},
	author = {Cestnik, Rok and Abel, Markus},
	doi = {10.1063/1.5096918/322115},
	eprint = {1904.03026},
	issn = {10541500},
	journal = {Chaos},
	month = {jun},
	number = {6},
	pages = {63128},
	pmid = {31266337},
	publisher = {American Institute of Physics Inc.},
	title = {{Inferring the dynamics of oscillatory systems using recurrent neural networks}},
	url = {/aip/cha/article/29/6/063128/322115/Inferring-the-dynamics-of-oscillatory-systems},
	volume = {29},
	year = {2019}
}

@article{Baranwal2022,
	author = {Baranwal, Mayank and Clark, Ryan L. and Thompson, Jaron and Sun, Zeyu and Hero, Alfred O. and Venturelli, Ophelia S.},
	doi = {10.7554/ELIFE.73870},
	issn = {2050084X},
	journal = {eLife},
	month = {jun},
	pmid = {35736613},
	publisher = {eLife Sciences Publications Ltd},
	title = {{Recurrent neural networks enable design of multifunctional synthetic human gut microbiome dynamics}},
	volume = {11},
	year = {2022}
}

@article{Samarasinghe2017,
	author = {Samarasinghe, S. and Ling, H.},
	doi = {10.1016/j.biosystems.2017.01.001},
	issn = {18728324},
	journal = {BioSystems},
	month = {mar},
	pages = {6--25},
	pmid = {28174135},
	publisher = {Elsevier Ireland Ltd},
	title = {{A system of recurrent neural networks for modularising, parameterising and dynamic analysis of cell signalling networks}},
	url = {https://pubmed.ncbi.nlm.nih.gov/28174135/ https://pubmed.ncbi.nlm.nih.gov/28174135/?utm_source=chatgpt.com},
	volume = {153-154},
	year = {2017}
}

@article{Jia2019,
	archivePrefix = {arXiv},
	arxivId = {1810.13075},
	author = {Jia, Xiaowei and Willard, Jared and Karpatne, Anuj and Read, Jordan and Zwart, Jacob and Steinbach, Michael and Kumar, Vipin},
	doi = {10.1137/1.9781611975673.63},
	eprint = {1810.13075},
	isbn = {9781611975673},
	journal = {SIAM International Conference on Data Mining, SDM 2019},
	pages = {558--566},
	publisher = {Society for Industrial and Applied Mathematics Publications},
	title = {{Physics guided RNNs for modeling dynamical systems: A case study in simulating lake temperature profiles}},
	url = {/doi/pdf/10.1137/1.9781611975673.63?download=true},
	year = {2019}
}

@article{Gauthier2021,
	archivePrefix = {arXiv},
	arxivId = {2106.07688},
	author = {Gauthier, Daniel J. and Bollt, Erik and Griffith, Aaron and Barbosa, Wendson A.S.},
	doi = {10.1038/S41467-021-25801-2;TECHMETA=119,129;SUBJMETA=1042,166,639,705,987;KWRD=COMPUTATIONAL+SCIENCE,ELECTRICAL+AND+ELECTRONIC+ENGINEERING},
	eprint = {2106.07688},
	issn = {20411723},
	journal = {Nature Communications},
	month = {dec},
	number = {1},
	pages = {1--8},
	pmid = {34548491},
	publisher = {Nature Research},
	title = {{Next generation reservoir computing}},
	url = {https://www.nature.com/articles/s41467-021-25801-2},
	volume = {12},
	year = {2021}
}

@article{Tanaka2019,
	archivePrefix = {arXiv},
	arxivId = {1808.04962},
	author = {Tanaka, Gouhei and Yamane, Toshiyuki and H{\'{e}}roux, Jean Benoit and Nakane, Ryosho and Kanazawa, Naoki and Takeda, Seiji and Numata, Hidetoshi and Nakano, Daiju and Hirose, Akira},
	doi = {10.1016/J.NEUNET.2019.03.005},
	eprint = {1808.04962},
	issn = {0893-6080},
	journal = {Neural Networks},
	month = {jul},
	pages = {100--123},
	pmid = {30981085},
	publisher = {Pergamon},
	title = {{Recent advances in physical reservoir computing: A review}},
	url = {https://www.sciencedirect.com/science/article/pii/S0893608019300784},
	volume = {115},
	year = {2019}
}

@article{Vidal-Saez2024,
	archivePrefix = {arXiv},
	arxivId = {2402.05243},
	author = {Vidal-Saez, Mar{\'{i}}a Sol and Vilarroya, Oscar and Garcia-Ojalvo, Jordi},
	doi = {10.1016/J.BBRC.2024.150301},
	eprint = {2402.05243},
	issn = {0006-291X},
	journal = {Biochemical and Biophysical Research Communications},
	month = {oct},
	pages = {150301},
	pmid = {38971000},
	publisher = {Academic Press},
	title = {{Biological computation through recurrence}},
	url = {https://www.sciencedirect.com/science/article/pii/S0006291X24008374},
	volume = {728},
	year = {2024}
}

@article{Vidal-Saez2025,
	author = {Vidal-Saez, Maria Sol and Garcia-Ojalvo, Jordi},
	doi = {10.1007/S12551-025-01295-W},
	issn = {18672469},
	journal = {Biophysical Reviews},
	month = {apr},
	number = {2},
	pages = {259--269},
	publisher = {Springer Science and Business Media Deutschland GmbH},
	title = {{Structural determinants of soft memory in recurrent biological networks}},
	url = {https://link.springer.com/article/10.1007/s12551-025-01295-w},
	volume = {17},
	year = {2025}
}

@article{Bourlard1988,
	author = {Bourlard, H. and Kamp, Y.},
	doi = {10.1007/BF00332918},
	issn = {03401200},
	journal = {Biological Cybernetics},
	month = {sep},
	number = {4-5},
	pages = {291--294},
	pmid = {3196773},
	publisher = {Springer-Verlag},
	title = {{Auto-association by multilayer perceptrons and singular value decomposition}},
	url = {https://link.springer.com/article/10.1007/BF00332918},
	volume = {59},
	year = {1988}
}

@article{Hinton2006,
	author = {Hinton, Geoffrey E. and Osindero, Simon and Teh, Yee Whye},
	doi = {10.1162/NECO.2006.18.7.1527},
	issn = {0899-7667},
	journal = {Neural Computation},
	month = {jul},
	number = {7},
	pages = {1527--1554},
	pmid = {16764513},
	publisher = {MIT Press},
	title = {{A Fast Learning Algorithm for Deep Belief Nets}},
	url = {https://dx.doi.org/10.1162/neco.2006.18.7.1527},
	volume = {18},
	year = {2006}
}

@article{Otto2019,
	archivePrefix = {arXiv},
	arxivId = {1712.01378},
	author = {Otto, Samuel E. and Rowley, Clarence W.},
	doi = {10.1137/18M1177846},
	eprint = {1712.01378},
	issn = {15360040},
	journal = {SIAM Journal on Applied Dynamical Systems},
	month = {mar},
	number = {1},
	pages = {558--593},
	publisher = {Society for Industrial and Applied Mathematics Publications},
	title = {{Linearly recurrent autoencoder networks for learning dynamics}},
	url = {/doi/pdf/10.1137/18M1177846?download=true},
	volume = {18},
	year = {2019}
}

@article{Sondak2021,
	archivePrefix = {arXiv},
	arxivId = {2011.07346},
	author = {Sondak, David and Protopapas, Pavlos},
	doi = {10.1103/PHYSREVE.104.034202},
	eprint = {2011.07346},
	issn = {24700053},
	journal = {Physical Review E},
	month = {sep},
	number = {3},
	pages = {034202},
	pmid = {34654102},
	publisher = {American Physical Society},
	title = {{Learning a reduced basis of dynamical systems using an autoencoder}},
	url = {https://journals.aps.org/pre/abstract/10.1103/PhysRevE.104.034202},
	volume = {104},
	year = {2021}
}

@article{Wehmeyer2018,
	archivePrefix = {arXiv},
	arxivId = {1710.11239},
	author = {Wehmeyer, Christoph and No{\'{e}}, Frank},
	doi = {10.1063/1.5011399/958887},
	eprint = {1710.11239},
	issn = {00219606},
	journal = {Journal of Chemical Physics},
	month = {jun},
	number = {24},
	pages = {241703},
	pmid = {29960344},
	publisher = {American Institute of Physics Inc.},
	title = {{Time-lagged autoencoders: Deep learning of slow collective variables for molecular kinetics}},
	url = {/aip/jcp/article/148/24/241703/958887/Time-lagged-autoencoders-Deep-learning-of-slow},
	volume = {148},
	year = {2018}
}

@article{Liu2021,
	author = {Liu, Suran and You, Yujie and Tong, Zhaoqi and Zhang, Le},
	doi = {10.3389/FGENE.2021.761629},
	issn = {16648021},
	journal = {Frontiers in Genetics},
	month = {oct},
	pages = {761629},
	publisher = {Frontiers Media S.A.},
	title = {{Developing an Embedding, Koopman and Autoencoder Technologies-Based Multi-Omics Time Series Predictive Model (EKATP) for Systems Biology research}},
	url = {www.frontiersin.org},
	volume = {12},
	year = {2021}
}

@article{Baig2023,
	author = {Baig, Yasa and Ma, Helena R. and Xu, Helen and You, Lingchong},
	doi = {10.1038/S41467-023-43455-0},
	issn = {20411723},
	journal = {Nature Communications},
	month = {dec},
	number = {1},
	pages = {1--17},
	pmid = {38049401},
	publisher = {Nature Research},
	title = {{Autoencoder neural networks enable low dimensional structure analyses of microbial growth dynamics}},
	url = {https://www.nature.com/articles/s41467-023-43455-0},
	volume = {14},
	year = {2023}
}

@article{Maizels2024,
	author = {Maizels, Rory J. and Snell, Daniel M. and Briscoe, James},
	doi = {10.1016/J.CELS.2024.04.004},
	issn = {2405-4712},
	journal = {Cell Systems},
	month = {may},
	number = {5},
	pages = {411--424.e9},
	pmid = {38754365},
	publisher = {Cell Press},
	title = {{Reconstructing developmental trajectories using latent dynamical systems and time-resolved transcriptomics}},
	url = {https://www.sciencedirect.com/science/article/pii/S2405471224001194?utm_source=chatgpt.com},
	volume = {15},
	year = {2024}
}

@article{Shrivastava2025,
	author = {Anika Shrivastava and Renu Rameshan and Samar Agnihotri},
    journal={Arxiv},
	doi = {10.5220/0013123700003912},
	issn = {21844321},
	month = {1},
	title = {Latent Space Characterization of Autoencoder Variants},
	url = {http://arxiv.org/abs/2412.04755},
	year = {2025},
}

@article{Asperti2021,
	author = {Andrea Asperti and Davide Evangelista and Elena Loli Piccolomini},
	doi = {10.1007/s42979-021-00702-9},
	issn = {2662-995X},
	issue = {4},
	journal = {SN Computer Science},
	month = {7},
	pages = {301},
	publisher = {Springer},
	title = {A Survey on Variational Autoencoders from a Green AI Perspective},
	volume = {2},
	url = {https://link.springer.com/10.1007/s42979-021-00702-9},
	year = {2021},
}

@article{Zeng2023,
	archivePrefix = {arXiv},
	arxivId = {2305.01090},
	author = {Zeng, Kevin and {De Jes{\'{u}}s}, Carlos E.P{\'{e}}rez and Fox, Andrew J. and Graham, Michael D.},
	doi = {10.1088/2632-2153/ad4ba5},
	eprint = {2305.01090},
	issn = {26322153},
	journal = {Machine Learning: Science and Technology},
	month = {may},
	number = {2},
	publisher = {Institute of Physics},
	title = {{Autoencoders for discovering manifold dimension and coordinates in data from complex dynamical systems}},
	url = {https://arxiv.org/pdf/2305.01090},
	volume = {5},
	year = {2023}
}

@article{Lopez2020,
	archivePrefix = {arXiv},
	arxivId = {2012.03448},
	author = {Lopez, Ryan and Atzberger, Paul J.},
	eprint = {2012.03448},
	issn = {16130073},
	journal = {CEUR Workshop Proceedings},
	month = {dec},
	publisher = {CEUR-WS},
	title = {{Variational Autoencoders for Learning Nonlinear Dynamics of Physical Systems}},
	url = {https://arxiv.org/pdf/2012.03448},
	volume = {2964},
	year = {2020}
}

@article{Williams2015,
	archivePrefix = {arXiv},
	arxivId = {1408.4408},
	author = {Williams, Matthew O. and Kevrekidis, Ioannis G. and Rowley, Clarence W.},
	doi = {10.1007/S00332-015-9258-5},
	eprint = {1408.4408},
	issn = {14321467},
	journal = {Journal of Nonlinear Science},
	month = {dec},
	number = {6},
	pages = {1307--1346},
	publisher = {Springer New York LLC},
	title = {{A Data–Driven Approximation of the Koopman Operator: Extending Dynamic Mode Decomposition}},
	url = {https://link.springer.com/article/10.1007/s00332-015-9258-5},
	volume = {25},
	year = {2015}
}

@article{Rowley2009,
	author = {Rowley, Clarence W. and Mezi, Igor and Bagheri, Shervin and Schlatter, Philipp and Henningson, Dan S.},
	doi = {10.1017/S0022112009992059},
	issn = {1469-7645},
	journal = {Journal of Fluid Mechanics},
	pages = {115--127},
	publisher = {Cambridge University Press},
	title = {{Spectral analysis of nonlinear flows}},
	url = {https://www.cambridge.org/core/journals/journal-of-fluid-mechanics/article/spectral-analysis-of-nonlinear-flows/311041E1027AE7FEE7DDA36AC9AD4270},
	volume = {641},
	year = {2009}
}

@article{Berkooz1993,
	author = {Berkooz, Gal and Holmes, Philip and Lumley, John L.},
	doi = {10.1146/ANNUREV.FL.25.010193.002543},
	issn = {00664189},
	journal = {Annual Review of Fluid Mechanics},
	month = {jan},
	number = {1},
	pages = {539--575},
	publisher = {Annual Reviews Inc.},
	title = {{The proper orthogonal decomposition in the analysis of turbulent flows}},
	volume = {25},
	year = {1993}
}

@article{Tu2014,
	archivePrefix = {arXiv},
	arxivId = {1312.0041},
	author = {Tu, Jonathan H. and Rowley, Clarence W. and Luchtenburg, Dirk M. and Brunton, Steven L. and Kutz, J. Nathan and Tu, Jonathan H. and Rowley, Clarence W. and Luchtenburg, Dirk M. and Brunton, Steven L. and Kutz, J. Nathan},
	doi = {10.3934/JCD.2014.1.391},
	eprint = {1312.0041},
	issn = {2158-2491},
	journal = {Journal of Computational Dynamics},
	month = {dec},
	number = {2},
	pages = {391--421},
	publisher = {Journal of Computational Dynamics},
	title = {{On dynamic mode decomposition:  Theory and applications}},
	url = {https://www.aimsciences.org/en/article/doi/10.3934/jcd.2014.1.391 https://www.aimsciences.org/en/article/doi/10.3934/jcd.2014.1.391?viewType=HTML https://www.aimsciences.org/article/doi/10.3934/jcd.2014.1.391},
	volume = {1},
	year = {2014}
}

@article{KarabiberCura2021,
	author = {{Karabiber Cura}, Ozlem and Akan, Aydin},
	doi = {10.1016/J.BBE.2020.11.002},
	issn = {0208-5216},
	journal = {Biocybernetics and Biomedical Engineering},
	month = {jan},
	number = {1},
	pages = {28--44},
	publisher = {Elsevier},
	title = {{Analysis of epileptic EEG signals by using dynamic mode decomposition and spectrum}},
	url = {https://www.sciencedirect.com/science/article/pii/S0208521620301303?utm_source=chatgpt.com},
	volume = {41},
	year = {2021}
}

@article{de2021modular,
  title={A modular approach for modeling the cell cycle based on functional response curves},
  author={De Boeck, Jolan and Rombouts, Jan and Gelens, Lendert},
  journal={PLoS computational biology},
  volume={17},
  number={8},
  pages={e1009008},
  year={2021},
  publisher={Public Library of Science San Francisco, CA USA}
}

@article{tyson2020dynamical,
  title={A dynamical paradigm for molecular cell biology},
  author={Tyson, John J and Novak, Bela},
  journal={Trends in cell biology},
  volume={30},
  number={7},
  pages={504--515},
  year={2020},
  publisher={Elsevier}
}

@article{Trapnell2014,
  title={The dynamics and regulators of cell fate decisions are revealed by pseudotemporal ordering of single cells},
  author={Trapnell, Cole and Cacchiarelli, Davide and Grimsby, Jonna and Pokharel, Prapti and Li, Shuqiang and Morse, Michael and Lennon, Niall J and Livak, Kenneth J and Mikkelsen, Tarjei S and Rinn, John L},
  journal={Nature Biotechnology},
  volume={32},
  number={4},
  pages={381--386},
  year={2014},
  publisher={Nature Publishing Group}
}

@article{Setty2016,
  title={Wishbone identifies bifurcating developmental trajectories from single-cell data},
  author={Setty, Manu and Tadmor, Michelle D and Reich-Zeliger, Shlomit and Angel, Oren and Salame, Tomer M and Kathail, Pooja and Choi, Karthik and Bendall, Sean and Friedman, Nir and Pe'er, Dana},
  journal={Nature Biotechnology},
  volume={34},
  number={6},
  pages={637--645},
  year={2016},
  publisher={Nature Publishing Group}
}

@article{Saelens2019,
  title={A comparison of single-cell trajectory inference methods},
  author={Saelens, Wouter and Cannoodt, Robrecht and Todorov, Helena and Saeys, Yvan},
  journal={Nature Biotechnology},
  volume={37},
  number={5},
  pages={547--554},
  year={2019}
}

@article{LaManno2018,
  title={RNA velocity of single cells},
  author={La Manno, Gioele and Soldatov, Ruslan and Zeisel, Amit and Braun, Emelie and Hochgerner, Hannah and Petukhov, Viktor and Lidschreiber, Michael and Kastriti, Maria E and L{\"o}nnerberg, Peter and Furlan, Alessandro and Fan, Jean and Borm, Lars E and Liu, Ziqi and van Bruggen, David and Guo, Jingyuan and He, Xinyi and Barker, Roger and Sundstr{\"o}m, Erik and Castelo-Branco, Goncalo and Cramer, Patrick and Adameyko, Igor and Linnarsson, Sten and Kharchenko, Peter V},
  journal={Nature},
  volume={560},
  number={7719},
  pages={494--498},
  year={2018}
}

@article{Schiebinger2019,
  title={Optimal-transport analysis of single-cell gene expression identifies developmental trajectories in reprogramming},
  author={Schiebinger, Geoffrey and Shu, Jian and Tabaka, Marcin and Cleary, Brian and Subramanian, Vikas and Solomon, Asha and Gould, Joshua and Liu, Shuxiao and Lin, Shixian and Berube, Paul and Lee, Leonard and Chen, Jina and Brumbaugh, Justin and Rigollet, Philippe and Hochedlinger, Konrad and Jaenisch, Rudolf and Regev, Aviv},
  journal={Cell},
  volume={176},
  number={4},
  pages={928--943.e22},
  year={2019}
}

@article{jansen2025danse,
  title={DANSE: a pipeline for dynamic modelling of time-series multi-omics data},
  author={Jansen Klomp, Lucas F and Yan, Xinqi and Snabel, Rebecca R and Veenstra, Gert Jan C and Meijer, Hil GE and Post, Janine N},
  journal={bioRxiv},
  pages={2025--07},
  year={2025},
  publisher={Cold Spring Harbor Laboratory}
}

@article{xu2025time,
  title={Time-series modeling with neural flow maps},
  author={Xu, Bingxian and Ho, Zoey E and Huang, Yitong},
  journal={bioRxiv},
  pages={2025--06},
  year={2025},
  publisher={Cold Spring Harbor Laboratory}
}

@article{cislo2025reconstructing,
  title={Reconstructing Waddington's Landscape from Data},
  author={Cislo, Dillon J and Del{\'a}s, M Joaquina and Briscoe, James and Siggia, Eric D},
  journal={bioRxiv},
  pages={2025--08},
  year={2025},
  publisher={Cold Spring Harbor Laboratory}
}

@article{fontaine2025dynamic,
  title={Dynamic landscape analysis of cell fate decisions: Predictive models of neural development from single-cell data},
  author={Fontaine, Marine and Delas, M Joaquina and Saez, Meritxell and Maizels, Rory J and Finnie, Elizabeth and Briscoe, James and Rand, David A},
  journal={bioRxiv},
  pages={2025--05},
  year={2025},
  publisher={Cold Spring Harbor Laboratory}
}

@article{tyson1985quantitative,
  title={A quantitative account of oscillations, bistability and traveling waves in the Belousov-Zhabotinsky reaction},
  author={Tyson, John J},
  journal={Oscillations and traveling waves in chemical systems},
  pages={93--144},
  year={1985},
  publisher={Wiley Interscience}
}

@article{field1974oscillations,
  title={Oscillations in chemical systems. IV. Limit cycle behavior in a model of a real chemical reaction},
  author={Field, Richard J and Noyes, Richard M},
  journal={The Journal of Chemical Physics},
  volume={60},
  number={5},
  pages={1877--1884},
  year={1974},
  publisher={American Institute of Physics}
}

@article{kolda2009tensor,
  title={Tensor decompositions and applications},
  author={Kolda, Tamara G and Bader, Brett W},
  journal={SIAM review},
  volume={51},
  number={3},
  pages={455--500},
  year={2009},
  publisher={SIAM}
}

@article{rombouts2025mechanistic,
  title   = {Mechanistic origins of temperature scaling in the early embryonic cell cycle},
  author  = {Rombouts, Jan and Tavella, Franco and Vandervelde, Alexandra and Phong, Connie and Ferrell, James E. Jr. and Yang, Qiong and Gelens, Lendert},
  journal = {Nature Communications},
  volume  = {16},
  pages   = {8045},
  year    = {2025},
  month   = aug,
  doi     = {10.1038/s41467-025-62918-0},
}

@article{stein2018enter,
  title={Enter the matrix: factorization uncovers knowledge from omics},
  author={Stein-O’Brien, Genevieve L and Arora, Raman and Culhane, Aedin C and Favorov, Alexander V and Garmire, Lana X and Greene, Casey S and Goff, Loyal A and Li, Yifeng and Ngom, Aloune and Ochs, Michael F and others},
  journal={Trends in Genetics},
  volume={34},
  number={10},
  pages={790--805},
  year={2018},
  publisher={Elsevier}
}

@article{devarajan2008nonnegative,
  title={Nonnegative matrix factorization: an analytical and interpretive tool in computational biology},
  author={Devarajan, Karthik},
  journal={PLoS computational biology},
  volume={4},
  number={7},
  pages={e1000029},
  year={2008},
  publisher={Public Library of Science San Francisco, USA}
}

@article{frigola2014variational,
  title={Variational Gaussian process state-space models},
  author={Frigola, Roger and Chen, Yutian and Rasmussen, Carl E},
  journal={Advances in neural information processing systems},
  volume={27},
  year={2014}
}

@phdthesis{duvenaud2014automatic,
author = {{Kristjanson Duvenaud}, David},
doi = {10.17863/CAM.14087},
month = {nov},
school = {University of Cambridge, Pembroke College},
title = {{Automatic model construction with Gaussian processes}},
url = {https://www.repository.cam.ac.uk/handle/1810/247281},
year = {2014}
}

@article{gao2008gaussian,
  title={Gaussian process modelling of latent chemical species: applications to inferring transcription factor activities},
  author={Gao, Pei and Honkela, Antti and Rattray, Magnus and Lawrence, Neil D},
  journal={Bioinformatics},
  volume={24},
  number={16},
  pages={i70--i75},
  year={2008},
  publisher={Oxford University Press}
}

@book{williams2006gaussian,
author = {Rasmussen, Carl Edward and Williams, Christopher K I},
doi = {10.7551/mitpress/3206.001.0001},
isbn = {9780262256834},
month = {nov},
publisher = {The MIT Press},
title = {{Gaussian Processes for Machine Learning}},
url = {https://doi.org/10.7551/mitpress/3206.001.0001 https://direct.mit.edu/books/book/2320/Gaussian-Processes-for-Machine-Learning},
year = {2005}
}

@incollection{suthaharan2016support,
  title={Support vector machine},
  author={Suthaharan, Shan},
  booktitle={Machine learning models and algorithms for big data classification: thinking with examples for effective learning},
  pages={207--235},
  year={2016},
  publisher={Springer}
}

@inproceedings{suykens2001nonlinear,
  title={Nonlinear modelling and support vector machines},
  author={Suykens, Johan AK},
  booktitle={IMTC 2001. proceedings of the 18th IEEE instrumentation and measurement technology conference. Rediscovering measurement in the age of informatics (Cat. No. 01CH 37188)},
  volume={1},
  pages={287--294},
  year={2001},
  organization={IEEE}
}

@article{zhang2023applications,
  title={Applications of transformer-based language models in bioinformatics: a survey},
  author={Zhang, Shuang and Fan, Rui and Liu, Yuti and Chen, Shuang and Liu, Qiao and Zeng, Wanwen},
  journal={Bioinformatics Advances},
  volume={3},
  number={1},
  pages={vbad001},
  year={2023},
  publisher={Oxford University Press}
}

@article{vaswani2017attention,
  title={Attention is all you need},
  author={Vaswani, Ashish and Shazeer, Noam and Parmar, Niki and Uszkoreit, Jakob and Jones, Llion and Gomez, Aidan N and Kaiser, {\L}ukasz and Polosukhin, Illia},
  journal={Advances in neural information processing systems},
  volume={30},
  year={2017}
}

@article{bronstein2021geometric,
  title={Geometric deep learning: Grids, groups, graphs, geodesics, and gauges},
  author={Bronstein, Michael M and Bruna, Joan and Cohen, Taco and Veli{\v{c}}kovi{\'c}, Petar},
  journal={arXiv preprint arXiv:2104.13478},
  year={2021}
}

@inproceedings{sanchez2018graph,
  title={Graph networks as learnable physics engines for inference and control},
  author={Sanchez-Gonzalez, Alvaro and Heess, Nicolas and Springenberg, Jost Tobias and Merel, Josh and Riedmiller, Martin and Hadsell, Raia and Battaglia, Peter},
  booktitle={International conference on machine learning},
  pages={4470--4479},
  year={2018},
  organization={PMLR}
}

@article{BBrunton2016,
	archivePrefix = {arXiv},
	arxivId = {1409.5496},
	author = {Brunton, Bingni W. and Johnson, Lise A. and Ojemann, Jeffrey G. and Kutz, J. Nathan},
	doi = {10.1016/J.JNEUMETH.2015.10.010},
	eprint = {1409.5496},
	issn = {0165-0270},
	journal = {Journal of Neuroscience Methods},
	month = {jan},
	pages = {1--15},
	pmid = {26529367},
	publisher = {Elsevier},
	title = {{Extracting spatial–temporal coherent patterns in large-scale neural recordings using dynamic mode decomposition}},
	url = {https://www.sciencedirect.com/science/article/pii/S0165027015003829?via%3Dihub},
	volume = {258},
	year = {2016}
}

@article{Balakrishnan2020,
	archivePrefix = {arXiv},
	arxivId = {2006.12726},
	author = {Balakrishnan, Shara and Hasnain, Aqib and Boddupalli, Nibodh and Joshy, Dennis M. and Egbert, Robert G. and Yeung, Enoch},
	doi = {10.23919/ACC45564.2020.9147230},
	eprint = {2006.12726},
	isbn = {9781538682661},
	issn = {07431619},
	journal = {Proceedings of the American Control Conference},
	month = {jul},
	pages = {3749--3756},
	publisher = {Institute of Electrical and Electronics Engineers Inc.},
	title = {{Prediction of fitness in bacteria with causal jump dynamic mode decomposition}},
	volume = {2020-July},
	year = {2020}
}

@article{Bourantas2014,
	author = {Bourantas, George C. and Ghommem, Mehdi and Kagadis, George C. and Katsanos, Konstantinos and Loukopoulos, Vassilis C. and Burganos, Vasilis N. and Nikiforidis, George C.},
	doi = {10.1118/1.4870976},
	issn = {00942405},
	journal = {Medical Physics},
	month = {may},
	number = {5},
	pages = {053301},
	pmid = {24784405},
	publisher = {John Wiley and Sons Ltd},
	title = {{Real-time tumor ablation simulation based on the dynamic mode decomposition method}},
	url = {/doi/pdf/10.1118/1.4870976 https://onlinelibrary.wiley.com/doi/abs/10.1118/1.4870976 https://aapm.onlinelibrary.wiley.com/doi/10.1118/1.4870976},
	volume = {41},
	year = {2014}
}

@article{Habibi2020,
	author = {Habibi, Milad and Dawson, Scott T. M. and Arzani, Amirhossein},
	doi = {10.3390/fluids5030111},
	issn = {2311-5521},
	journal = {Fluids},
	month = {jul},
	number = {3},
	pages = {111},
	publisher = {Multidisciplinary Digital Publishing Institute},
	title = {{Data-Driven Pulsatile Blood Flow Physics with Dynamic Mode Decomposition}},
	url = {https://www.mdpi.com/2311-5521/5/3/111/htm https://www.mdpi.com/2311-5521/5/3/111},
	volume = {5},
	year = {2020}
}

@article{Skantze2023,
	author = {Skantze, Viktor and Jirstrand, Mats and Brunius, Carl and Sandberg, Ann Sofie and Landberg, Rikard and Wallman, Mikael},
	doi = {10.3389/FNUT.2023.1304540},
	issn = {2296861X},
	journal = {Frontiers in Nutrition},
	month = {jan},
	pages = {1304540},
	publisher = {Frontiers Media SA},
	title = {{Data-driven analysis and prediction of dynamic postprandial metabolic response to multiple dietary challenges using dynamic mode decomposition}},
	url = {https://github.com/},
	volume = {10},
	year = {2023}
}

@article{Chen2012,
	author = {Chen, Kevin K. and Tu, Jonathan H. and Rowley, Clarence W.},
	doi = {10.1007/S00332-012-9130-9},
	issn = {09388974},
	journal = {Journal of Nonlinear Science},
	month = {dec},
	number = {6},
	pages = {887--915},
	publisher = {Springer},
	title = {{Variants of dynamic mode decomposition: Boundary condition, Koopman, and fourier analyses}},
	url = {https://link.springer.com/article/10.1007/s00332-012-9130-9},
	volume = {22},
	year = {2012}
}

@article{Hemati2017,
	author = {Hemati, Maziar S. and Rowley, Clarence W. and Deem, Eric A. and Cattafesta, Louis N.},
	doi = {10.1007/S00162-017-0432-2},
	issn = {14322250},
	journal = {Theoretical and Computational Fluid Dynamics},
	month = {aug},
	number = {4},
	pages = {349--368},
	publisher = {Springer New York LLC},
	title = {{De-biasing the dynamic mode decomposition for applied Koopman spectral analysis of noisy datasets}},
	url = {https://link.springer.com/article/10.1007/s00162-017-0432-2},
	volume = {31},
	year = {2017}
}

@article{Jiang2022,
	author = {Jiang, Wentao and Wu, Ziyu and Gao, Zheng and Wan, Mimi and Zhou, Min and Mao, Chun and Shen, Jian},
	doi = {10.1021/ACSNANO.2C06104/ASSET/IMAGES/LARGE/NN2C06104_0019.JPEG},
	issn = {1936086X},
	journal = {ACS Nano},
	month = {oct},
	number = {10},
	pages = {15705--15733},
	pmid = {36226996},
	publisher = {American Chemical Society},
	title = {{Artificial Cells: Past, Present and Future}},
	url = {/doi/pdf/10.1021/acsnano.2c06104?ref=article{\_}openPDF},
	volume = {16},
	year = {2022}
}

@article{Buddingh2017,
	author = {Buddingh', Bastiaan C. and {Van Hest}, Jan C.M.},
	doi = {10.1021/ACS.ACCOUNTS.6B00512},
	issn = {15204898},
	journal = {Accounts of Chemical Research},
	month = {apr},
	number = {4},
	pages = {769--777},
	pmid = {28094501},
	publisher = {American Chemical Society},
	title = {{Artificial Cells: Synthetic Compartments with Life-like Functionality and Adaptivity}},
	url = {/doi/pdf/10.1021/acs.accounts.6b00512?ref=article{\_}openPDF},
	volume = {50},
	year = {2017}
}

@article{Velten2023,
	author = {Velten, Britta and Stegle, Oliver},
	doi = {10.1038/s41592-023-01992-y},
	issn = {1548-7091},
	journal = {Nature Methods},
	month = {oct},
	number = {10},
	pages = {1462--1474},
	pmid = {37710019},
	publisher = {Nature Publishing Group},
	title = {{Principles and challenges of modeling temporal and spatial omics data}},
	url = {https://www.nature.com/articles/s41592-023-01992-y},
	volume = {20},
	year = {2023}
}

@article{Bressan2023,
	author = {Bressan, Dario and Battistoni, Giorgia and Hannon, Gregory J.},
	doi = {10.1126/SCIENCE.ABQ4964},
	issn = {10959203},
	journal = {Science},
	month = {aug},
	number = {6657},
	pmid = {37535749},
	publisher = {American Association for the Advancement of Science},
	title = {{The dawn of spatial omics}},
	url = {/doi/pdf/10.1126/science.abq4964},
	volume = {381},
	year = {2023}
}

@article{Specht2017,
	author = {Specht, Elizabeth A. and Braselmann, Esther and Palmer, Amy E.},
	doi = {10.1146/ANNUREV-PHYSIOL-022516-034055;},
	issn = {15451585},
	journal = {Annual Review of Physiology},
	month = {feb},
	pages = {93--117},
	pmid = {27860833},
	publisher = {Annual Reviews Inc.},
	title = {{A Critical and Comparative Review of Fluorescent Tools for Live-Cell Imaging}},
	volume = {79},
	year = {2017}
}

@article{Hickey2022,
	author = {Hickey, Shane M. and Ung, Ben and Bader, Christie and Brooks, Robert and Lazniewska, Joanna and Johnson, Ian R.D. and Sorvina, Alexandra and Logan, Jessica and Martini, Carmela and Moore, Courtney R. and Karageorgos, Litsa and Sweetman, Martin J. and Brooks, Douglas A.},
	doi = {10.3390/CELLS11010035;},
	issn = {20734409},
	journal = {Cells},
	month = {jan},
	number = {1},
	pmid = {35011596},
	publisher = {MDPI},
	title = {{Fluorescence microscopy—an outline of hardware, biological handling, and fluorophore considerations}},
	volume = {11},
	year = {2022}
}

@article{Balasubramanian2023,
	author = {Balasubramanian, Harikrushnan and Hobson, Chad M. and Chew, Teng-Leong and Aaron, Jesse S.},
	doi = {10.1038/s42003-023-05468-9},
	issn = {2399-3642},
	journal = {Communications Biology},
	month = {oct},
	number = {1},
	pages = {1096},
	pmid = {37898673},
	publisher = {Nature Publishing Group},
	title = {{Imagining the future of optical microscopy: everything, everywhere, all at once}},
	url = {https://www.nature.com/articles/s42003-023-05468-9},
	volume = {6},
	year = {2023}
}

@article{Iwata2020,
	archivePrefix = {arXiv},
	arxivId = {2012.06191},
	author = {Iwata, Tomoharu and Kawahara, Yoshinobu},
	doi = {10.3934/jcd.2022029},
	eprint = {2012.06191},
	issn = {21582505},
	journal = {Journal of Computational Dynamics},
	month = {dec},
	number = {2},
	pages = {268--280},
	publisher = {American Institute of Mathematical Sciences},
	title = {{Neural Dynamic Mode Decomposition for End-to-End Modeling of Nonlinear Dynamics}},
	url = {https://arxiv.org/pdf/2012.06191},
	volume = {10},
	year = {2020}
}

@article{Bakarji2022,
	archivePrefix = {arXiv},
	arxivId = {2202.04643},
	author = {Bakarji, Joseph and Callaham, Jared and Brunton, Steven L. and Kutz, J. Nathan},
	doi = {10.1038/s43588-022-00355-5},
	eprint = {2202.04643},
	issn = {26628457},
	journal = {Nature Computational Science},
	month = {feb},
	number = {12},
	pages = {834--844},
	publisher = {Springer Nature},
	title = {{Dimensionally Consistent Learning with Buckingham Pi}},
	url = {https://arxiv.org/pdf/2202.04643},
	volume = {2},
	year = {2022}
}

@article{Yuan2021,
	author = {Yuan, Bo and Shen, Ciyue and Luna, Augustin and Korkut, Anil and Marks, Debora S. and Ingraham, John and Sander, Chris},
	doi = {10.1016/j.cels.2020.11.013},
	issn = {24054712},
	journal = {Cell Systems},
	month = {feb},
	number = {2},
	pages = {128--140.e4},
	pmid = {33373583},
	publisher = {Cell Press},
	title = {{CellBox: Interpretable Machine Learning for Perturbation Biology with Application to the Design of Cancer Combination Therapy}},
	url = {https://www.cell.com/action/showFullText?pii=S2405471220304646 https://www.cell.com/action/showAbstract?pii=S2405471220304646 https://www.cell.com/cell-systems/abstract/S2405-4712(20)30464-6 http://www.ncbi.nlm.nih.gov/pubmed/33373583 https://linkinghub.elsevier.com/retrieve/pii/S2405471220304646},
	volume = {12},
	year = {2021}
}

@article{Kuwahara2024,
	author = {Kuwahara, Bruce and Bauch, Chris T.},
	doi = {10.1016/j.heliyon.2024.e25363},
	issn = {24058440},
	journal = {Heliyon},
	month = {feb},
	number = {4},
	pages = {e25363},
	publisher = {Elsevier},
	title = {{Predicting Covid-19 pandemic waves with biologically and behaviorally informed universal differential equations}},
	url = {https://www.sciencedirect.com/science/article/pii/S240584402401394X https://linkinghub.elsevier.com/retrieve/pii/S240584402401394X},
	volume = {10},
	year = {2024}
}

@article{Rojas-Campos2023,
    journal={Arxiv},
	archivePrefix = {arXiv},
	arxivId = {2310.16804},
	author = {Rojas-Campos, Adrian and Stelz, Lukas and Nieters, Pascal},
	eprint = {2310.16804},
	month = {oct},
	title = {{Learning COVID-19 Regional Transmission Using Universal Differential Equations in a SIR model}},
	url = {https://arxiv.org/pdf/2310.16804},
	year = {2023}
}

@article{Kana2021,
	author = {Kana, Omar and Bhattacharya, Sudin},
	doi = {10.1096/FASEBJ.2021.35.S1.03839},
	issn = {1530-6860},
	journal = {The FASEB Journal},
	month = {may},
	number = {S1},
	publisher = {John Wiley {\&} Sons, Ltd},
	title = {{Single Cell Universal Differential Equations: A Machine Learning Framework for Extracting Ordinary Differential Equations for Gene Regulatory Network Inference}},
	url = {https://onlinelibrary.wiley.com/doi/full/10.1096/fasebj.2021.35.S1.03839 https://onlinelibrary.wiley.com/doi/abs/10.1096/fasebj.2021.35.S1.03839 https://faseb.onlinelibrary.wiley.com/doi/10.1096/fasebj.2021.35.S1.03839},
	volume = {35},
	year = {2021}
}

@article{Buckner2024,
	archivePrefix = {arXiv},
    journal={Arxiv},
	arxivId = {2410.09233},
	author = {Buckner, Jack H. and Meunier, Zechariah D. and Arroyo-Esquivel, Jorge and Fitzpatrick, Nathan and Greiner, Ariel and McManus, Lisa C. and Watson, James R.},
	eprint = {2410.09233},
	month = {oct},
	title = {{Recovering complex ecological dynamics from time series using state-space universal dynamic equations}},
	url = {https://arxiv.org/pdf/2410.09233},
	year = {2024}
}

@article{Boddupalli2023,
	author = {Boddupalli, N. and Matchen, T. and Moehlis, J.},
	doi = {10.1063/5.0134464},
	issn = {1089-7682},
	journal = {Chaos (Woodbury, N.Y.)},
	month = {aug},
	number = {8},
	pmid = {38060788},
	publisher = {American Institute of Physics Inc.},
	title = {{Symbolic regression via neural networks.}},
	url = {/aip/cha/article/33/8/083150/2907845/Symbolic-regression-via-neural-networks http://www.ncbi.nlm.nih.gov/pubmed/38060788},
	volume = {33},
	year = {2023}
}

@article{AhmadiDaryakenari2024,
	archivePrefix = {arXiv},
	arxivId = {2310.01433},
	author = {{Ahmadi Daryakenari}, Nazanin and {De Florio}, Mario and Shukla, Khemraj and Karniadakis, George Em},
	doi = {10.1371/journal.pcbi.1011916},
	editor = {Fariselli, Piero},
	eprint = {2310.01433},
	isbn = {1111111111},
	issn = {1553-7358},
	journal = {PLOS Computational Biology},
	month = {mar},
	number = {3},
	pages = {e1011916},
	pmid = {38470870},
	publisher = {Public Library of Science},
	title = {{AI-Aristotle: A physics-informed framework for systems biology gray-box identification}},
	url = {https://journals.plos.org/ploscompbiol/article?id=10.1371/journal.pcbi.1011916 https://dx.plos.org/10.1371/journal.pcbi.1011916},
	volume = {20},
	year = {2024}
}

@article{Voina2024,
	archivePrefix = {arXiv},
    journal={Arxiv},
	arxivId = {2410.02079},
	author = {Voina, Doris and Brunton, Steven and Kutz, J. Nathan},
	eprint = {2410.02079},
	month = {oct},
	title = {{Deep Generative Modeling for Identification of Noisy, Non-Stationary Dynamical Systems}},
	url = {https://arxiv.org/pdf/2410.02079},
	year = {2024}
}

@article{Bakarji2023,
	archivePrefix = {arXiv},
	arxivId = {2201.05136},
	author = {Bakarji, Joseph and Champion, Kathleen and {Nathan Kutz}, J. and Brunton, Steven L.},
	doi = {10.1098/rspa.2023.0422},
	eprint = {2201.05136},
	issn = {1364-5021},
	journal = {Proceedings of the Royal Society A: Mathematical, Physical and Engineering Sciences},
	month = {aug},
	number = {2276},
	title = {{Discovering governing equations from partial measurements with deep delay autoencoders}},
	url = {http://arxiv.org/abs/2201.05136 https://royalsocietypublishing.org/doi/10.1098/rspa.2023.0422},
	volume = {479},
	year = {2023}
}

@article{Phillips2015,
	author = {Phillips, Rob},
	doi = {10.1016/j.tcb.2015.10.007},
	issn = {09628924},
	journal = {Trends in Cell Biology},
	month = {dec},
	number = {12},
	pages = {723--729},
	pmid = {26584768},
	publisher = {Elsevier Ltd},
	title = {{Theory in Biology: Figure 1 or Figure 7?}},
	url = {https://www.cell.com/action/showFullText?pii=S0962892415001944 https://www.cell.com/action/showAbstract?pii=S0962892415001944 https://www.cell.com/trends/cell-biology/abstract/S0962-8924(15)00194-4 https://linkinghub.elsevier.com/retrieve/pii/S0962892415001944},
	volume = {25},
	year = {2015}
}

@article{Paluch2015,
	author = {Paluch, Ewa K.},
	doi = {10.1016/j.tcb.2015.08.001},
	issn = {09628924},
	journal = {Trends in Cell Biology},
	month = {dec},
	number = {12},
	pages = {711--713},
	pmid = {26437593},
	publisher = {Elsevier Current Trends},
	title = {{After the Greeting: Realizing the Potential of Physical Models in Cell Biology}},
	url = {https://www.sciencedirect.com/science/article/pii/S0962892415001476?via%3Dihub https://linkinghub.elsevier.com/retrieve/pii/S0962892415001476},
	volume = {25},
	year = {2015}
}

@article{Mogilner2006,
	author = {Mogilner, Alex and Wollman, Roy and Marshall, Wallace F.},
	doi = {10.1016/j.devcel.2006.08.004},
	issn = {15345807},
	journal = {Developmental Cell},
	month = {sep},
	number = {3},
	pages = {279--287},
	pmid = {16950120},
	publisher = {Cell Press},
	title = {{Quantitative Modeling in Cell Biology: What Is It Good for?}},
	url = {https://www.sciencedirect.com/science/article/pii/S1534580706003509?via%3Dihub https://linkinghub.elsevier.com/retrieve/pii/S1534580706003509},
	volume = {11},
	year = {2006}
}

@article{Goldstein2018,
	archivePrefix = {arXiv},
	arxivId = {1807.11336},
	author = {Goldstein, Raymond E.},
	doi = {10.7554/eLife.40018},
	eprint = {1807.11336},
	issn = {2050-084X},
	journal = {eLife},
	month = {jul},
	pmid = {30033910},
	title = {{Are theoretical results ‘Results'?}},
	url = {https://elifesciences.org/articles/40018},
	volume = {7},
	year = {2018}
}

@article{Gunawardena2014,
	author = {Gunawardena, Jeremy},
	doi = {10.1186/1741-7007-12-29},
	issn = {1741-7007},
	journal = {BMC Biology},
	month = {dec},
	number = {1},
	pages = {29},
	pmid = {24886484},
	publisher = {BioMed Central Ltd.},
	title = {{Models in biology: ‘accurate descriptions of our pathetic thinking'}},
	url = {https://bmcbiol.biomedcentral.com/articles/10.1186/1741-7007-12-29},
	volume = {12},
	year = {2014}
}

@article{Shiguihara2021,
	author = {Shiguihara, Pedro and Lopes, Alneu De Andrade and Mauricio, David},
	doi = {10.1109/ACCESS.2021.3105520},
	issn = {2169-3536},
	journal = {IEEE Access},
	pages = {117639--117648},
	publisher = {Institute of Electrical and Electronics Engineers Inc.},
	title = {{Dynamic Bayesian Network Modeling, Learning, and Inference: A Survey}},
	url = {https://ieeexplore.ieee.org/document/9514854/},
	volume = {9},
	year = {2021}
}

@article{Kitson2023,
	archivePrefix = {arXiv},
	arxivId = {2109.11415},
	author = {Kitson, Neville Kenneth and Constantinou, Anthony C. and Guo, Zhigao and Liu, Yang and Chobtham, Kiattikun},
	doi = {10.1007/s10462-022-10351-w},
	eprint = {2109.11415},
	issn = {0269-2821},
	journal = {Artificial Intelligence Review},
	month = {aug},
	number = {8},
	pages = {8721--8814},
	publisher = {Springer Nature},
	title = {{A survey of Bayesian Network structure learning}},
	url = {https://link.springer.com/article/10.1007/s10462-022-10351-w https://link.springer.com/10.1007/s10462-022-10351-w},
	volume = {56},
	year = {2023}
}

@article{Cebrian-Lacasa2024,
	archivePrefix = {arXiv},
    journal = {arXiv},
	arxivId = {2412.16094},
	author = {Cebri{\'{a}}n-Lacasa, Daniel and Pi{\~{n}}eros, Liliana and Vanderbeke, Arno and Ruiz-Reyn{\'{e}}s, Daniel and Wouters, Thibeau and Goryachev, Andrew B. and Frolov, Nikita and Gelens, Lendert},
	eprint = {2412.16094},
	month = {dec},
	title = {{Spiral waves speed up cell cycle oscillations in the frog cytoplasm}},
	url = {https://arxiv.org/pdf/2412.16094 http://arxiv.org/abs/2412.16094},
	year = {2024}
}

@book{Brunton2022book,
  title={Data-driven science and engineering: Machine learning, dynamical systems, and control},
  author={Brunton, Steven L and Kutz, J Nathan},
  year={2022},
  publisher={Cambridge University Press}
}

@article{Winfree1984,
	author = {Winfree, A. T.},
	doi = {10.1021/ed061p661},
	issn = {0021-9584},
	journal = {Journal of Chemical Education},
	month = {aug},
	number = {8},
	pages = {661},
	publisher = {Division of Chemical Education},
	title = {{The prehistory of the Belousov-Zhabotinsky oscillator}},
	url = {/doi/pdf/10.1021/ed061p661?ref=article_openPDF https://pubs.acs.org/doi/abs/10.1021/ed061p661},
	volume = {61},
	year = {1984}
}

@article{Tyson1977,
author = {Tyson, John J.},
doi = {10.1063/1.433997},
issn = {0021-9606},
journal = {The Journal of Chemical Physics},
month = {feb},
number = {3},
pages = {905--915},
publisher = {American Institute of PhysicsAIP},
title = {{Analytic representation of oscillations, excitability, and traveling waves in a realistic model of the Belousov–Zhabotinskii reaction}},
url = {https://aip.scitation.org/doi/abs/10.1063/1.433997 http://aip.scitation.org/doi/10.1063/1.433997},
volume = {66},
year = {1977}
}

@article{Field1972,
	author = {Field, Richard J. and K\"{o}r\"{o}s, Endre and Noyes, Richard M.},
	doi = {10.1021/ja00780a001},
	issn = {15205126},
	journal = {Journal of the American Chemical Society},
	number = {25},
	pages = {8649--8664},
	title = {{Oscillations in Chemical Systems. II. Thorough Analysis of Temporal Oscillation in the Bromate–Cerium–Malonic Acid System}},
	volume = {94},
	year = {1972}
}

@article{Novak1993,
author = {Novak, Bela and Tyson, John J.},
doi = {10.1006/jtbi.1993.1179},
issn = {10958541},
journal = {Journal of Theoretical Biology},
month = {nov},
number = {1},
pages = {101--134},
publisher = {Academic Press},
title = {{Modeling the cell division cycle: M-phase trigger, oscillations, and size control}},
volume = {165},
year = {1993}
}

@article{Parra-Rivas2023,
author = {Parra-Rivas, Pedro and Ruiz-Reyn{\'{e}}s, Daniel and Gelens, Lendert},
doi = {10.1091/mbc.E22-11-0527},
editor = {Mogilner, Alex},
issn = {1059-1524},
journal = {Molecular Biology of the Cell},
month = {may},
number = {6},
publisher = {The American Society for Cell Biology},
title = {{Cell cycle oscillations driven by two interlinked bistable switches}},
url = {https://www.molbiolcell.org/doi/10.1091/mbc.E22-11-0527},
volume = {34},
year = {2023}
}

@article{Tyson2022,
	author = {Tyson, John J.},
	doi = {10.1042/BCJ20210370},
	issn = {0264-6021},
	journal = {Biochemical Journal},
	month = {jan},
	number = {2},
	pages = {185--206},
	pmid = {35098993},
	publisher = {Portland Press},
	title = {{From the Belousov–Zhabotinsky reaction to biochemical clocks, traveling waves and cell cycle regulation}},
	url = {/biochemj/article/479/2/185/230720/From-the-Belousov-Zhabotinsky-reaction-to https://portlandpress.com/biochemj/article/479/2/185/230720/From-the-Belousov-Zhabotinsky-reaction-to},
	volume = {479},
	year = {2022}
}

@book{epstein1998introduction,
  title={An introduction to nonlinear chemical dynamics: oscillations, waves, patterns, and chaos},
  author={Epstein, Irving R and Pojman, John A},
  year={1998},
  publisher={Oxford university press}
}

@book{tyson2013belousov,
  title={The belousov-zhabotinskii reaction},
  author={Tyson, John J},
  volume={10},
  year={2013},
  publisher={Springer Science \& Business Media}
}

@article{gyorgyi1990mechanistic,
  title={Mechanistic details of the oscillatory Belousov-Zhabotinskii reaction},
  author={Gyorgyi, Laszlo and Tur{\'a}nyi, Tam{\'a}s and Field, Richard J},
  journal={Journal of physical chemistry},
  volume={94},
  number={18},
  pages={7162--7170},
  year={1990},
  publisher={ACS Publications}
}

@article{rombouts2021dynamic,
  title={Dynamic bistable switches enhance robustness and accuracy of cell cycle transitions},
  author={Rombouts, Jan and Gelens, Lendert},
  journal={PLoS computational biology},
  volume={17},
  number={1},
  pages={e1008231},
  year={2021},
  publisher={Public Library of Science San Francisco, CA USA}
}

@article{tyson2015bistability,
  title={Bistability, oscillations, and traveling waves in frog egg extracts},
  author={Tyson, John J and Novak, Bela},
  journal={Bulletin of mathematical biology},
  volume={77},
  number={5},
  pages={796--816},
  year={2015},
  publisher={Springer}
}

@InProceedings{Takens1981,
author={Takens, Floris},
editor={Rand, David and Young, Lai-Sang},
title={{Detecting strange attractors in turbulence}},
booktitle={{Dynamical Systems and Turbulence, Warwick 1980}},
year={1981},
publisher={{Springer Berlin Heidelberg}},
address={{Berlin, Heidelberg}},
pages={{366--381}},
isbn={978-3-540-38945-3},
}

@article{Nolet2020,
author = {Nolet, Felix E. and Vandervelde, Alexandra and Vanderbeke, Arno and Pi{\~{n}}eros, Liliana and Chang, Jeremy B. and Gelens, Lendert},
doi = {10.7554/ELIFE.52868},
issn = {2050084X},
journal = {eLife},
month = {may},
pmid = {32452767},
publisher = {eLife Sciences Publications Ltd},
title = {{Nuclei determine the spatial origin of mitotic waves}},
volume = {9},
year = {2020}
}

@article{Newton1729,
  title={Sir Isaac Newton's Mathematical Principles of Natural Philosophy and His System of the World},
  journal={translated by Andrew Motte},
  author={Newton, Isaac and Cajori, Florian and Motte, Andrew},
  year={1729}
}

@article{Granger1969,
author = {Granger, C. W. J.},
doi = {10.2307/1912791},
issn = {00129682},
journal = {Econometrica},
month = {aug},
number = {3},
pages = {424},
publisher = {JSTOR},
title = {{Investigating Causal Relations by Econometric Models and Cross-spectral Methods}},
volume = {37},
year = {1969}
}

@article{Schmidt2009,
author = {Schmidt, Michael and Lipson, Hod},
doi = {10.1126/science.1165893},
issn = {0036-8075},
journal = {Science},
month = {apr},
number = {5923},
pages = {81--85},
pmid = {19342586},
publisher = {American Association for the Advancement of Science},
title = {{Distilling Free-Form Natural Laws from Experimental Data}},
url = {https://www.science.org/doi/abs/10.1126/science.1165893 https://www.science.org/doi/10.1126/science.1165893},
volume = {324},
year = {2009}
}

@article{Park2023,
author = {Park, Se Ho and Ha, Seokmin and Kim, Jae Kyoung},
doi = {10.1038/s41467-023-39983-4},
issn = {2041-1723},
journal = {Nature Communications},
month = {jul},
number = {1},
pages = {4287},
publisher = {Nature Publishing Group},
title = {{A general model-based causal inference method overcomes the curse of synchrony and indirect effect}},
url = {https://www.nature.com/articles/s41467-023-39983-4},
volume = {14},
year = {2023}
}

@article{Rackauckas2020,
archivePrefix = {arXiv},
arxivId = {2001.04385},
author = {Rackauckas, Christopher and Ma, Yingbo and Martensen, Julius and Warner, Collin and Zubov, Kirill and Supekar, Rohit and Skinner, Dominic and Ramadhan, Ali and Edelman, Alan},
eprint = {2001.04385},
month = {jan},
title = {{Universal Differential Equations for Scientific Machine Learning}},
url = {http://arxiv.org/abs/2001.04385},
year = {2020},
journal = {ArXiv},
doi={10.48550/arXiv.2001.04385}
}

@article{Cranmer2019,
   author = {Miles D. Cranmer and Rui Xu and Peter Battaglia and Shirley Ho},
   journal = {Preprint at ArXiv},
   month = {9},
   doi= {10.48550/arXiv.1909.05862},
   title = {{Learning Symbolic Physics with Graph Networks}},
   url = {http://arxiv.org/abs/1909.05862},
   year = {2019},
}

@article{Saint-Antoine2020,
archivePrefix = {arXiv},
arxivId = {1911.04046},
author = {Saint-Antoine, Michael M. and Singh, Abhyudai},
doi = {10.1016/j.copbio.2019.12.002},
eprint = {1911.04046},
issn = {09581669},
journal = {Current Opinion in Biotechnology},
month = {jun},
pages = {89--98},
pmid = {31927423},
publisher = {Elsevier Current Trends},
title = {{Network inference in systems biology: recent developments, challenges, and applications}},
url = {https://linkinghub.elsevier.com/retrieve/pii/S0958166919301399},
volume = {63},
year = {2020}
}

@article{Brunton2016,
	author = {Steven L Brunton and Joshua L Proctor and J Nathan Kutz},
	doi = {10.1073/pnas.1517384113},
	issn = {0027-8424},
	issue = {15},
	journal = {Proceedings of the National Academy of Sciences},
	month = {4},
	pages = {3932-3937},
	title = {{Discovering governing equations from data by sparse identification of nonlinear dynamical systems}},
	volume = {113},
	url = {https://www.pnas.org/content/pnas/113/15/3932.full.pdf https://www.ncbi.nlm.nih.gov/pmc/articles/PMC4839439 https://pnas.org/doi/full/10.1073/pnas.1517384113},
	year = {2016},
}

@article{Reinbold2021,
author = {Reinbold, Patrick A. K. and Kageorge, Logan M. and Schatz, Michael F. and Grigoriev, Roman O.},
doi = {10.1038/s41467-021-23479-0},
issn = {2041-1723},
journal = {Nature Communications},
month = {dec},
number = {1},
pages = {3219},
title = {{Robust learning from noisy, incomplete, high-dimensional experimental data via physically constrained symbolic regression}},
url = {http://www.nature.com/articles/s41467-021-23479-0},
volume = {12},
year = {2021}
}

@article{Reinbold2020,
author = {Reinbold, Patrick A. K. and Gurevich, Daniel R. and Grigoriev, Roman O.},
doi = {10.1103/PhysRevE.101.010203},
issn = {2470-0045},
journal = {Physical Review E},
month = {jan},
number = {1},
pages = {010203},
title = {{Using noisy or incomplete data to discover models of spatiotemporal dynamics}},
url = {https://link.aps.org/doi/10.1103/PhysRevE.101.010203},
volume = {101},
year = {2020}
}

@article{Hirsh2022,
author = {Hirsh, Seth M. and Barajas-Solano, David A. and Kutz, J. Nathan},
doi = {10.1098/rsos.211823},
issn = {2054-5703},
journal = {Royal Society Open Science},
month = {feb},
number = {2},
title = {{Sparsifying priors for Bayesian uncertainty quantification in model discovery}},
url = {https://royalsocietypublishing.org/doi/10.1098/rsos.211823},
volume = {9},
year = {2022}
}

@article{Fasel2022,
	author = {U. Fasel and J. N. Kutz and B. W. Brunton and S. L. Brunton},
	doi = {10.1098/rspa.2021.0904},
	issn = {1364-5021},
	issue = {2260},
	journal = {Proceedings of the Royal Society A: Mathematical, Physical and Engineering Sciences},
	month = {4},
	title = {{Ensemble-SINDy: Robust sparse model discovery in the low-data, high-noise limit, with active learning and control}},
	volume = {478},
	url = {https://royalsocietypublishing.org/doi/10.1098/rspa.2021.0904},
	year = {2022},
}

@article{Sandoz2023,
archivePrefix = {arXiv},
arxivId = {2208.10826},
author = {Sandoz, Antoine and Ducret, Verena and Gottwald, Georg A. and Vilmart, Gilles and Perron, Karl},
doi = {10.1098/RSPA.2022.0556},
eprint = {2208.10826},
issn = {14712946},
journal = {Proceedings of the Royal Society A},
month = {jan},
number = {2269},
publisher = {
The Royal Society
},
title = {{SINDy for delay-differential equations: application to model bacterial zinc response}},
url = {https://royalsocietypublishing.org/doi/10.1098/rspa.2022.0556},
volume = {479},
year = {2023}
}

@article{Anjos2023,
author = {dos Anjos, Lucas and Naozuka, Gustavo Taiji and Volpatto, Diego Tavares and Godoy, Wesley Augusto Conde and Costa, Michel Iskin da Silveira and Almeida, Regina C.},
doi = {10.1016/J.ECOINF.2023.102168},
issn = {1574-9541},
journal = {Ecological Informatics},
month = {nov},
pages = {102168},
publisher = {Elsevier},
title = {{A new modelling framework for predator-prey interactions: A case study of an aphid-ladybeetle system}},
volume = {77},
year = {2023}
}

@article{Gutierrez-Vilchis2023,
author = {Gutierrez-Vilchis, Abel and Perfecto-Avalos, Yocanxochitl and Garcia-Gonzalez, Alejandro},
doi = {10.1109/EMBC40787.2023.10341078},
issn = {26940604},
journal = {Annual International Conference of the IEEE Engineering in Medicine and Biology Society. IEEE Engineering in Medicine and Biology Society. Annual International Conference},
month = {jul},
pages = {1--4},
pmid = {38083503},
title = {{Modeling bacteria pairwise interactions in human microbiota by Sparse Identification of Nonlinear Dynamics (SINDy)}},
volume = {2023},
year = {2023}
}

@article{Messenger2024,
	author = {Messenger, Daniel and Dwyer, Greg and Dukic, Vanja},
	doi = {10.1098/RSIF.2024.0376},
	issn = {17425662},
	journal = {Journal of the Royal Society Interface},
	month = {dec},
	number = {221},
	publisher = {The Royal Society},
	title = {{Weak-form inference for hybrid dynamical systems in ecology}},
	url = {https://royalsocietypublishing.org/doi/10.1098/rsif.2024.0376},
	volume = {21},
	year = {2024}
}

@article{Mangan2016,
arxivId = {1605.08368},
author = {Mangan, Niall M. and Brunton, Steven L. and Proctor, Joshua L. and Kutz, J. Nathan},
doi = {10.1109/TMBMC.2016.2633265},
eprint = {1605.08368},
issn = {2372-2061},
journal = {IEEE Transactions on Molecular, Biological and Multi-Scale Communications},
month = {jun},
number = {1},
pages = {52--63},
publisher = {Institute of Electrical and Electronics Engineers Inc.},
title = {{Inferring Biological Networks by Sparse Identification of Nonlinear Dynamics}},
url = {http://ieeexplore.ieee.org/document/7809160/},
volume = {2},
year = {2016}
}

@article{Brummer2023,
author = {Brummer, Alexander B. and Xella, Agata and Woodall, Ryan and Adhikarla, Vikram and Cho, Heyrim and Gutova, Margarita and Brown, Christine E. and Rockne, Russell C.},
doi = {10.3389/fimmu.2023.1115536},
issn = {1664-3224},
journal = {Frontiers in Immunology},
month = {may},
pages = {1115536},
pmid = {37256133},
publisher = {Frontiers Media S.A.},
title = {{Data driven model discovery and interpretation for CAR T-cell killing using sparse identification and latent variables}},
url = {http://www.ncbi.nlm.nih.gov/pubmed/37256133 http://www.pubmedcentral.nih.gov/articlerender.fcgi?artid=PMC10226275 https://www.frontiersin.org/articles/10.3389/fimmu.2023.1115536/full},
volume = {14},
year = {2023}
}

@article{Zhang2019,
author = {Zhang, Linan and Schaeffer, Hayden},
doi = {10.1137/18M1189828},
issn = {1540-3459},
journal = {Multiscale Modeling and Simulation},
month = {jan},
number = {3},
pages = {948--972},
title = {{On the Convergence of the SINDy Algorithm}},
url = {https://epubs.siam.org/doi/10.1137/18M1189828},
volume = {17},
year = {2019}
}

@article{Delahunt2022,
archivePrefix = {arXiv},
arxivId = {2111.04870},
author = {Delahunt, Charles B. and Kutz, J. Nathan},
doi = {10.1109/ACCESS.2022.3159335},
eprint = {2111.04870},
issn = {21693536},
journal = {IEEE Access},
pages = {31210--31234},
publisher = {Institute of Electrical and Electronics Engineers Inc.},
title = {{A Toolkit for Data-Driven Discovery of Governing Equations in High-Noise Regimes}},
volume = {10},
year = {2022}
}

@article{Cortiella2023,
archivePrefix = {arXiv},
arxivId = {2201.12683},
author = {Cortiella, Alexandre and Park, K. C. and Doostan, Alireza},
doi = {10.1115/1.4054573},
eprint = {2201.12683},
issn = {1530-9827},
journal = {Journal of Computing and Information Science in Engineering},
month = {may},
number = {1},
pages = {1--34},
publisher = {American Society of Mechanical Engineers (ASME)},
title = {{A Priori Denoising Strategies for Sparse Identification of Nonlinear Dynamical Systems: A Comparative Study}},
url = {https://dx.doi.org/10.1115/1.4054573 http://arxiv.org/abs/2201.12683 https://asmedigitalcollection.asme.org/computingengineering/article/doi/10.1115/1.4054573/1141034/A-Priori-Denoising-Strategies-for-Sparse},
volume = {23},
year = {2022}
}

@article{Messenger2021,
archivePrefix = {arXiv},
arxivId = {2005.04339},
author = {Messenger, Daniel A. and Bortz, David M.},
doi = {10.1137/20M1343166},
eprint = {2005.04339},
issn = {1540-3459},
journal = {Multiscale Modeling \& Simulation},
month = {jan},
number = {3},
pages = {1474--1497},
publisher = {Society for Industrial and Applied Mathematics},
title = {{Weak SINDy: Galerkin-Based Data-Driven Model Selection}},
url = {https://epubs.siam.org/doi/10.1137/20M1343166},
volume = {19},
year = {2021}
}

@article{Raissi2019,
   author = {M. Raissi and P. Perdikaris and G.E. Karniadakis},
   doi = {10.1016/j.jcp.2018.10.045},
   issn = {00219991},
   journal = {Journal of Computational Physics},
   month = {2},
   pages = {686-707},
   title = {{Physics-informed neural networks: A deep learning framework for solving forward and inverse problems involving nonlinear partial differential equations}},
   volume = {378},
   url = {https://linkinghub.elsevier.com/retrieve/pii/S0021999118307125},
   year = {2019},
}

@article{Lagergren2020BINN,
   author = {John H. Lagergren and John T. Nardini and Ruth E. Baker and Matthew J. Simpson and Kevin B. Flores},
   doi = {10.1371/journal.pcbi.1008462},
   editor = {Inna Lavrik},
   isbn = {1111111111},
   issn = {1553-7358},
   issue = {12},
   journal = {PLOS Computational Biology},
   month = {12},
   pages = {e1008462},
   pmid = {33259472},
   publisher = {Public Library of Science},
   title = {{Biologically-informed neural networks guide mechanistic modeling from sparse experimental data}},
   volume = {16},
   url = {https://dx.plos.org/10.1371/journal.pcbi.1008462},
   year = {2020},
}

@article{vapnik1996support,
  title={Support vector method for function approximation, regression estimation and signal processing},
  author={Vapnik, Vladimir and Golowich, Steven and Smola, Alex},
  journal={Advances in neural information processing systems},
  volume={9},
  year={1996}
}

\onecolumngrid
\clearpage

\begin{center}
  \textbf{\large Supplementary Information}
\end{center}
\setcounter{section}{0}
\renewcommand{\thesection}{Supplementary Note~\arabic{section}}

% Reset figure/table counters and redefine them with 'S'
\setcounter{figure}{0}
\renewcommand{\thefigure}{S\arabic{figure}}

\setcounter{table}{0}
\renewcommand{\thetable}{S\arabic{table}}

% (Optional) For equations as well
\setcounter{equation}{0}
\renewcommand{\theequation}{S\arabic{equation}}

\setcounter{infobox}{0}
\renewcommand{\theinfobox}{S\Roman{infobox}}

\section{Literature-based benchmarking of data-driven dynamical methods}

As there is no unified benchmarking framework that allows one to directly compare different methods listed in Table~\ref{tab:data_driven_methods_overview}, we
performed a structured, literature-informed quantitative comparison of forecasting accuracy, interpretability, and dynamical characterization across methods, drawing on benchmarks reported in prior studies. 
Here we provide a list of literature, models used, and methods evaluated in Table~\ref{tab:overview_benchmarks}.

\begin{table*}[h]
\begin{ruledtabular}
\caption{Overview of literature providing quantitative benchmarks for methods listed in Table~\ref{tab:data_driven_methods_overview}.}
\label{tab:overview_benchmarks}
\begin{tabular}{l p{0.9cm} p{5cm} l}
\textbf{Manuscript} & \textbf{Year} & \textbf{Models / Data} & \textbf{Methods}\\
\hline
Herdenau \textit{et al.} & 2025 & Lorenz and other strange attractors from \textit{dysts} \cite{Gilpin2023} & GR \\
Hadrien \textit{et al.} &  & Lorenz, CSTR, other ODE systems & SINDy, DMD, RC, SVR \\
Kayshap \textit{et al.} & 2024 & Lorenz and other ODE systems & SINDy, GPR, SR, NN architectures \\
North \textit{et al.} & 2023 & ODE and PDE systems (e.g. Burgers equation) & SINDy, NeuralODE, PINN \\
Brum \textit{et al.} & 2025 & Lorenz and SIR models & SINDy, AI-Feynman, SR \\
Brunton, Kutz \textit{et al.} & 2016 & Lorenz, Lotka--Volterra, ODE and PDE systems & SINDy, DMD, eDMD, HAVOK, SINDy+AE \\
Strasser \textit{et al.} & 2025 & Lorenz system and robot arm dynamics & FFNN, RNN, NN architectures \\
Ahmad \textit{et al.} & 2025 & CSTR and chemical process models & RNN, (N)ARMAX \\
Chadebec \textit{et al.} & 2022 & ODE and PDE systems (Pythae library) & AE, VAE \\
Prokop \textit{et al.} & 2024 & FitzHugh--Nagumo and other ODE systems & SINDy and extensions \\
Prokop \textit{et al.} & 2025 & FitzHugh--Nagumo and other ODE systems & CLINE, SINDy, SR \\
Saelens \textit{et al.} & 2019 & Synthetic and real single-cell transcriptomics datasets & Pseudotime and trajectory inference methods (over 40 tools)\\

\end{tabular}
\end{ruledtabular}
\end{table*}

\section{Two illustrative oscillators in chemistry and biology}

\subsection*{The Belousov–Zhabotinsky (BZ) reaction and the Oregonator model}\label{eqn:BZ}
The Belousov–Zhabotinsky (BZ) reaction is a prototypical non-equilibrium chemical oscillator in which oxidation and reduction reactions between bromate, malonic acid, and metal-ion catalysts produce periodic color changes and traveling wave patterns \cite{Winfree1984,Field1972,gyorgyi1990mechanistic,tyson2013belousov,Tyson1977,epstein1998introduction}.  
The ``Oregonator'' model by Field and Noyes \cite{field1974oscillations}, later cast in dimensionless form by Tyson \cite{tyson1985quantitative}, distills the reaction mechanism into three interacting intermediates: $u = [\mathrm{HBrO_2}]$, $v = [\mathrm{Br^-}]$, and $w = [\mathrm{Ce^{4+}}]$.  
In compact form:
\begin{equation}
\begin{split}
u_t &= p \,[u(1-u) - v(u-q)], \\
v_t &= p'\,[uv + g v - 2f w], \\
w_t &= u - w,
\end{split}
\label{eq:oregonator_supp}
\end{equation}
where $p,\,p',\,q,\,f,\,g$ are nondimensional parameters governing rates of autocatalysis, inhibition, and catalyst recycling.  
This reduced model reproduces key experimental features, including relaxation oscillations and spiral waves in reaction–diffusion extensions \cite{morris2009belousov-zhabotinsky-reaction}.

\subsection*{The embryonic cell-cycle oscillator}\label{eqn:CCO}
In early embryonic cycles of \textit{Xenopus}, cyclin B is continuously synthesized and forms an active complex with cyclin-dependent kinase 1 (Cdk1), which triggers mitotic entry. Active Cdk1 then activates the anaphase-promoting complex/cyclosome (APC/C), promoting cyclin B degradation and driving mitotic exit \cite{newport1984regulation,king1996proteolysis,pomerening2005systems,yang2013cdk1}.  
This delayed negative feedback produces robust oscillations in cyclin B–Cdk1 activity, visible experimentally as biochemical and spatial mitotic waves \cite{chang2013mitotic,Nolet2020,afanzar2020nucleus,Puls2024,deneke2016waves,Cebrian-Lacasa2024}.  

A reduced model by Parra-Rivas \textit{et al.} \cite{Parra-Rivas2023} captures the essential feedback structure with three variables: $u = [\text{cyclin B}]$, $v = [\text{Cdk1}_a]$, and $w = [\text{APC/C}]$.  
\begin{align}
\begin{split}
u_t &= k_s - (d_1 + d_2 w)u, \\
v_t &= e_v^{-1}\!\left(\frac{s_u}{s_y}u - v + b_v s_v v^2 - s_v^2 v^3\right), \\
w_t &= e_w^{-1}\!\left(\frac{s_{uv}}{s_w}uv - w + b_w s_w w^2 - s_w^2 w^3\right),
\end{split}
\label{eq:CCO_supp}
\end{align}
where $k_s$ is the cyclin B synthesis rate, $d_1$ and $d_2$ are basal and APC/C-mediated degradation rates, and $e_v$, $e_w$ represent separation of timescales between fast (Cdk1, APC/C) and slow (cyclin) variables.  
Nonlinear terms ($b_v$, $b_w$) produce bistability, and the feedback $u \to v \to w \dashv u$ generates oscillations resembling experimental Cdk1 activity traces \cite{ferrell2011modeling,Novak1993,tyson2015bistability,de2021modular,rombouts2021dynamic}.  

\subsection*{Shared design principles}
Despite differences in molecular identity, the BZ and cell-cycle oscillators share universal dynamical motifs: (i) a negative feedback loop, (ii) a delay or slow intermediate, and (iii) nonlinear activation producing bistability or ultrasensitivity \cite{novak2008design,Tyson2022}.  
These minimal ingredients suffice to generate self-sustained oscillations, demonstrating how similar dynamical architectures can arise in both chemical and biological systems.

\section{Supplementary analyses of data-driven dynamical methods}
\subsection{Example of Koopman lifting}
\label{box:koopman_example}
% \refstepcounter{infobox}  
% \begin{tcolorbox}[float=h,  enhanced jigsaw,breakable,arc=2mm,boxrule=0.6pt,colback=blue!5!white,colframe=blue!75!black,  title={Box~\theinfobox: Example of Koopman linearization}]
  % \refstepcounter{infobox}  \label{box:koopman_example}

To illustrate how Koopman theory transforms a nonlinear system into a linear one by lifting with polynomial functions \cite{Iacob2023}, consider one of the simplest two-dimensional nonlinear system by  Ref.~\cite{Brunton2022}:
\begin{equation}
    \begin{split}
        x_{t,1} &= \mu x_1, \\
        x_{t,2} &= \delta (x_2 - x_1^2)
    \end{split}
    \label{eq:koopman_simple_system}
\end{equation}
with $\delta < \mu < 0$.  
The quadratic terms $x_1^2$ and $x_1$ prevent direct diagonalization in $(x_1, x_2)$ coordinates.  

If we augment the state with a new observable $y_3 = x_1^2$ and define $\mathrm{y} = (x_1, x_2, y_3)$, the system becomes
\begin{equation}
    \mathrm{y}_t =
    \begin{pmatrix}
        \mu & 0 & 0 \\
        0 & \delta & -\delta \\
        0 & 0 & 2\mu
    \end{pmatrix} \mathrm{y},
\end{equation}
which is fully linear in the augmented coordinates.  
Here, the nonlinear interaction has been absorbed into an extra dimension, revealing a structure that can be analyzed with standard linear tools.  
% \textbf{Perhaps we need to change this if this is straight out of the Brunton book?}
% \end{tcolorbox}
\subsection{A primer on regression}
\label{box:regression_primer}
% \begin{tcolorbox}[float=h,
%   enhanced jigsaw,breakable,arc=2mm,boxrule=0.6pt,
%   colback=blue!5!white,colframe=blue!75!black,
%   title={Box~\theinfobox: A primer on regression}]
  % \refstepcounter{infobox}\label{box:regression_primer}

\begin{figure*}[h]
\centering
\includegraphics[width=0.5\linewidth]{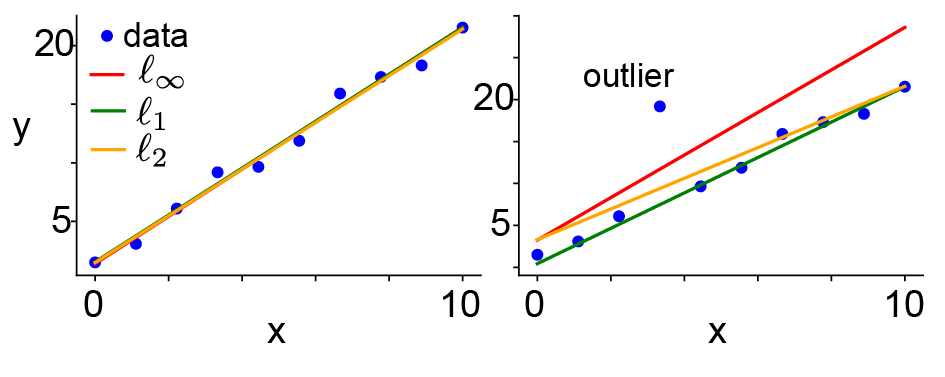}
\caption{\label{sfig:primer_regression}Fitting a linear model using different error norms to clean data and to data with an outlier.}
\end{figure*}

Regression methods minimize the mismatch between data and model predictions, typically using error norms such as  
\begin{align}
  E_{2} &= \left(\tfrac{1}{n}\sum_k |f(x_k) - y_k|^2\right)^{1/2} \quad \text{(least squares)},\\
  E_{1} &= \tfrac{1}{n}\sum_k |f(x_k) - y_k| \quad \text{(mean absolute error)},\\
  E_{\infty} &= \max_k |f(x_k) - y_k| \quad \text{(maximum error)} .
\end{align}

Each metric emphasizes different aspects: $\ell_2$ minimizes average error and is most common; $\ell_1$ is more robust to outliers; $\ell_\infty$ targets worst-case deviations, relevant in control.  

For example, fitting $f(x_k) = \beta_2 x_k + \beta_1$ to clean data yields similar fits across all norms (see Fig~\ref{sfig:primer_regression}).  
With an outlier, $\ell_2$ is distorted, $\ell_\infty$ focuses on the outlier, and $\ell_1$ remains comparatively robust.  
Thus, regression outcomes depend not only on model structure $f$ but also on the chosen error metric, which balances accuracy, robustness, and worst-case guarantees.

% \end{tcolorbox}
\subsection{Sparse regression with SINDy applied to Oregonator data}
\label{box:SINDy_oregonator}
% \begin{tcolorbox}[float=h,
%   enhanced jigsaw,breakable,arc=2mm,boxrule=0.6pt,
%   colback=blue!5!white,colframe=blue!75!black,
%   title={ Box~\theinfobox: Polynomial regression with SINDy applied to oregonator data}]

%\captionsetup{type=figure,labelformat=empty}%
\begin{figure*}[h]
\centering
  \includegraphics[width=0.9\linewidth]{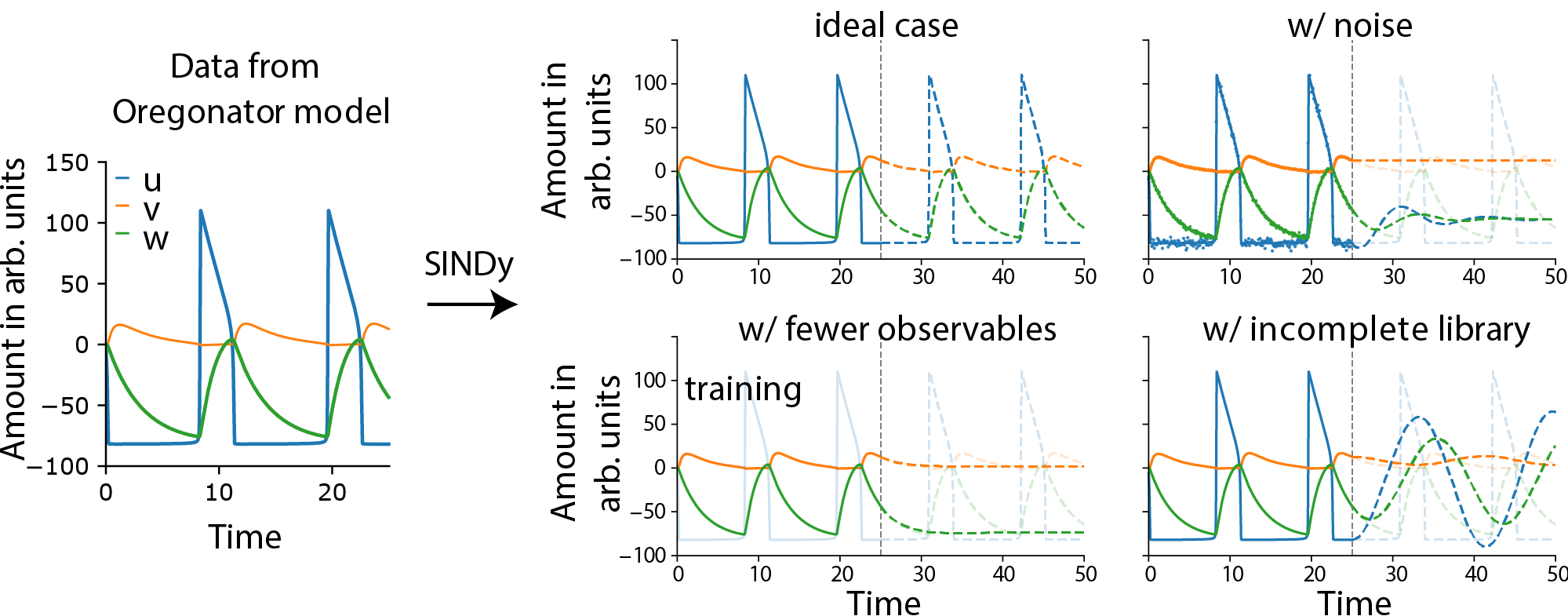}
  \caption{\label{sfig:eval_sindy} Applying the SINDy method to data from Oregonator model. When SINDy is applied to high-quality and complete data, SINDy successfully identifies the underlying model equation. However, when facing realistic scenarios such as noise in measured data, access to fewer observables or an incomplete term library, SINDy struggles to capture the underlying dynamics.}
\end{figure*}

To illustrate the opportunities and pitfalls of regression-based approaches, we examine the Sparse Identification of Nonlinear Dynamics (SINDy) method on data from the Oregonator model of the Belousov–Zhabotinsky reaction~\cite{Tyson1977} (see Fig.~\ref{sfig:eval_sindy}). This system is a widely used testbed because it is nonlinear, oscillatory, and has a known ground-truth model.

\textbf{Ideal conditions.} We first simulate the Oregonator with high temporal resolution ($\Delta t=0.001$) and no added noise. The term library contains all monomials up to order two, $\Theta = [u, v, w, u^2, v^2, w^2, uv, uw, vw]$,
and we use ridge regression with a cutoff $\xi_{\text{thres}}=0.01$. Under these conditions, SINDy exactly recovers the original model:
\begin{align}
	\begin{split}
		u_t &= -42.241 u -8198.303 v -0.500 u^2 -99.979 uv,\\
		v_t &= -0.592 u -83.999 v + 0.600 w  -1.000 uv, \\
		w_t &= 0.333 u -0.333 w,
	\end{split}
\end{align}
and simulations of the inferred equations reproduce the true dynamics beyond the training data range.

\textbf{Adding noise.} With 5\% Gaussian noise, identification fails: the inferred equations no longer reproduce the oscillations. The reason is that SINDy performs regression on the estimated derivatives $X_t$, which strongly amplifies measurement noise in $X$. Numerous strategies have been proposed to mitigate this issue, including smoothed differentiation schemes and noise-aware extensions of SINDy~\cite{Delahunt2022,Cortiella2023,Prokop2024,Messenger2021,Fasel2022}. Nevertheless, high-quality, temporally resolved data remain a strict requirement in practice.

\textbf{Partial observability.} In real experiments, not all system variables are measurable. For the BZ reaction, only $v$ and $w$ are typically accessible. Restricting SINDy to these two variables with a second-order library fails to reproduce the oscillatory dynamics. Even when extending the library up to the 8\textsuperscript{th} order, the method cannot approximate the true dynamics. From a Koopman perspective, this failure occurs because the necessary lifting functions are missing: without the unobserved state $u$, the system cannot be properly linearized in observable space.

\textbf{Incomplete libraries.} Finally, when all three state variables are available but the library only includes first-order terms $\Theta = [u,v,w]$, the inferred model again fails. Without nonlinear terms, the algorithm lacks the expressive power to approximate the Koopman operator.

\subsection{A primer on neural networks}
\label{box:nn_basics}
% \begin{tcolorbox}[float=h,
%   enhanced jigsaw,breakable,arc=2mm,boxrule=0.6pt,
%   colback=blue!5!white,colframe=blue!75!black,
%   title={Box~\theinfobox: A primer on neural networks}]
%   label=box:nn_basics
% ]
\begin{figure*}[h]
\centering
\includegraphics[width=0.4\linewidth]{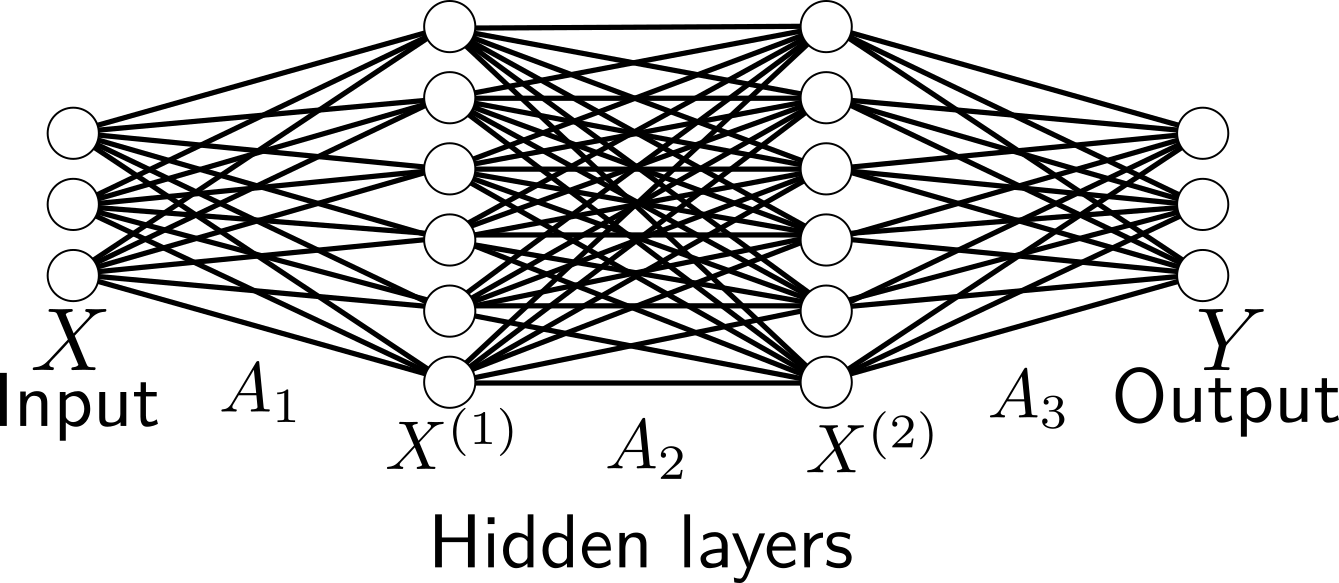}

\caption{\label{sfig:multilayernn}Example of a multilayer FFNN with an input and output layer with hidden nonlinearly activated layers in between.}
\end{figure*}

The basic architecture of a neural network (NN) consists of input nodes, hidden layers, and output nodes. Here we show an example of a fully connected feed-forward neural network with two hidden layers (see Fig~\ref{sfig:multilayernn}).

In the linear case, each layer corresponds to a matrix multiplication:
\begin{subequations}
\begin{align}
X^{(1)} &= A_1 X, \\
X^{(2)} &= A_2 X^{(1)}, \\
Y &= A_3 X^{(2)},
\end{align}
\end{subequations}
leading to the composite mapping
\begin{equation}
Y = A_3 A_2 A_1 X, 
\end{equation}
or, more generally,
\begin{equation}
Y = \prod_{j=1}^l A_j X.
\end{equation}
Since the composition of linear functions is linear, such networks cannot represent nonlinear systems.  
Nonlinear activation functions overcome this limitation:
\begin{subequations}
\begin{align}
f_a(x) &= \frac{1}{1+e^{-x}} && \text{(sigmoid)}, \\
f_a(x) &= \tanh(x) && \text{(tanh)}, \\
f_a(x) &= \max(0,x) && \text{(ReLU)}.
\end{align}
\end{subequations}
With these, the network becomes a nested nonlinear function:
\begin{equation}
Y = f_{a,l}\big(A_l \cdot f_{a,l-1}(A_{l-1} \cdots f_{a,1}(A_1 X))\big).
\end{equation}

Training consists of optimizing the weights $A_l$ by minimizing a loss function, typically
\begin{equation}
\mathcal{L} = \left\|\textit{NN}(X) - Y\right\|^2_2,
\end{equation}
with regularization (e.g., $\ell_1$, $\ell_2$) to prevent overfitting.  
The universal approximation theorem guarantees that sufficiently large NNs can approximate any continuous function \cite{Hornik1989}.  

\subsection{A feed-forward neural network to approximate Oregonator data}
\label{box:ffnn_example}
% \begin{tcolorbox}[float=h,
%   enhanced jigsaw,breakable,arc=2mm,boxrule=0.6pt,
%   colback=blue!5!white,colframe=blue!75!black,
%   title={Box~\theinfobox: A feed-forward neural network to approximate Oregonator data}
% ]

\begin{figure*}[h]
\centering
\includegraphics[width=0.9\linewidth]{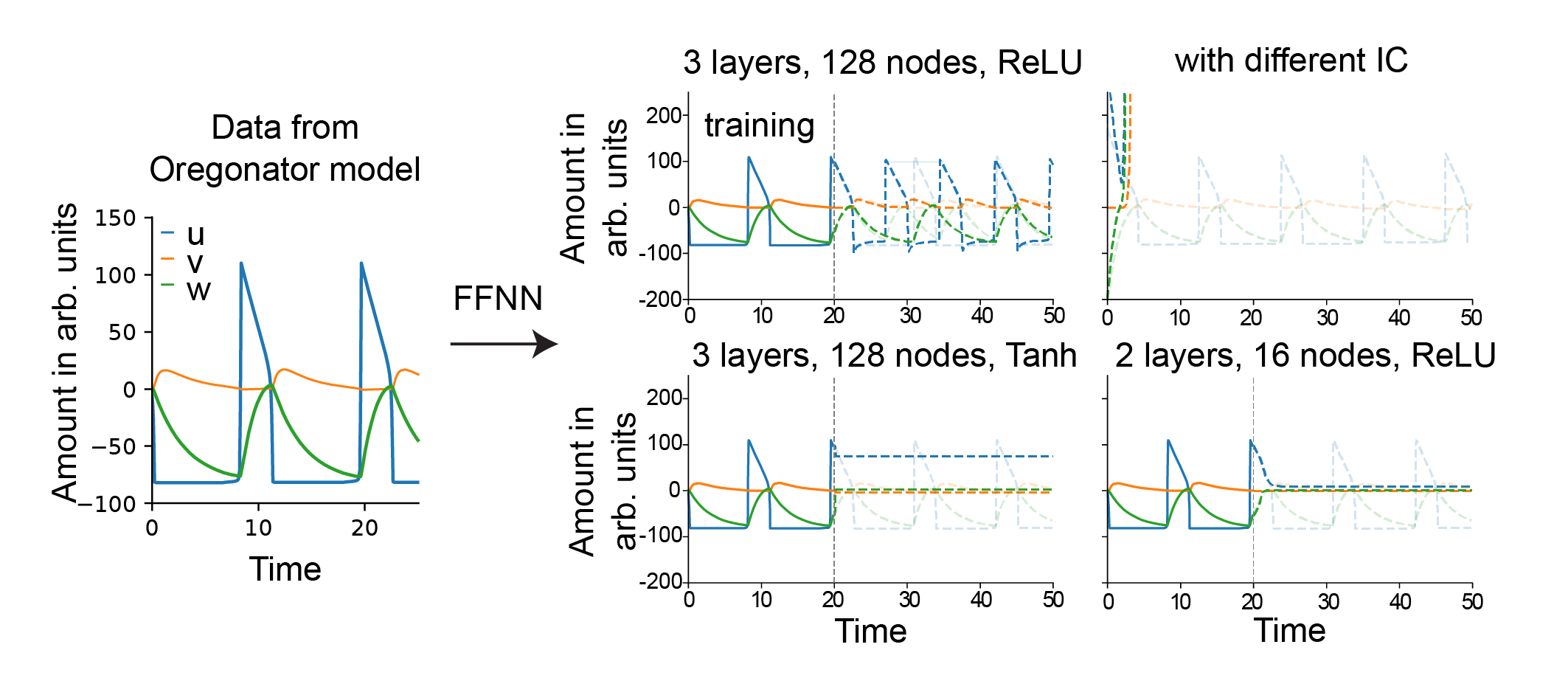}
\caption{\label{sfig:FFNN_Oregonator} Using a FFNN to capture dynamical behavior of the Oregonator model. When high-quality data is used to train the FFNN, it is able to reproduce the dynamics. However, this success is non-generalizable as it fails when facing unknown initial conditions and architecture dependent.}
\end{figure*}

To illustrate opportunities and pitfalls of FFNNs, we train a network on synthetic data from the Oregonator model aiming to reproduce the oscillatory dynamics (see Fig.~\ref{sfig:FFNN_Oregonator}). 

\textbf{Ideal conditions.} Training a network with three hidden layers (128 nodes each, ReLU activation) on high-resolution data from the Oregonator model we are able to reproduce the qualitative dynamical behavior Within the training domain.

\textbf{Different initial conditions.} However, it fails to generalize to unseen initial conditions that lie outside the training domain. 
Thus, the network is only able to reproduce the dynamics within a narrow set of conditions, requiring additional measurements to capture the general dynamics of a system.

\textbf{Different architectural choices.} Additionally, using different architectural choices affects the ability of the network to reproduce the dynamics. 
Using a different activation energy (ReLU to Tanh) or reducing the size of the network disallows to capture any of the oscillatory features of the Oregonator.

\section{Explicit Koopman operator inference using eDMD}
\label{box:edmd_example}
% \begin{tcolorbox}[float=h,
%   enhanced jigsaw,breakable,arc=2mm,boxrule=0.6pt,
%   colback=blue!5!white,colframe=blue!75!black,
%   title={Box~\theinfobox: Extended DMD on Oregonator data}
% ]

\begin{figure*}[h]
\includegraphics[width=0.9\linewidth]{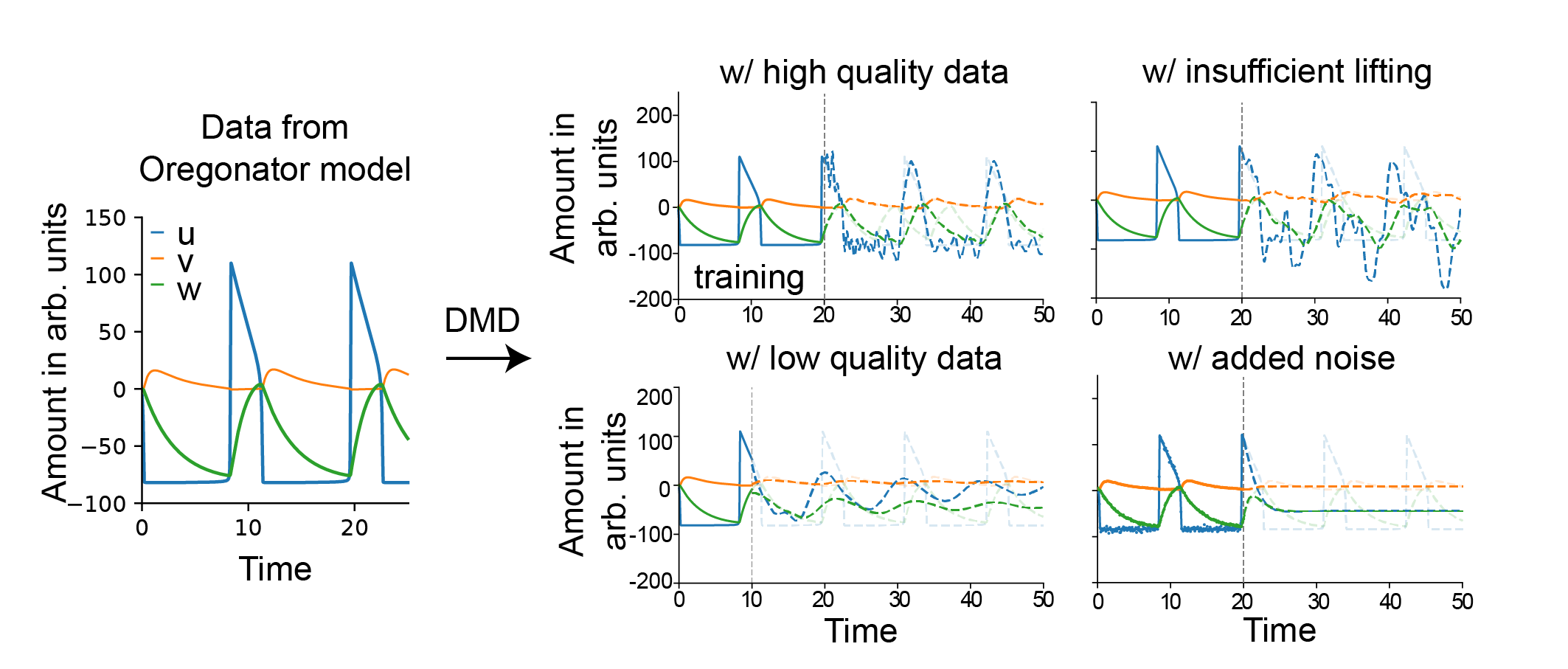}
\caption{\label{sfig:dmd_oregonator}Capturing the dynamics of the Oregonator using DMD. When data quality is high and sufficient lifting is applied, the DMD is able to reproduce the oscillation in the Oregonator model. However, when insufficient lifting is applied, data-quality decreased or noise added to data, DMD fails to recreate the dynamics.}
\end{figure*}

We evaluate decomposition approaches on Oregonator trajectories using eDMD evaluating its ability to reproduce the oscillatory behavior (see Fig.~\ref{sfig:dmd_oregonator}).

\textbf{Ideal conditions.} When applying eDMD to high-resolution data covering multiple periods and sufficient lifting (polynomial lifting up to sixth order with 67 dimensions used), eDMD captures the qualitative dynamical behavior.

\textbf{Insufficient lifting.} However, when not enough lifting functions are provided, e.g. with polynomial lifting up to 3 order that includes all terms included in the Oregonator model, eDMD fails to reproduce the oscillations.

\textbf{Low quality data.} When applying eDMD on just a few periods with low resolution, a scenario often encountered when observing biological systems, and despite a sufficient amount of lifting dimensions, eDMD fails. 

\textbf{Added noise.} The ability to reproduce dynamics with eDMD is limited when providing noisy, high-quality data with the exact same set up as in ideal conditions. The introduction of noise hinders the identification of relevant dynamical modes in the measured data.

\end{document}